%% file: main.tex
\documentclass[twocolumn, secnumarabic, nobibnotes, aps, prx
, superscriptaddress]{revtex4-2}

\usepackage[utf8]{inputenc}
\usepackage[T1]{fontenc}
\usepackage{amsmath, amssymb, amsfonts, amsthm}
\usepackage{hyperref}
\usepackage{cleveref}
\usepackage{tikz}
\usepackage{booktabs}
\usepackage{subcaption}
\usepackage{csquotes}
\usepackage[linesnumbered,ruled,vlined]{algorithm2e}
\usepackage{physics}
\usepackage{makecell}
\usepackage{enumitem}

\begin{document}

\title{Superconducting qubits in the millions: the potential and limitations of modularity}

\author{S. N. Saadatmand}
\email[Corresponding author: ]{nsaadatmand@rigetti.com}
\affiliation{Rigetti Computing, 775 Heinz Avenue, Berkeley, California 94710, USA}

\author{Tyler L. Wilson}
\affiliation{Rigetti Computing, 775 Heinz Avenue, Berkeley, California 94710, USA}

\author{Mark J. Hodson}
\affiliation{Rigetti Computing, 775 Heinz Avenue, Berkeley, California 94710, USA}

\author{Mark Field}
\affiliation{Rigetti Computing, 775 Heinz Avenue, Berkeley, California 94710, USA}

\author{Simon J. Devitt}
\affiliation{InstituteQ, 02150 Espoo, Finland}
\affiliation{Centre for Quantum Software and Information, University of Technology Sydney, Sydney, NSW 2007, Australia}

\author{Madhav Krishnan Vijayan}
\affiliation{Centre for Quantum Software and Information, University of Technology Sydney, Sydney, NSW 2007, Australia}

\author{Alan Robertson}
\affiliation{Centre for Quantum Software and Information, University of Technology Sydney, Sydney, NSW 2007, Australia}

\author{Thinh P. Le}
\affiliation{Centre for Quantum Software and Information, University of Technology Sydney, Sydney, NSW 2007, Australia}

\author{Jannis Ruh}
\affiliation{Centre for Quantum Software and Information, University of Technology Sydney, Sydney, NSW 2007, Australia}

\author{Alexandru Paler}
\affiliation{Aalto University, 02150 Espoo, Finland}
\affiliation{InstituteQ, 02150 Espoo, Finland}

\author{Arshpreet Singh Maan}
\affiliation{Aalto University, 02150 Espoo, Finland}
\affiliation{InstituteQ, 02150 Espoo, Finland}

\author{Ioana Moflic}
\affiliation{Aalto University, 02150 Espoo, Finland}
\affiliation{InstituteQ, 02150 Espoo, Finland}

\author{Athena Caesura}
\affiliation{Zapata AI Inc., Boston, MA 02110 USA}

\author{Josh Y. Mutus}
\affiliation{Rigetti Computing, 775 Heinz Avenue, Berkeley, California 94710, USA}

\begin{abstract}
The development of fault-tolerant quantum computers (FTQCs) is receiving increasing attention within the quantum computing community. Like conventional digital computers, FTQCs, which utilize error correction and millions of physical qubits, have the potential to address some of humanity's grand challenges. However, accurate estimates of the tangible scale of future FTQCs, based on transparent assumptions, are uncommon. How many physical qubits are necessary to solve a practical problem intractable for classical hardware? What costs arise from distributing quantum computation across multiple machines? This paper presents an architectural model of a potential FTQC based on superconducting qubits, divided into discrete modules and interconnected via coherent links. We employ a resource estimation framework and software tool to assess the physical resources required to execute specific quantum algorithms compiled into their graph-state form and arranged onto a modular superconducting hardware architecture. Our tool can predict the size, power consumption, and execution time of these algorithms based on explicit assumptions about the system's physical layout, thermal load, and modular connectivity. We assess the resources needed for quantum computation examples that serve as building blocks of proposed applications, quantifying the architectural bottlenecks and trade-offs that remain to be addressed to deliver utility.
\end{abstract}

\date{\today}
\maketitle

\section{Introduction}

Recent studies suggest that future idealized FTQCs with 1-10 million physical qubits can address classically intractable problems in integer factoring \cite{Gidney2025factoring} and molecular chemistry \cite{Lee_2021,Goings_2022}, completing their assigned tasks in a matter of days. At the same time, the size, connectivity, and fidelity of today's noisy physical qubit devices are continuously improving, and now support some of the fundamental building blocks of fault tolerance \cite{deLeon2021,Field2023,Acharya2025}. These emerging platforms employ quantum error correction (QEC) codes \cite{Fowler2012,Campbell2017,Litinski2019magicstate,Litinski2019GOSC,Webster2022,Acharya2023,Ustun2024,Acharya2025} to reduce logical errors at the cost of additional resources: more space, meaning physical qubits, and more time needed to operate the code.

The complexity of fault-tolerant (FT) algorithms, including Shor's factoring \cite{Shor1997}, quantum signal processing \cite{low2017optimal} (QSP), and ground state energy estimation \cite{Lin2022}, is well understood. However, research on the magnitude and scaling of the physical resources required to execute such computations is limited. Furthermore, the overhead and space-time resource requirements for physical hardware are even less well understood and depend on platform and micro-architecture choices. Scaling tangible resources such as energy, space, and time, based on specific assumptions about a potential machine, is crucial for evaluating the feasibility of methods for developing FTQC.

Fault-tolerant resource estimation tools currently exist, but are restricted to using the 
\enquote{$T$-counting} method \cite{Litinski2019GOSC, paler2019opensurgery, LatticeSurgery2022, beverland2022, azure_estimation} only. These footprint-style estimation tools synthesize the logical circuit into Clifford+$T$ gates (or other universal gate sets), rescale the space-time volume according to the desired level of parallelization, and derive physical space-time quantities based on the $T$-count and circuit depth, along with assumptions regarding $T$-distillation
and hardware timescales. This methodology lacks targeted estimates specific to Quantum Processing Unit (QPU) layouts, provides insufficient estimates for surface code distance, and may overestimate the upper limits of space-time resources.

Here, we present a transparent, FT architectural model for square-lattice superconducting-qubit systems that uses the surface code~\cite{Fowler2012, Litinski2019GOSC, mcewen2023relaxing, forlivesi2023, Orourke2024}. This model supports the essential elements of FTQC: logical Clifford gates via lattice-surgery-based approaches~\cite{Horsman2012-ua, Brown2017, Litinski2018latticesurgery, Fowler2019, Litinski2019GOSC}, magic- or $T$-state distillation (MSD) \cite{Litinski2019magicstate, Litinski2019GOSC, Gidney2024cultivation} for non-Clifford gates, and queuing and routing of specific resources. We also establish explicit assumptions about the size of a single module and the penalties for executing an FT algorithm that exceeds its capacity, allowing us to quantify the impact of coherently distributing a quantum computation.

We also present application-specific resource estimates for the proposed FT superconducting qubit architecture. These estimates are derived from a graph-state method detailed in \cref{sec:graph-formalism}. This framework aims to assist the design and feasibility assessment of next-generation FTQCs while quantifying the trade-offs associated with hardware implementations. We also discuss an accompanying open-source software called Rigetti Resource Estimations~\cite{RRE} (RRE), which we used to generate all the results presented in 
\cref{sec:results} for a selection of test algorithms outlined in 
\cref{sec:test_cases}. We will explore potential future work in 
\cref{sec:future} and provide a conclusion in 
\cref{sec:next_steps}.

\section{A primer for algorithm execution, architecture designs, and resource estimation on FTQCs}
\label{sections/graph-state-architecture}
\input{sections/primer-intro}

\subsection{Mathematical description of the graph-state formalism}
\label{sec:graph-formalism}
\input{sections/graph-formalism}

\subsection{High-level macro-architecture for modular graph state processing}
\label{sec:logical-architecture}
\input{sections/logical-architecture}

\subsection{Superconducting hardware micro-architecture and lower-level components}
\label{sec:hardware-architecture}
\input{sections/hardware-architecture}

\subsection{Modular graph-state-based compilation and resource estimations}
\label{sections/graph-state-compilation}
\input{sections/graph-state-compilation}

\section{Test algorithms}
\label{sec:test_cases}
\input{sections/testcases}

\section{Resource estimation results}
\label{sec:results} 
\input{sections/results}

\section{Future work}
\label{sec:future}
\input{sections/future}

\section{Conclusion}
\label{sec:next_steps} 

This paper proposes a fault-tolerant architecture based on modular quantum processing units that uses superconducting qubits, the surface code, and coherent inter-module connections. We provide quantitative estimates for the scale and runtime of various quantum computations. Our software tool compiles large-scale quantum circuits into a graph state, schedules them on a model of our logical architecture, and, from this, estimates the physical resources and time required for test cases derived from the existing literature. This model-driven approach delineates \textit{how} a superconducting platform could accomplish the target computation, including details on the scheduled execution of graph state preparation and consumption, as well as hardware-level justifications for some of the most critical size and power considerations. We have demonstrated that such a fault-tolerant quantum computer could perform scientifically interesting computations, such as a $20\times20$ Fermi-Hubbard time evolution, using 5.2M physical qubits in less than two days. However, the value of such a computation remains an open question. When utilized in the context of a complete high-temperature superconductivity study, the number of computations and their probability of success increase the estimated computation time to the order of years \cite{Agrawal2024}. More work is needed to co-optimize this FTQC architecture with algorithms of interest, with distributed computation across multiple modules presenting as the major bottleneck in this study.

This work has enabled research toward the development of real quantum computing platforms that provide the advantages of fault-tolerant operations. It serves as a software testbed for evaluating the impact of intrinsic algorithmic, compilation, decoding, and physical-layer proposals, as well as system-architecture proposals. Moreover, this allows for \emph{direct} per-algorithm comparison of different modalities by assessing run times, spatial, and energy resource costs. Rapid iteration in a simulated environment is a proven method for accelerating technological development, and we hope that our efforts to develop this tooling contribute positively to this endeavor.

\section{Acknowledgments}

The views, opinions, and/or findings expressed are those of the author(s). They should not be interpreted as representing the official views or policies of the Department of Defense or the U.S. Government. This research was developed with funding from the Defense Advanced Research Projects Agency under Agreement HR00112230006. Jannis Ruh was supported by the Sydney Quantum Academy, Sydney, NSW, Australia.

\section{DATA AVAILABILITY}

The data that supports the findings of this article cannot be made publicly available. The data are available upon reasonable request from the authors. Codes, scripts, and sample configuration files required to reproduce similar results are accessible via the RRE software \cite{RRE}.

\bibliography{references.bib}

\appendix

\section{Glossary of default RRE outputs}
\input{sections/RRE-outputs}
\label{supp:RRE-outputs} 

\section{Widgetization via Subcircuit Dependency Graph}
\input{sections/widgetization}
\label{supp:decomp}

\section{A unitariness verification protocol for RRE outputs}
\input{sections/unitary-verification}
\label{supp:unitary-verification}

\section{Noise modeling and logical error rate scaling}
\label{supp:noise-and-scaling}
\input{sections/noise-and-scaling}

\end{document}

%% file: sections/primer-intro.tex
The execution of algorithms on an FTQC fundamentally differs from merely executing the gates of a quantum circuit in the order they appear. By definition, FTQCs depend on the underlying QEC scheme they use - see, for example, \cite{Fowler2012,Litinski2019magicstate,Litinski2019GOSC,Acharya2023,Ustun2024,Acharya2025}. FTQCs employ noisy physical qubits, encoded within an error-correction scheme, to create logical qubits with exponentially suppressed error rates. 

The error-correcting scheme we consider here is the surface code \cite{Fowler2012,Litinski2019GOSC,mcewen2023relaxing,forlivesi2023}, implemented on square arrays of physical qubits. In an array, physical qubits are categorized as \textit{data} or \textit{ancilla} qubits, with each ancilla qubit coupled to four data qubits and vice versa. Physical ancilla qubits enable repeated composite parity measurements on their four neighboring data qubits in the form $X_1X_2X_3X_4$ or $Z_1Z_2Z_3Z_4$. These composite parity measurements are performed by entangling the physical data qubits with the ancilla via \textit{CNOT}s and measuring the ancilla in either the Pauli $X$ or $Z$ basis. Physical errors on the physical qubits in these arrays result in patterns of bit flips among the parity measurements, known as \textit{syndromes}. Tracking the outcomes of the parity measurements and the resulting error syndromes, and mapping them to the most probable error event, is known as \textit{decoding}. In practice, corrections can be applied after decoding through judicious syndrome tracking and software corrections after running the algorithm (potentially incurring time delays due to decoder latency).

One price we pay for this encoding is that the resulting logical qubits can perform only a limited set of FT operations, specifically the Clifford set. At least one non-Clifford operation must be engineered into the FTQC to achieve universality and execute arbitrary algorithms fault-tolerantly. We consider the $T$ gates (a $\pi/4~R_z$ rotation) as the non-Clifford operations choice in this paper. However, forcing an error-correcting code to produce an operation outside its native subspace (such as a $T$ gate) involves significant overhead, necessitating the dedication of substantial portions of the FTQC to MSD or, more recently, cultivation \cite{Litinski2019magicstate,Litinski2019GOSC,Gidney2024cultivation} to produce $T$ gates.

\begin{figure}[!h]
    \centering
    \includegraphics[width=1.0\linewidth]{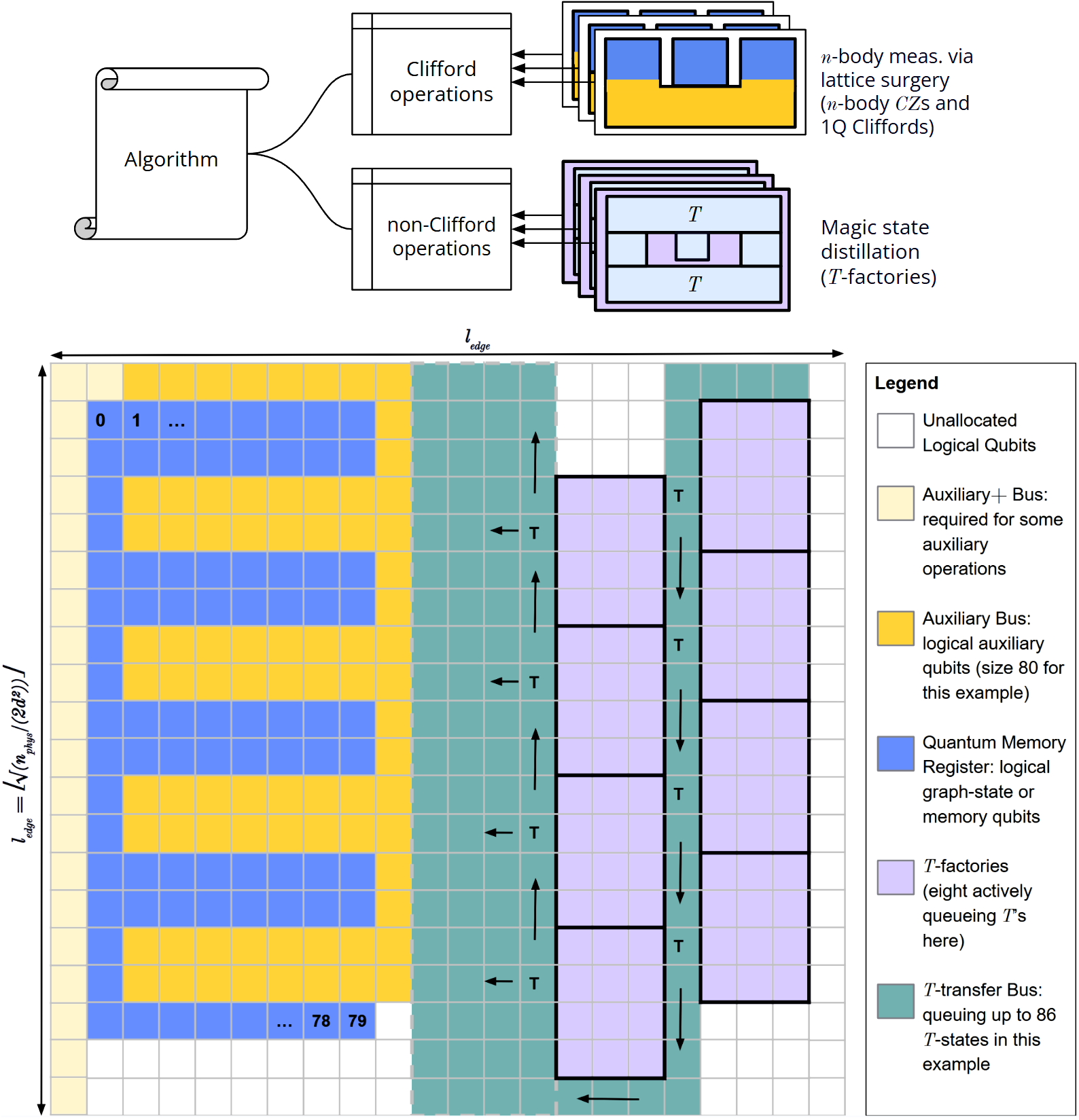}
    \caption{(\textit{top}) Fault-tolerant algorithm execution involves segmenting the application into logical Clifford and non-Clifford portions. Clifford operations are performed using a set of measurements on logical memory qubits, facilitated by auxiliary buses. Because measurement outcomes are probabilistic, classical feedback and feedforward are incorporated into operation schedules. Non-Clifford portions are implemented using $T$-states distilled by $T$-factories. (\textit{bottom}) This results in our proposed logical micro-architecture, detailed in \cref{sec:hardware-architecture}. The architecture includes graph preparation, consumption, local $T$-factories, $T$-state and Bell-state queuing, with some parallelization. We arrange logical qubits on a square lattice of rotated surface code patches of distance-$d$ made of physical qubits. A left portion of the module is allocated for the logical memory (blue tiles). It also contains an \enquote{auxiliary bus} of logical auxiliary qubits needed for graph preparation, consumption, $T$, and Bell-state routing. Yellow tiles represent the auxiliary bus, followed by lighter-yellow \enquote{auxiliary$+$ bus} required for routing. To its left, a separate bus is designated for continuous queuing of $T$-states distilled in an array of $T$-factories on the opposite side of the module (teal tiles) — see also Parameters 6 and 8 of \cref{supp:RRE-outputs}. This architecture can be distributed across multiple modules, provided they are coherently connected via the auxiliary bus and each module consumes local $T$-states.}
    \label{fig:log-layout}
\end{figure}

Due to the constraints imposed by the error-correcting code, we ultimately decompose the algorithm into a sequence of Clifford+$T$ gates. This separation of functional operations leads to a specialization of the fundamental architecture: certain sections of the FTQC are dedicated solely to performing the original logical Clifford operations through \textit{m}ulti-\textit{p}roduct (-qubit) \textit{P}auli measurement \textit{o}perations or MPPOs — a process based on the lattice surgery~\cite{Horsman2012-ua, Brown2017, Litinski2018latticesurgery, Fowler2019, Litinski2019GOSC}. For example, MPPOs can be understood as applying $n$-body $CZ$ gates and single-qubit Clifford gates to the quantum state, thereby leading to a formalism that we later explore. Additionally, there are designated areas for generating and queuing $T$ states produced via MSD. Our proposed logical layout for this operational model is presented schematically in \cref{fig:log-layout}.

We adapt this logical-operation model and the fixed yet extensible micro-architecture of \cref{fig:log-layout} to a physical layout of superconducting qubits with square-lattice connectivity. We estimate the execution of selected algorithms on this architecture. To this end, we generate schedules of quantum operations using a graph-state-based approach, as proposed in \cite{hein2004multiparty, anders2006fast, vijayan2024, Ruh2024, Liu2023, Raussendorf2001, raussendorf2001computational, raussendorf2003measurement} and further refined below. This approach yields several abstractions that facilitate distributing the algorithm across fixed hardware. Specifically, we can manipulate graphs or subgraphs to tailor the overall algorithm's execution within an FTQC composed of distinct yet interconnected modules.

One can implement the original quantum algorithm 
using the algorithm-specific graph (ASG) formalism~\cite{Paler2017,vijayan2024,Bowen2025} and associated schedules of operations~\cite{Liu2023,Ruh2024}, which apply only 
logical measurements and classical feedforward. The ASG formalism is a resource-efficient variant of the Measurement-Based Quantum Computing (MBQC) framework~\cite{Raussendorf2001,NIELSEN2006}. MBQC is particularly appealing for distributed FTQC because it requires only an algorithm-specific resource, such as the graph states and measurements on logical qubits. We will explore and utilize a specific form of MBQC as outlined in \cite{vijayan2024} for compilation purposes. We prepare the graph-state resource using lattice-surgery-based processes to entangle logical qubits in a memory register, facilitated by a single bus of logical auxiliary qubits. Due to the probabilistic nature of quantum measurements, we must track the set of measurement results classically using a process called Pauli frame tracking \cite{Knill2005,Paler2014,Riesebos2017,Ruh2024}. Next, we move locally-distilled $T$-states, through patch deformations\cite{Litinski2019GOSC,Horsman2012-ua}, to the register of logical memory qubits to apply the non-Clifford portion in conjunction with the Pauli Frame corrections of the algorithm. Finally, we apply any corrections flagged by syndromes tracked by the decoder, resulting from physical errors in the hardware. We present a mathematical description of the ASG formalism in \cref{sec:graph-formalism}. 

We need to arrange the elements of the graph-state-based algorithm execution across physical components at scale. In large-scale classical computations, there are limits to the amount of computation that can be performed on a single processor within a given time frame. These limits stem from the processing power that can be integrated into a single chip, ultimately constrained by physical limits and manufacturing yields. Consequently, computations are often distributed across multiple processors and machines. This distribution incurs penalties, such as higher latency or reduced bandwidth, but it enables greater overall computational power. We can draw an analogy to distributed FTQC, which involves a network of interconnected compute modules. Manufacturing yields limit the quantum computing power of each module. These operational (thermal) and manufacturing constraints restrict the number of physical qubits that can be incorporated into a single QPU housed within a refrigeration module. This prompted us to propose a logical macro-architecture composed of several modules, as described in \cref{sec:logical-architecture}. 

In contrast to the flexible connectivity and relatively slow speeds between compute modules, QPUs based on superconducting qubits offer fixed connectivity and faster gate speeds. In \cref{sec:hardware-architecture}, we describe the hardware micro-architecture, explain how it can be customized to accommodate specific hardware constraints, and provide a physical rationale for fixing the number of physical qubits in a single compute module to be around one million. We propose sparsely and coherently interconnecting the modules, which may result in reduced fidelity, lower bandwidth, or slower entangling gate speeds. We will refer to the \enquote{coherent interconnect} between modules with the surface code distance of $d$ as a single \enquote{pipe.} We will abstract the transduction efficiency as an interaction time, characterized by the \enquote{intermodule tock,} and quantify how this time impacts the algorithm's overall runtime. We assume that the pipes share identical noise characteristics and code distance with the rest of FTQC.

We present our compilation and estimation methodology for RRE, based on user inputs, in \cref{section:t-gate-error}. Further details on the RRE output glossary, our circuit decomposition strategy, and the verification of RRE outputs are provided separately in \cref{supp:RRE-outputs}, \cref{supp:decomp}, and \cref{supp:unitary-verification}, respectively.

Setting aside the technical details of our proposed logical layouts, it is important to note that the fixed intra-module micro-architecture in 
\cref{fig:log-layout} and the inter-module macro-architecture in 
\cref{sec:logical-architecture} were not tailored to any specific application. Instead, we adapted the overall architecture to the strengths and limitations of our superconducting hardware and the graph-state-based algorithm execution. We will then evaluate its performance on selected applications. In 
\cref{fig:log-layout}, the bilinear pattern in the memory and auxiliary buses arises from the ease of scheduling the full FT algorithm using graph states. $T$-factories are kept local to modules and connected via another bus primarily because practical algorithms require a large number of $T$ operations, and coherent interconnects are slow for $T$-teleportations. However, we assume that coherent interconnects are effective for less frequent Bell-state-based teleportation of the graph state. This is why we propose the distributed macro-architecture described in 
\cref{sec:logical-architecture}; the graph state would not fit into a single module for practical algorithms. Future design improvements could include alternative memory and bus structures, such as city-style patch layouts and more efficient distinct QEC codes for memories, to reduce reliance on many coherent interconnects, which may be more suitable for certain applications, as suggested by our results in 
\cref{sec:results}.

%% file: sections/graph-formalism.tex
A mathematical graph, denoted by $\mathcal{G}$, is defined by its vertices and edges. Given a graph, $\mathcal{G}$, the associated quantum (Clifford) state, called a \textit{graph state}, is constructed by assigning each vertex in $\mathcal{G}$ to a logical qubit initialized to $\ket{+}$ and each edge in $\mathcal{G}$ to an entangling $CZ$ operation that links the logical qubits at the corresponding vertices. We often refer to the logical qubit associated with a vertex as a \enquote{node} and use \enquote{graph} and \enquote{graph state} interchangeably for brevity. The nodes undergo non-Pauli-basis measurements corresponding to the original circuit's non-Clifford gates and some $X$-basis measurements to facilitate teleporting unknown inputs into the graph state \cite{vijayan2024}. 

The ASG formalism provides measurement schedules for two main stages to fully execute the FT algorithm: one schedule details the preparation of the graph through lattice surgery (corresponding to the original nontrivial Cliffords, first stage, in \cref{fig:log-layout}(\emph{top})) and another outlines the necessary single-qubit measurements in non-Clifford bases (corresponding to the original non-Cliffords, second stage, in \cref{fig:log-layout}(\emph{top})). We also need to track Pauli frame corrections in the software. We denote the nested set of stabilizer measurement schedules that \enquote{prepare} the graph in the first stage as $S^{\text{prep}}_\mathcal{G}$ and refer to it as the preparation schedule for the graph state. $S^{\text{prep}}_\mathcal{G}$ specifies how to initialize $\mathcal{G}$, the graph state, based on the original entangling operations and other step-by-step requirements~\cite{Liu2023,scheduler}.
Conversely, we denote the nested set of non-Clifford measurement schedules that \enquote{consume} the graph state in the second stage as $S^{\text{consump}}_\mathcal{G}$ and refer to it as the consumption schedule. $S^{\text{consump}}_\mathcal{G}$ specifies how to perform logical $R_Z(\theta)$-basis measurements (for arbitrary non-Clifford angle $\theta$) relevant to executing the original non-Cliffords~\cite{Ruh2024,jabalizer,Cabaliser}.
Both schedules consist of subsets or subschedules that specify the order of measurements, with each subset containing the nodes designated for simultaneous measurements \cite{scheduler,jabalizer,vijayan2024}. Furthermore, the essential stages of MSD, routing $T$-states, and decoding can occur at either stage or be postponed, depending on the requirements.

At a high level, the ASG formalism requires the FT algorithm to be ingested as single- and two-qubit gates. We employ a straightforward method for decomposing, compiling, and executing the algorithm, segmenting it into equal-width, recurring subcircuits along the time axis. These subcircuits are called \enquote{widgets,} have their own (sub)graphs, and reconstruct the original algorithm when \enquote{stitched} together (see also~\cref{section:t-gate-error}). We refer to this decomposition process as \enquote{widgetization,} detailed in \cref{supp:decomp}. We will examine the graph properties of these widgets and the scheduling process for preparing and consuming them on the proposed hardware architecture.

%% file: sections/logical-architecture.tex
We consider a macro-architectural model that, at its core, uses superconducting qubits on square lattices and the logical layout shown in \cref{fig:log-layout}(bottom). This enables us to estimate the resources required to execute the FT algorithm and understand the associated trade-offs. To this end, we consider a physical layout for FTQC consisting of sets of \textit{rotated} surface code patches, which are tiled together to create all intra-module components. For example, a group of patches forms a graph-state memory register, while others are designated for the $T$-transfer bus. We assume that each patch contains $2d^2$ physical qubits. Although the rotated surface code requires only $2d^2-1$ physical qubits~\cite{Litinski2019GOSC,mcewen2023relaxing,forlivesi2023,Orourke2024}, we adopt a physical size of $2d^2$ for all patches to streamline our approximate scaling calculations, ignoring non-dominant spatial terms and tiling details.

Executing the quantum algorithm based on the ASG formalism and time-direction widgetization inherently yields a macro-architecture that partitions the hardware into sets of modules that execute widgets in an interleaved fashion. The question is how to lay out modules and assign them to potentially interleaving operations. In steady state, the execution of the quantum algorithm comprises \textit{three} main steps: 1) preparation or creation of the graph, which initializes a quantum state using the original nontrivial Clifford and entangling operations, 2) consumption of the graph, which executes the original non-Cliffords, and 3) following every consumption step, the handover of the graph through Bell-pair teleportations or by stitching output nodes from the preceding widget to the input nodes of the current widget.

\begin{figure}[!h]
    \centering
    \includegraphics[width=1.0\linewidth]{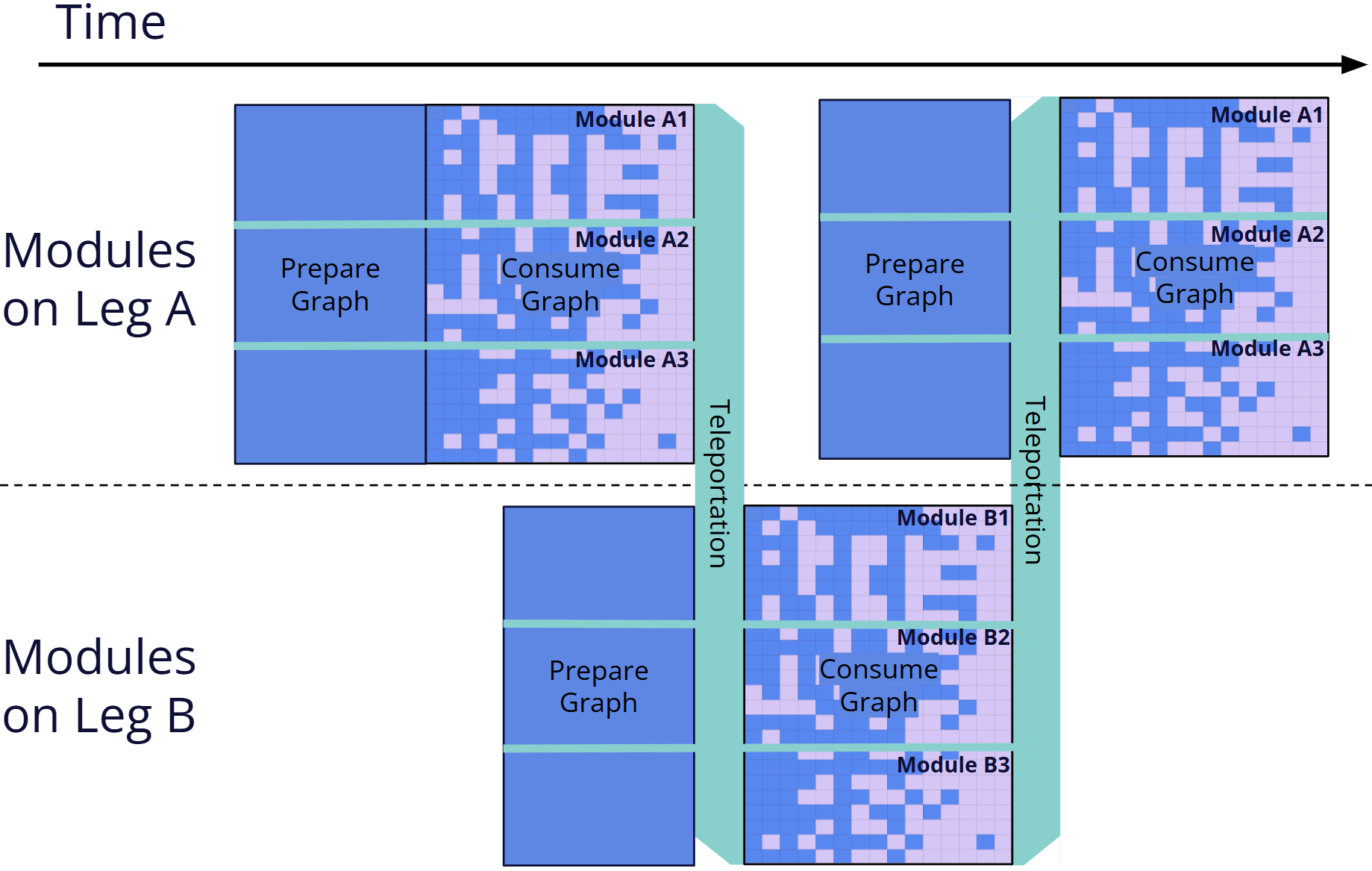}
    \caption{The high-level graph state processing cycle: the algorithm is divided into time-sliced widgets that are interleaved across two module sets in the time direction. Each set, consisting of three modules in this example, interleaves graph-state preparation with graph-state consumption. Graph-state preparation consists solely of Clifford operations (blue), while consumption utilizes $T$-state resources (purple). Teleportation moves logical state between module sets, from the output of one graph-state consumption to the input of the next. Teleportation is also used to facilitate operations that cross module boundaries within the set.}
    \label{fig:graph-proc-cycle}
\end{figure}

\begin{figure}[!h]
    \centering
    \includegraphics[width=0.9\linewidth]{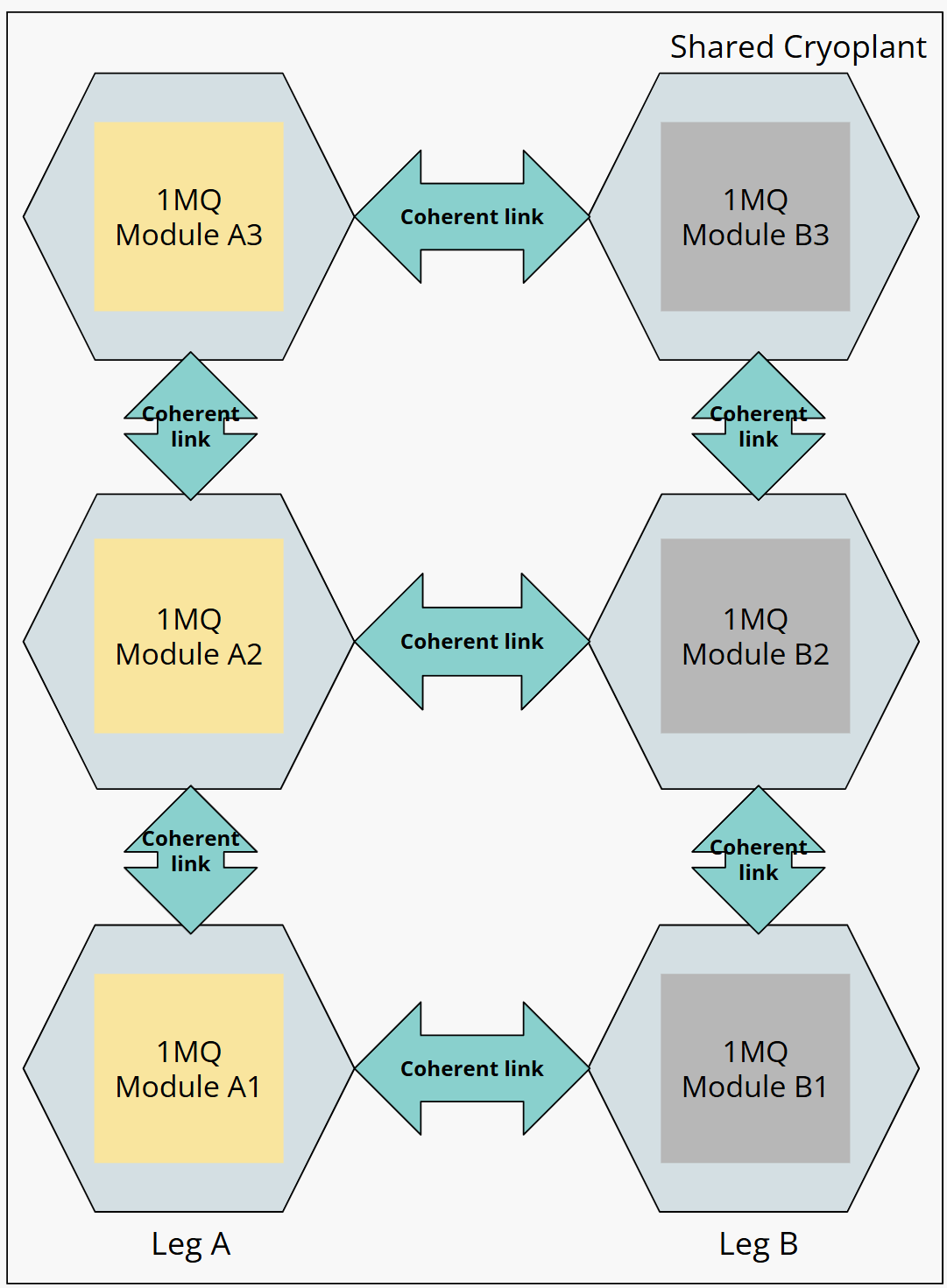}
    \caption{The high-level macro-architecture consists of two sets of 1-million physical qubit (1MQ) modules arranged like the legs of a ladder. Following the example in \cref{fig:graph-proc-cycle}, each leg contains three modules, each following the layout of \cref{fig:log-layout}(bottom). Portions of the algorithm are divided and executed solely within modules on the same leg, where intra-module interactions with high connectivity and short gate times dominate. All modules communicate via width-$d$ \enquote{coherent interconnects} (or links) at a lower rate and with reduced connectivity, used only for graph preparation (vertical interconnects) and state handover (horizontal interconnects).}
    \label{fig:2_modules}
\end{figure}  

Step 2 for the current widget and Step 1 for the widget can occur concurrently during graph state processing. This interleaved execution structure enables us to explore a degree of parallelism in a distributed FTQC. We propose distributing the interleaving preparation and consumption steps across two physically distinct sets of modules positioned along the two legs of a ladder structure. Teleportation may be used to mediate graph processing operations spanning multiple modules, but these should be minimized in scheduling. The graph processing cycle is streamlined because each set of modules alternates between only preparing or consuming the graph at each step. In contrast, the other sets of modules do the opposite on the adjacent widget, as illustrated in 
\cref{fig:graph-proc-cycle}. During all preparation and consumption steps, the FTQC performs MSD within the modules and fills a dedicated $T$-transfer bus as required. The timescales for $T$-state distillation and inter-module operations are typically one order of magnitude worse than those for other surface code operations, including graph preparations, so MSD remains local to individual modules to improve computation speed. Only the consumption stage requires the FTQC to apply the $T$ states; since graph preparations and MSD are performed independently, we need to account for possible preparation, MSD, and other delays in each widget's consumption stage. For the consumption steps, some graph-state nodes require a single $T$-basis measurement, while others require multiple $T$-basis measurements to execute non-Clifford rotations in the original algorithm. Subsequently, the FTQC teleports the current graph's output nodes to the newly prepared widget's input nodes. This cycle continues until the FTQC processes all widgets in the complete algorithm (see also~\cref{section:t-gate-error}).

The architecture must be scalable to accommodate algorithms that require more space-time resources than a single module or time step. Furthermore, MSD is arguably the most resource-intensive component of a universal FTQC, and we assume that each module has an equal number of $T$-factories supplying $T$-states locally. To address both issues, we propose a ladder-style macro-architecture illustrated in 
\cref{fig:2_modules} (see \cref{fig:log-layout} for the intra-module layout). The quantum memory register will be shared equally among all modules on the same ladder leg at any given time. Here, we can use the faster native connectivity of a lattice of superconducting qubits within a module for graph consumption and MSD, while reduced connectivity or slower interactions are used for some graph preparation operations and for teleporting the output nodes of graph states on one leg of the ladder to the input nodes on the other leg.

We can estimate the time required to execute the system-wide operations schedule by accounting for all local and inter-module operations. To account for inter-module operations, we abstract the characteristics of the teleportation interaction between modules into a single time, referred to as the \textit{inter-module handover time}, and set it to $t_\text{inter}=1~\mu$s. We incorporate this characteristic time to measure the cost of scheduled operations that cross a module boundary and to estimate the total FT handover time.

%% file: sections/hardware-architecture.tex
We propose a logical micro-architecture concept (see \cref{fig:log-layout}) that informs a system-level model, which, at a minimum, must include the following components:
\begin{enumerate}[start=0]
    \item At its core, a lattice of rotated surface-code physical qubits, used as logical patches to build the following distinct subsystems.
    \item A register of logical graph-state or memory qubits for graph preparation and consumption, 
    \item Continuous buses of logical auxiliary qubits that couple with, connect to, and operate on logical memory qubits while routing $T$-states for graph consumption. 
    \item Another continuous bus of logical auxiliary qubits for queuing and moving $T$-states must maintain a broad enough connection to the preceding auxiliary qubits to transfer $T$-states to them on demand.
    \item $T$-factories.
    \item Quantum coherent interconnects between modules to enable connections along the auxiliary bus portions.
\end{enumerate}

Section~\ref{sections/graph-state-architecture} highlighted the critical mechanism that executes quantum algorithms in the ASG formalism: a series of non-Clifford rotations and MPPOs on logical graph-state qubits, supported by auxiliary buses, perform logical operations. Consequently, this architecture requires cryostat modules to connect solely via the buses of logical auxiliary qubits. This significantly simplifies the hardware, as dense, long-range connectivity between modules is no longer necessary. The essential requirement is that the number of coherent interconnects along the buses must equal the code distance, $d$; adding additional connections will enhance overall computation for specific applications. We will quantify this trade-off in \cref{sec:results}.

Here, we outline the system's physical specifications to ensure the necessary resources to execute the FT algorithm are available. We start with the physical qubit layout. To provide the most relevant estimates for current hardware, we base our projections on a superconducting qubit architecture consisting of tunable transmon qubits coupled via tunable couplers \cite{Yan_2018}. This architecture has been implemented with high fidelity by several groups~\cite{supremacy,Xu_2023, Wu_2021,robledomoreno2024chemistry}. Here, we lay out the superconducting transmon qubits in a square lattice, coupled to one another via tunable couplers. This square lattice directly corresponds to the surface code patches, allowing us to relate physical qubit numbers to the code distance easily.

\begin{figure}[htpb]
    \centering
    \includegraphics[width=0.5\textwidth]{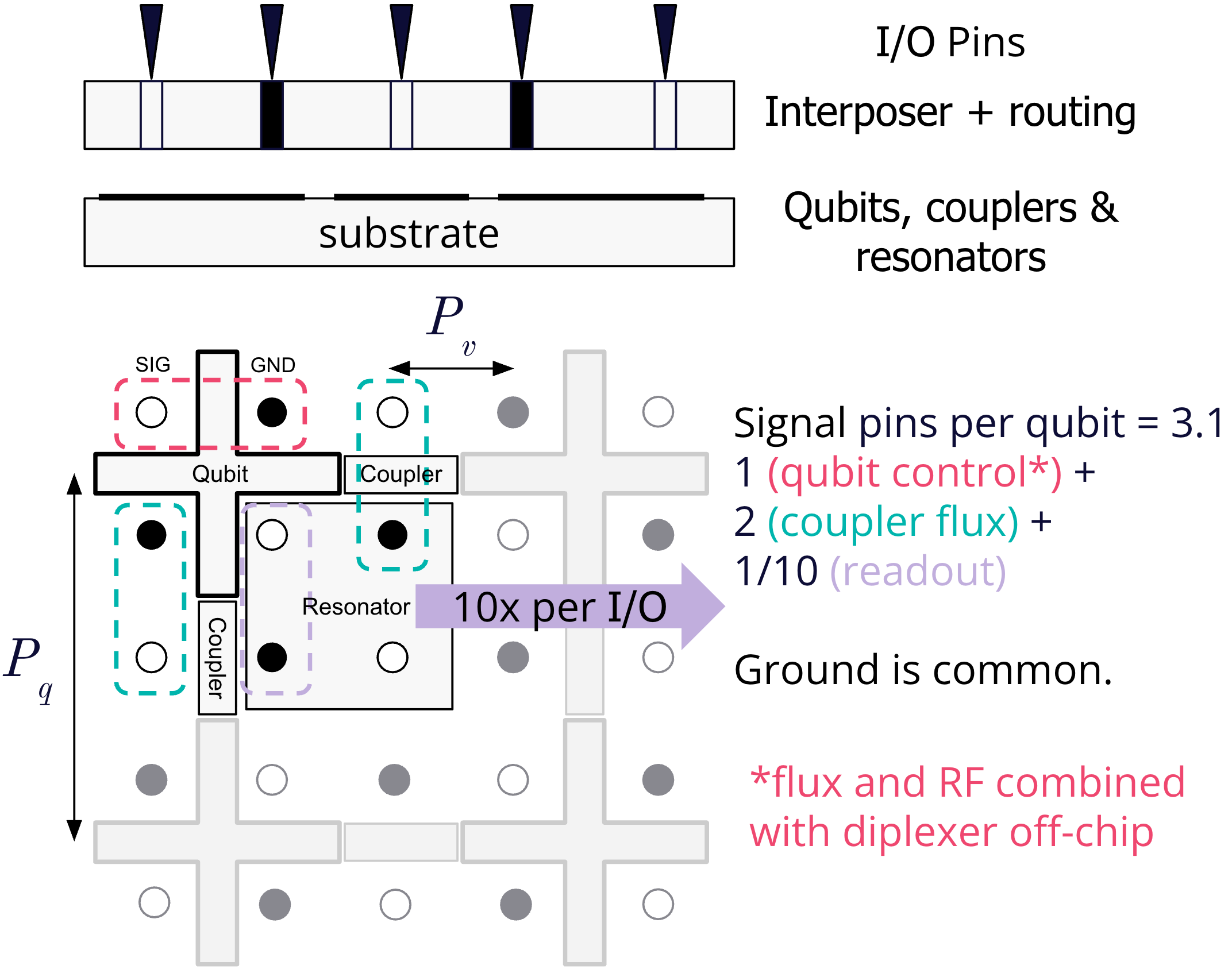}
    \caption{A schematic layout for the required components to couple superconducting qubits through tunable couplers in a surface code layout. In this example, we demonstrated the essential components: qubits, readout resonators, tunable couplers, and IO pins. Each physical qubit and coupler requires a single control line, their respective ground connections, and an additional RF line for multiplexed readout shared among 10 qubits. This limits the physical qubit pitch to match that of the signal lines. A qubit-to-qubit pitch at \textit{three} times the signal line pitch has sufficient IO pins to accommodate this.}
    \label{fig:vias}
\end{figure}

To determine the number of physical qubits per module and the overall size of the QPU, we must assess the qubit pitch. In this tunable coupled transmon architecture, the unit cell consists of a transmon and two tunable couplers. Accounting for this, each physical qubit requires $3.1$ control signals at the chip, allocated to:
\begin{itemize}
    \item 1x qubit control (flux bias for frequency tuning and RF drive for gates, combined externally to the chip via a diplexer\cite{manenti2021full}),
    \item 2x coupler bias lines,
    \item 0.1x RF line for readout resonators shared among 10 qubits for multiplexed dispersive readout\cite{sank2024characterization}.
\end{itemize}
While most QPUs employ perimeter wire bonding for individual control lines, several proposals and demonstrations focus on addressing physical qubits individually using through-substrate vias (TSVs)\cite{mallek2021fabrication, Rosenberg_2017}.

By adopting the TSVs method, QPUs require one signal pin per physical qubit and one per tunable coupler. Each signal line needs at least one ground line for return currents. As schematically illustrated in \cref{fig:vias}, achieving a sufficient number of signal and ground connections is possible if the physical qubit pitch $P_q$ is three times the via pitch $P_v$. Ultimately, the transmon, coupler, and TSVs can all be fabricated with a pitch of less than $100~\mu\text{m}$. However, the bottleneck occurs at the input and output: each signal must interface with the wiring and, eventually, with the control electronics. Any significant fan-out near the physical qubit chip would be costly; therefore, we assume a compliant joint to establish a reusable interconnect with a printed circuit board. With this assumption, a $P_v = 333~\mu\text{m}$ interposer would result in a $P_q = 1~\text{mm}$. This leads to a physical qubit density of one million qubits/m$^2$ assumed in our architectural model.

Fabricating a single device containing hundreds of thousands or even millions of physical qubits may seem unrealistic by today's standards, but we can assess where the fundamental limitations lie. Most features in a superconducting qubit circuit extend beyond the micron scale, where existing processes for flat-panel displays can easily handle critical dimensions. Consequently, the distributed and lumped-element microwave circuitry and TSVs could likely be produced on substrates at the m$^2$ scale. However, the most fundamental component of the superconducting transmon qubit is the Josephson junction (JJ) \cite{martinis2004superconducting}, which is often defined using electron-beam lithography \cite{Kreikebaum_2020}. In this context, yield and manufacturing precision are crucial for frequency targeting and other factors. While recent advances have improved the tunability of JJs, leading to more scalable manufacturing methods \cite{pappas2024alternating, vandamme2024highcoherence}, adapting an appropriate JJ fabrication process at the m$^2$ scale may prove unattainable.

\begin{figure}[!h]
    \centering
    \includegraphics[width=1\linewidth]{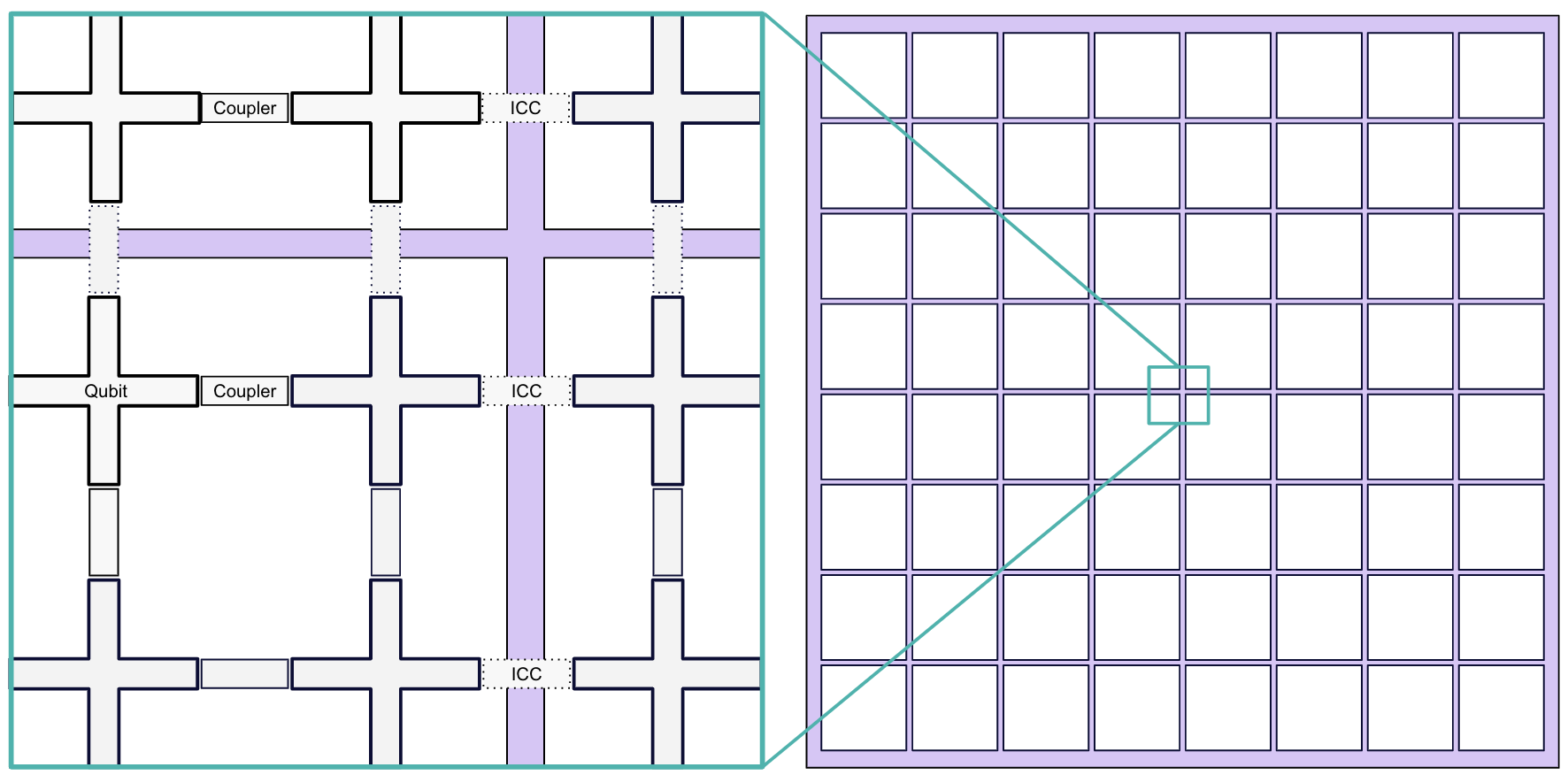}
    \caption{Physical qubit chips are tiled on a carrier. Tunable couplers manage two-qubit gates on physical qubits within the chip. ICC refers to \enquote{inter-chip coupler.}}
    \label{fig:carrier}
\end{figure}

In contrast, tiling superconducting dies on a single monolithic carrier is feasible, necessitating the execution of high-fidelity two-qubit gates across the interface between adjacent physical qubit dies. This concept has been demonstrated in \cite{Field2023}, where a functionally similar tunable coupler~\cite{sete2021floating}, commonly used in high-performing superconducting transmon qubit systems, can bridge the gap between chips. Based on this, we develop our architectural model on physical qubit chips tiled on a common carrier, featuring native four-fold connectivity across the entire carrier \cref{fig:carrier}. While other yield concerns exist with multi-chip modules, as noted in \cite{smith2022scaling}, we assume that dies are pre-screened and defect-free in this paper.

\subsubsection{Decoder and noise modeling}
\label{sec:MWPM}

To conduct physically grounded, device-specific resource estimates, we have developed an error-scaling pipeline that integrates the effects of decoding and the noise properties of superconducting systems. This pipeline converts a device-level noise model into the physical-to-logical error scaling law used in our resource counts. As part of our pipeline, we selected the square planar \textit{iSWAP} surface code \cite{mcewen2023relaxing} as implemented by the \verb|midout| package \cite{midout}. To generate input noise models for the QEC simulations, we create physical-device-inspired two-qubit correlated noise models for our $iSWAP$ gates. We then apply the resulting terms in a Pauli error model to simulate the noise process in a standard Clifford Tableau simulator, \verb|Stim| \cite{gidney2021stim}. For simplicity in this initial workflow, we use a Minimum Weight Perfect Matching (MWPM) decoder implemented in \verb|pymatching| \cite{pymatch}.

We assume a power-law physical-to-logical error scaling law for a single cycle of the surface code \cite{Fowler2012,Devitt2014}:
\begin{equation}
    p_C = \kappa\left (\frac{p}{p_\text{thresh}} \right)^{(d + 1)/2},
\label{eq:log_to_phys}
\end{equation}
where $p_C$ is the per-cycle logical error rate and $p$ denotes the homogeneous, mean, or dominant physical error rate. $\kappa$ and $p_\text{thresh}$ are fitting parameters. As part of our MWPM decoder simulations and device noise modeling, we sweep through $d$-values for one- and two-qubit error channels, which led to $p_C[\text{MWPM}] \approx 0.009 \left (\frac{p}{0.016} \right )^{\frac{d + 1}{2}}$, i.e., $\kappa[\text{MWPM}]=0.009$ and $p_\text{thresh}[\text{MWPM}]=0.016$. The scaling coefficients, although consistent with familiar values for surface codes, are illustrative and depend on the chosen simulation configuration. The resource estimates that follow are commensurate with this choice. Though simplistic, this choice of noise model and decoder was made to demonstrate how RRE allows the user to specify the coefficients appropriate to their device and decoder - see \cref{supp:noise-and-scaling} for additional noise modeling details leading to the coefficients above, as well as \cref{sec:future} for an alternative approach using distinct scaling coefficients.

\subsubsection{Intra-module micro-architecture dimensions}
\label{sec:intra-arch}

We aim to estimate the overall resources required to execute the algorithm on the FTQC depicted in 
\cref{fig:2_modules}, comprising two sets of modules. To this end, we present estimates for the spatial dimensions of the intra-module micro-architecture, which is identical across all modules and used for all resource estimates, illustrated in 
\cref{fig:log-layout}.

Each module must manage graph-state preparation, consumption, local MSD, $T$, and Bell-state routing. Within each module, the FTQC executes the algorithm based on the ASG formalism through intra- and inter-module operations on a wrapped logical memory register (blue tiles), mediated by continuous combs of logical auxiliary qubits (yellow and light yellow tiles in \cref{fig:log-layout}). The blue quantum memory chain is part of the complete memory register shared among all $n^\text{per-leg}_\text{modules}$ modules on the same leg of the macro-architecture. At any moment, auxiliary bus portions on the same leg can communicate through $d$-wide coherent interconnects. The yellow auxiliary bus mediates logical measurements from the schedules on most of the blue logical memory qubits. In contrast, the light-yellow auxiliary$+$ bus performs the same task on the otherwise inaccessible corner memory tiles and routes the Bell pairs required for the graph handover. We assign two connected auxiliary and supplementary buses to simplify resource-scaling calculations, maintain symmetry in intra-module patterns, and highlight their roles; these auxiliary bus elements are otherwise identical. More efficient configurations may exist for embedding intra-module components, depending on problem requirements — \cref{fig:log-layout} was selected solely for its simplicity and scalability.

The quantum memory register across all modules on the same leg has a logical size of $\tilde{n}_\text{logical}$, totaling $2\tilde{n}_\text{logical}$ patches for the macro-architecture. Modules have a designated area comprising at least $2\lceil \Tilde{n}_\text{logical} / n^\text{per-leg}_\text{modules} \rceil$ patches for the blue memory register and yellow auxiliary bus portions. We allocate one complete column and one connected tile to the auxiliary$+$ bus ($l_\text{edge}+1$ patches on the far left side in \cref{fig:log-layout}), then assign the memory register and auxiliary bus portions to one complete edge of an individual module. The memory register and auxiliary bus portions form a symmetric bilinear comb pattern repeated throughout the modules. The bilinear pattern occupies as much space as necessary to support graph state processing on $\lceil \Tilde{n}_\text{logical} / n^\text{per-leg}_\text{modules} \rceil$ memory nodes following the schedules. The last module structures, which may require fewer active patches for the memory register and auxiliary bus portions, remain unchanged for consistency. 

The FTQC performs MPPOs of $\Tilde{S}^\mathcal{G}_\text{prep}$ and $T$-basis measurements of $\Tilde{S}^\mathcal{G}_\text{consump}$ on the blue logical memory qubits via the yellow and light yellow bus patches. Therefore, the buses must be connected and continuous, and they must have clear access to memory patches as required by the schedules. The bilinear memory register and auxiliary bus pattern must be at least \emph{three} tiles wide to support its comb-like pattern, enabling complete graph-state processing. The logical width of the bilinear pattern per module (in the direction perpendicular to the boundaries of the bilinear pattern and the $T$-transfer bus) is given by 
\begin{equation}
\begin{split}
 l_{\text{qbus}} = \max\left( \left\lceil \frac{\left\lceil \frac{{\tilde n}_{\text{logical}}}{n_{\text{per-leg}}^{\text{modules}}} \right\rceil}{ 2 \left\lfloor \frac{l_{\text{edge}} - 2}{4} \right\rfloor + 1} \right\rceil, 3 \right)~.
\end{split}
\end{equation}
The unallocated space, the blank patches in \cref{fig:log-layout}, is not explicitly assigned but can still serve as additional buses for $O(1)$ operations.

Depending on the required accuracy for the complete FT algorithm and other architectural configurations, we select $n_\text{T-factories}$ copies of a suitable $T$-factory (also known as a $T$-distillery, magic-state distillery, or distillation widget) per module for MSD, which generates purified $T$-states at all steps as needed for consumption operations - see also \cite{Litinski2019magicstate, Gidney2024cultivation} and \cref{section:t-gate-error}. We place the $T$-factories on the opposite side of the bilinear pattern. The number of $T$-factories per module is
\begin{equation}
\begin{split}
    n_\text{T-factories} = 
    \left \lfloor \frac{l_\text{edge}-1}{\lceil\frac{L_\text{width}}{\sqrt{2}d} \rceil} \right \rfloor 
    n^\text{col}_\text{T-factories} \geq 1~, 
\end{split}
\end{equation}
and each has a physical size $L_{\text{length}} \times L_{\text{width}}$. Furthermore, the number of $T$-factory columns (purple in \cref{fig:log-layout}) in each module is given by 
\begin{equation}
\begin{split}
  n^\text{col}_\text{T-factories} = \left \lfloor
  \frac{l_\text{edge} - l_\text{qbus} - 1}{\lceil \frac{L_\text{length}}{\sqrt{2}d} \rceil + 1}
  \right \rfloor~.
  \label{eq:distill_columns}
\end{split}
\end{equation}

The quantum memory and auxiliary portions also serve as the endpoint for graph consumption. $T$-state teleportations are \textit{not} permitted between modules, whether they are on the same or different legs. This means all $T$-state distillation, movements (as exemplified by arrows in \cref{fig:log-layout}), and consumption occur locally within a module. During a graph-consumption step, the local generation, queue, and transfer of $T$-states continue to support all $T$- and $R_z$-basis measurements on graph-state nodes corresponding to the subschedule operations. We introduced a specific mid-way auxiliary system, called a $T$-transfer bus or buffer (teal in \cref{fig:log-layout}), to streamline this process.
Once we allocate the quantum memory and auxiliary portions, we can arrange the $T$-transfer bus in another comb-like pattern wrapped around $T$-factories to provide a continuous interface and maximize $n_\text{T-factories}$.
The $T$-transfer bus is responsible for actively queuing the magic states distilled by $n_\text{T-factories}$ $T$-factories, up to its logical length given by 
\begin{equation}
\begin{split}
 l_\text{transfer-bus} = &\left(l_\text{edge} - l_\text{qbus} - n^\text{col}_\text{T-factories} \left\lceil \frac{L_\text{length}}{\sqrt{2}d} \right\rceil\right) l_\text{edge} \\ 
 &+ n^\text{col}_\text{T-factories} \left\lceil \frac{L_\text{length}}{\sqrt{2}d} \right\rceil~.
\end{split}
\end{equation}

The $T$-transfer bus includes a central buffer area
depicted in \cref{fig:log-layout}. This buffer connects to the auxiliary qubits in the bilinear pattern along one entire edge of the module. In the simplest form of $T$-routing for graph consumption, the architecture ensures that the $T$-transfer bus can concurrently service at most $n'=\min(n_\text{T-factories},n^\text{row}_\text{qbus})$ nodes of the auxiliary bus, which is a conservative lower limit on the number of simultaneous $T$-states accessible. The FTQC moves the queued distilled $T$-states to the required patch of the auxiliary buses, strictly local to the module, via lattice-surgery-based patch deformation operations\cite{Litinski2019GOSC,Horsman2012-ua}, which cost $O(1)$ in the unit of the standard period of $d$ code cycles, sometimes referred to as a \enquote{tock} \cite{Litinski2019GOSC}. In our FT resource estimates, we assume these operations incur no additional temporal costs, i.e., they can be performed in parallel with other logical operations or with negligible delays. Here, $n^\text{row}_\text{qbus} = \left \lfloor \frac{\left\lceil {\tilde n}_{\text{logical}}/n_{\text{per-leg}}^{\text{modules}} \right\rceil+1}{l_\text{qbus}} \right \rfloor$. Note that there can be at most $\left\lceil {\tilde n}_{\text{logical}}/n_{\text{per-leg}}^{\text{modules}} \right\rceil$ nodes requiring $T$-basis measurements simultaneously at any given time. Upon allocating all essential components, there may be none or multiple partially or fully unallocated columns of logical qubits left on the right side; \cref{fig:log-layout} shows a single leftover column as an example.

\subsubsection{Computation of hardware's thermal resources}
\label{sec:thermal}

In addition to space-time volume data, including algorithmic execution times, physical space, and logical qubit counts, RRE can identify the hardware resources required to execute specific algorithms. These include counts of signal lines, functional chip area, and the number of amplifiers, among others. Heat loads are a significant concern for superconducting qubits operating at milliKelvin temperatures. Thus, we also estimate the power consumption, dissipation, and energy needed to run the FTQC.

Cryogenic energy consumption calculations for the proposed architecture are based on per-line thermal loads. These loads are calculated assuming an all-superconducting signaling solution with input and output lines below 4\,K, conventional dispersive readout using a Traveling Wave Parametric Amplifier (TWPA), and High-Electron-Mobility Transistors (HEMTs). We assume each physical qubit requires a microwave line for XY control and a flux bias line for tuning the qubit frequency. Each pair of physical qubits includes a tunable coupler \cite{chu2021coupler} that requires a single flux line, as shown in \cref{fig:vias}. Dispersive readout is expected to read out 10 physical qubits multiplexed in parallel. Heat loads are currently dominated by HEMT dissipation and static heat loads, so only those are considered. Example numbers are provided in RRE based on estimates from \cite{Krinner_2019}. 

The power consumed by the overall system is sensitive to the efficiency of cooling the FTQC. We scale overall power consumption using rough estimates for the power required to operate 4\,K cryo-plants (single efficiency factor of 500\,kW of power for 500\,W of cooling at 4\,K) and dilution refrigerators (single efficiency factor of 1\,kW of power for 1\,$\mu$W of cooling at 20\,mK), as outlined in the published work \cite{Martinez_2010} and manufacturer specifications - see also Parameter 49 of \cref{supp:RRE-outputs}. We anticipate that these estimates will evolve as they depend on the detailed design of the signal chain.

%% file: sections/graph-state-compilation.tex
We assume the user has a fixed description of their logical and hardware architecture, represented by a set of configuration parameters, denoted by $\{\text{config}\}$. The user also provides a circuit that implements a unitary operator $U$ acting on $n_\text{input}$ logical qubits of the FT algorithm:
\begin{equation}
\begin{split}
    \Big| \text{output} \Big\rangle =
    &U \ket{ \text{input} := s_0=0, s_1=0, \cdots, s_{n_\text{input}-1}=0 }.
\end{split}
\end{equation}
In practice, $U$ consists of a sequence of quantum instructions that may include tens of millions of logical gates. We always start with an all-zero $\ket{ \text{input}}$ state to simplify calculations. 

One can always transpile the initial circuit to express it solely in terms of logical Cliffords, $T$, $T^\dagger$, and arbitrary-angle non-Clifford $R_Z$ gates, i.e.~$U\big(\{\text{Clifford},T,T^\dagger,R_z(\theta)\}\big)$. We denote the exact number of original logical $T$ and $T^\dagger$ gates, excluding those implicitly included within the arbitrary-angle rotations, as $n^{T}_\text{init}$. Moreover, $n^{R_z}_\text{init}$ represents the exact number of arbitrary-angle non-Clifford $R_Z$ gates in $U$, including those implicitly in the logical gates (excluding $T$ and $T^\dagger$ operations counted separately in $n^{T}_\text{init}$). We assume FT execution of $U\big(\{\text{Clifford},T,T^\dagger,R_z(\theta)\}\big)$ using rotated surface code patches, as described in \cite{Litinski2019GOSC, forlivesi2023, Orourke2024}, with a code distance $d$. In our approach, the decomposition of non-Clifford rotations takes place after compilation during the consumption stages of the graphs (see~\cref{fig:graph-proc-cycle}). This process is sometimes referred to as gate synthesis \cite{GridSynth, Kliuchnikov2022}. We denote the diamond norm precision for gate syntheses as $\varepsilon$.

We consider an MBQC framework that uses ASG compilation for circuit execution. One advantage of this approach is that it does not require initializing the entire graph state, $\mathcal{G}$, on FTQC at all times~\cite{Ruh2024, Elman2024}. Instead, we split circuit execution into multiple time steps, each containing a subcircuit or widget with its own (sub)graph. We identify the parts of the graph state needed at any moment by constructing three complete schedules by sequentially adding sub-schedules across time steps: the preparation schedule for the widgetized algorithm, labeled as $\Tilde{S}^\mathcal{G}_\text{prep}$. The consumption schedule for the widgetized algorithm will be labeled as $\Tilde{S}^\mathcal{G}_\text{consump}$. Additionally, a related schedule specifies the measurement bases used by FTQC during the consumption stage, which we refer to as $\Tilde{S}^\mathcal{G}_\text{meas}$.

We verify that the widgetized schedules $\{\Tilde{S}^\mathcal{G}_\text{prep}, \Tilde{S}^\mathcal{G}_\text{consump}, \Tilde{S}^\mathcal{G}_\text{meas} \}$ provide a faithful representation of the preparation and consumption of the quantum state $\mathcal{G}$, in accordance with $U$. This is equivalent to any valid set of $\{S^\mathcal{G}_\text{prep}, S^\mathcal{G}_\text{consump}, S^\mathcal{G}_\text{meas} \}$ when time-sliced widgetization is not used. Previous studies~\cite{vijayan2024,Litinski2019GOSC,Liu2023,scheduler,Ruh2024,jabalizer,Cabaliser} show that valid arrangements of such schedules can effectively prepare~\cite{Liu2023} and consume~\cite{Ruh2024} the graph, highlighting their connection to its spatial and temporal structure. Our focus will be on introducing the notation for $\{\Tilde{S}^\mathcal{G}_\text{prep}, \Tilde{S}^\mathcal{G}_\text{consump}, \Tilde{S}^\mathcal{G}_\text{meas}\}$.

$\Tilde{S}^\mathcal{G}_\text{prep}$ is a time-ordered list of $n_\text{widgets}$ preparation sub-schedules, each corresponding to a widget. The FTQC must follow these sub-schedules to initialize subgraphs in several sub-steps using $n$-body $CZ$ and single-qubit Cliffords~\cite{vijayan2024, Liu2023, scheduler} that correspond to lattice surgery's MPPOs for $X$ and $Z$-stabilizers~\cite{Horsman2012-ua, Brown2017, Litinski2018latticesurgery, Fowler2019, Litinski2019GOSC}. Specifically, $\Tilde{S}^\mathcal{G}_\text{prep} = \{\Tilde{S}^{g_0}_\text{prep},\dots,\Tilde{S}^{g_{n_\text{widgets}-1}}_\text{prep}\}$. Each preparation sub-schedule may comprise multiple sub-steps, represented as $\Tilde{S}^{g_i}_\text{prep} = \{\Tilde{S}^{g^0_i}_\text{prep},\dots,\Tilde{S}^{g^{J(i)}_i}_\text{prep}\}~\forall~i\in\{0,\dots,n_\text{widgets}-1\}$. Furthermore, a $\Tilde{S}^{g^j_i}_\text{prep}$ can contain several non-overlapping MPPOs~\cite{Liu2023, scheduler}. The MPPOs within a sub-step can be executed simultaneously using the auxiliary buses.

Similarly, $\Tilde{S}^\mathcal{G}_\text{consump}$ is a sequence of $n_\text{widgets}$ sets, with each set serving as the consumption sub-schedule for a corresponding widget. To implement the original non-Cliffords, the FTQC must follow these sub-schedules, performing logical arbitrary-angle $R_Z(\theta)$-basis measurements on the relevant nodes in multiple sub-steps. In detail, $\Tilde{S}^\mathcal{G}_\text{consump} = \{\Tilde{S}^{g_0}_\text{consump},\dots,\Tilde{S}^{g_{n_\text{widgets}-1}}_\text{consump}\}$. Each of these consumption sub-schedules may contain several sub-steps, represented as $\Tilde{S}^{g_i}_\text{consump} = \{\Tilde{S}^{g^0_i}_\text{consump},\dots,\Tilde{S}^{g^{K(i)}_i}_\text{consump}\}~\forall~i\in\{0,\dots,n_\text{widgets}-1\}$. A specific $\Tilde{S}^{g^k_i}_\text{consump}$ can contain multiple non-overlapping measurement lists, as detailed in \cite{Ruh2024,Cabaliser}. Measurements in non-overlapping lists within a sub-step can be executed simultaneously. Meanwhile, $\Tilde{S}_\text{meas}^G$ is a connected sequence that specifies the rotational basis ($T$ or $R_Z(\theta)$) required for node measurements across the corresponding sub-steps of $\Tilde{S}^\mathcal{G}_\text{consump}$. This method distinguishes two categories of node measurements during the consumption stage: those that require $T$-basis measurements, consuming a single distilled $T$ state, and those that require $R_Z(\theta)$-basis measurements, consuming multiple distilled $T$ states following gate-synthesis of $R_Z(\theta)$ to a Clifford+$T$ sequence.

The number of nodes in $\mathcal{G}$, called the \textit{graph size} and denoted by $N$, is typically much larger than the number of input logical qubits, $N \gg n_\text{input}$. This is because we always add logical auxiliary qubits to implement an MBQC of a given quantum algorithm. Moreover, $N$ is always greater than $n_\text{input}$~\cite{Cabaliser,vijayan2024,Elman2024}. However, due to the scheduling requirements above, only a relatively small subset of logical qubits, each representing a graph node, is required at any given step to execute the unitary operation $U$. We denote the maximum number of logical qubits needed at any step as $\tilde{n}^\mathcal{G}_\text{logical}$, i.e., the maximum quantum memory required. Typically, $n_\text{input} < \tilde{n}^\mathcal{G}_\text{logical} < N$~\cite{Cabaliser,Ruh2024}.

One can create distinct graph states with similar attributes from different input states, but the same $U$. We construct $\mathcal{G}$ by \emph{stitching} graphs from each time step for resource estimations. To admit arbitrary input nodes into these ASGs, we always initialize all nodes in the all-zero input state, $\ket{0,0,\dots,0}$, and then replace the input nodes with $n_\text{input}$ copies of $(\ket{0+}+\ket{1-})^{\otimes n_\text{input}}$, attaching the appropriate auxiliary nodes. This process allows us to teleport from the output nodes of one (sub)graph to the input nodes of another in the consecutive time step~\cite{Cabaliser, vijayan2024}, which we use to verify and stitch graph states.

To summarize, we can achieve \emph{exact} resource estimates for $U$ by providing an MBQC-based set of primitives for the non-widgetized algorithm:
\begin{equation}
\begin{split}
    S_\text{est}&(U, \{\text{config}\}) := \\
    &\{d, \varepsilon, n^{T}_\text{init}, n^{R_z}_\text{init}, n^\mathcal{G}_\text{logical}, S^{\mathcal{G}}_\text{prep}, S^{\mathcal{G}}_\text{consump}, S^{\mathcal{G}}_\text{meas}\},
\label{eq:est_set}
\end{split}
\end{equation}
In this equation, the required user inputs are explicitly specified on the left-hand side, whereas the right-hand side yields the corresponding set of primitives for resource estimations. It becomes extremely challenging to construct $S_\text{est}$ for a practical $U$ containing tens of millions of gates and considerable circuit depth, whether for resource estimation or hardware execution. Therefore, we proposed a time-sliced widgetization of $U$ that yields an approximate resource estimation set $\tilde{S}_\text{est}$. Below, we will provide a step-by-step recipe for creating and rigorously verifying $\tilde{S}_\text{est}$.

\subsubsection{RRE software workflow}

Our FT resource estimation tool, RRE~\cite{RRE}, was extensively developed and used to generate the results presented in this work. The RRE estimation strategy follows a workflow that translates a logical circuit into an ASG, maps it onto the logical layout shown in \cref{fig:log-layout}, and then assesses the required physical resources. This section provides a brief overview of RRE's step-by-step workflow, while detailed explanations of selected steps appear in subsequent sections.

\begin{enumerate}
    \item \textit{Widgetized logical circuit and other parameter inputs}: The user provides an FT algorithm, $U$, as a custom \texttt{JSON} file, which can contain a single OpenQASM 2.0 circuit or be widgetized into multiple OpenQASM 2.0 subcircuits, crucial for large-scale applications. Refer to \cref{sub:widgitizing} for the widgetization strategy we employed. To estimate resources, in addition to supplying a logical circuit, the user must also set a set of system architectural parameters, $\{\text{config}\}$, including physical error rate, $p_{\text{logcell}}$-to-$p_{\text{physical}}$ scaling coefficients (based on decoder and device models), cycle times, and the desired total failure rate budget, $p_{\text{algo-fail}}$. RRE users can select preconfigured $\{\text{config}\}$ and modify them through an input \texttt{YAML} file.

    \item \textit{Transpilation}: RRE parses the OpenQASM 2.0 circuit(s) and counts the initial $T$ gates, $n^{T}_\text{init}$, and all arbitrary-angle rotation gates, $n^{R_z}_\text{init}$, in the FT algorithm, as needed for later steps. It replaces any unsupported unitary in the circuit(s) with \textit{exact} gate equivalents when the downstream fault-tolerant compiler does not support it. See \cref{sub:transpilation} for details.

    \item \textit{Graph state processing and compilation}: Cabaliser (\cref{sub:cabalizer} and \cite{Cabaliser}) is used to compile the transpiled circuits into ASGs. The graph objects are cached and reusable. Even if the architecture changes, the portable graph objects can be used to estimate resource requirements for that algorithm without reprocessing the graph states.
    
    \item \textit{Substrate scheduling}: The Substrate Scheduler is called to generate preparation schedules for ASGs, as detailed in \cref{sub:scheduler} and \cite{scheduler}.
    
    \item \textit{Find the required precision and distance iteratively}: Given $p_{\text{algo-fail}}$, the graphs, and the preparation schedules, iterate through a Lookup Widget table to find $\varepsilon$, $d$, and the optimal $T$-factory (with output logical error rate $p_\text{out}$) that satisfies $\varepsilon < p_{\text{logical}}$ and $p_\text{out} < p_\text{logical}$. $\varepsilon$ determines the length of the Clifford+$T$ chain, $L_\varepsilon$, for arbitrary angle transformations needed at the final measurement points of the consumption schedules.
    
    \item \textit{Identify space complexity}: RRE determines the total physical qubit count to be allocated to the hardware, $n_\text{physical}$, which scales with $\Tilde{n}_\text{logical}$, the maximum node degree of the graph, our proxy for the number of logical qubits required. In practice, $\Tilde{n}_\text{logical}$ will only provide an upper bound on the required logical qubit count.
    
    \item \textit{Identify time complexity}: Given $d$ and the characteristic timescales set in the system's architectural configurations, $\text{config}$, find $\{t_\text{tock}^\text{quantum}, t_\text{tock}^\text{decoding}, t_\text{tock}^\text{intermodule}\}$.
    
    \item \textit{Find full \texttt{physHW} object and hardware times}: RRE uses the tocks, schedules, and other architectural configurations from $\text{config}$ (cryostat sizes, line, coupler, and power dissipation details) to identify an FT hardware object, \texttt{physHW}, including the number of couplers, modules, area, memory, energy usage, etc. \texttt{physHW} fully identifies all outputs including total FT hardware time, decoding core numbers, and total power and energy consumption. Hardware execution times are computed from MPPOs across auxiliary buses for local and inter-module interactions.
    
    \item \textit{Display results}: Tabulate and print resource estimation outputs to the console and a \texttt{CSV} file as requested. All 48 default RRE output parameters are detailed in \cref{supp:RRE-outputs}.
\end{enumerate}

\subsubsection{Input circuit widgetization}
\label{sub:widgitizing}

Due to the large scales involved in a practical input circuit, a decomposition strategy is essential to avoid memory bottlenecks on RRE. Luckily, quantum algorithms usually consist of repeating blocks, making their decomposition easier.

\begin{figure*}
    \centering
    \includegraphics[width=0.99\linewidth]{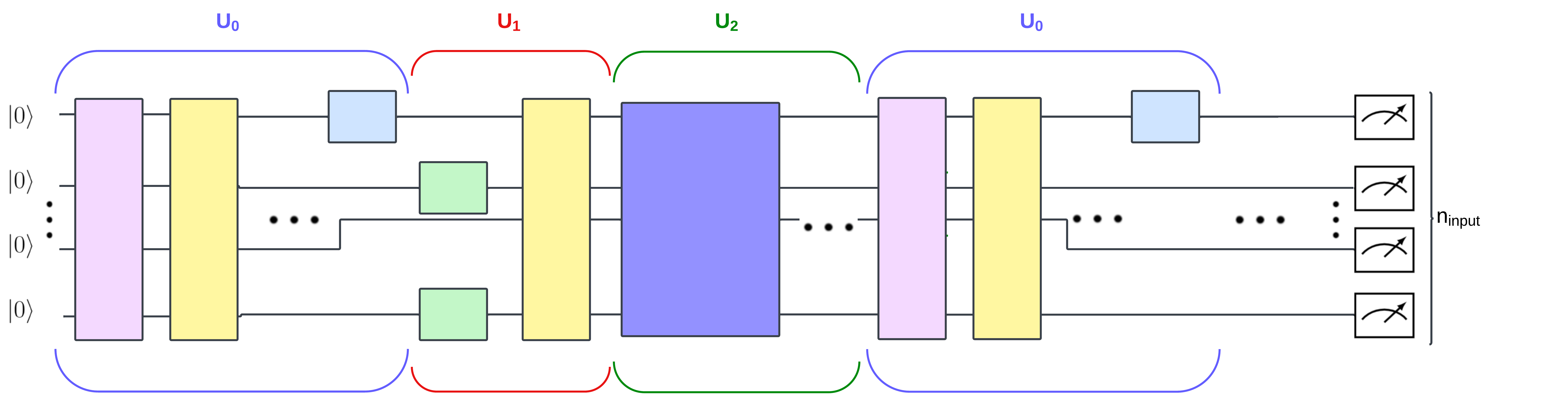}
    \caption{\emph{Input circuit decomposition strategy}: This visualization demonstrates how we \enquote{widgetize} a complete logical algorithm, $U$, into $n_{\text{widgets}}$ time-sliced widgets, represented as $U_{n_{\text{widgets}}-1} \dots U_1U_0 \equiv U$. Each widget has the same width, $n_{\text{input}}$, but varying depth. Some widgets may be repeated in the time direction (for example, $U_3 \equiv U_0$ above), allowing us to compile only $n^*_{\text{widgets}}$ distinct subcircuits.}
    \label{fig:widgetization}
\end{figure*}

Our decomposition strategy slices the unitary operator $U$ along the time direction into $n_\text{widgets}$ widgets (subcircuits) of fixed width $n_\text{input}$. Each widget may have a distinct, emergent depth. Crucially, no entangling gates connect adjacent subcircuits. We can then execute the complete algorithm as a sequence of these subcircuits, expressed as $U = U_{n_\text{widgets}-1} \dots U_1 U_0$, where 
\begin{equation}
\begin{split}
    U_i \ket{\text{state}}_i = \ket{\text{state}}_{i+1}~\text{for}~i\in\{0,\cdots,n_\text{widgets}-1\}~.
\end{split}
\end{equation}
Here, $\ket{\text{state}}_0  \equiv \ket{\text{input}}$ and $\ket{\text{state}}_{n_\text{widgets}} \equiv \ket{\text{output}}$.

In a typical quantum algorithm, many widgets may be equivalent and repeated along the time direction. In such cases, we need to compile only the set of distinct widgets, for example, $[U_0, U_4, \dots, U_{n_\text{widgets}-7}]$. We use $n^*_\text{widgets}$ to represent the number of distinct widgets in the set $[U_0, U_1, \dots, U_{n_\text{widgets}-1}]$ ($n^*_{\text{widgets}} \leq n_{\text{widgets}}$). We refer to this process as \emph{circuit widgetization}, illustrated in \cref{fig:widgetization}. We base our choice of widgetization strategy on the creation of subcircuit dependency graphs. This approach allows us to compile large circuits into elementary operations and unitarily recombine compiled widgets, avoiding excessive memory overhead. Compilation is essential for executing the algorithm and for accurate resource estimates. Additionally, this strategy helps keep the number of distinct widgets $n^*_\text{widgets}$ relatively low, facilitates implementation, and justifies using time slicing with a fixed $n_\text{input}$. However, this particular widgetization choice is likely not optimized to minimize space-time costs. We present further details on our widgetization strategy in \cref{supp:decomp}.

\subsubsection{Stitching strategy}
\label{sub:stictching}

Following the widgetization process, we can construct an approximate version of $S_{\text{est}}$. We propose the widgetization-based set $\Tilde{S}_{\text{est}}$ as a good approximation of $S_{\text{est}}$ for resource estimations, defined as follows:
\begin{equation}
\begin{split}
    \Tilde{S}_\text{est} &:=  \\
    &\{d, \varepsilon, n^{T}_\text{init}, n^{R_z}_\text{init}, \Tilde{{n}}^{\mathcal{G}}_\text{logical}, \Tilde{S}^{\mathcal{G}}_\text{prep}, \Tilde{S}^{\mathcal{G}}_\text{consump}, \Tilde{S}^{\mathcal{G}}_\text{meas} \} \\ 
    &\approx S_\text{est}~.
\label{eq:approx_set}
\end{split}
\end{equation}
Recall that $n^{T}_\text{init}$ and $n^{R_Z}_\text{init}$ are the exact counts of explicit $\{T,T^\dagger\}$ and all arbitrary-angle $R_Z$ operations in $U$. In this widgetization-based estimation procedure, we first obtain estimation primitives for each widget based on its (sub)graph. We then obtain the overall circuit resource estimates as follows:
\begin{equation}
\begin{split}
    &\Tilde{n}_\text{logical}^{\mathcal{G}} := \max_{g_i}(n^{g_i}_\text{logical}) \approx n_\text{logical}^{\mathcal{G}}, \\
    &\Tilde{S}^{\mathcal{G}}_\text{prep} := \cup_{i=0}^{n_{\text{widgets}}-1}S^{g_i}_\text{prep} \equiv S^{\mathcal{G}}_\text{prep}, \\
    &\Tilde{S}^{\mathcal{G}}_\text{consump} := \cup_{i=0}^{n_\text{widgets}-1}S^{g_i}_\text{consump} \equiv S^{\mathcal{G}}_\text{consump}, \\
    &\Tilde{S}^{\mathcal{G}}_\text{meas} := \cup_{i=0}^{n_\text{widgets}-1}S^{g_i}_\text{meas}  \equiv S^{\mathcal{G}}_\text{meas}~.
\label{eq:stitching}
\end{split}
\end{equation}
Here, $g_i$ represents the graph state compiled from the $i$th widget. 

Equation \ref{eq:stitching} formalizes our stitching strategy and outlines the key assumptions underlying our estimates. It identifies how we combine the individual graphs and scheduling lists to approximate the complete $\mathcal{G}$, as well as $S^{\mathcal{G}}_{\text{prep}}$, $S^{\mathcal{G}}_{\text{consump}}$, and $S^{\mathcal{G}}_{\text{meas}}$. While an FTQC must physically stitch the widgets (through teleportations) to execute the complete circuit, this is unnecessary for resource estimation. Consequently, we reduced the computational cost of obtaining resource estimates by computing the widgetized quantities in \cref{eq:stitching} through software.

\subsubsection{Pre-processing the input subcircuits}
\label{sub:transpilation}

To prepare for graph compilation, we must parse each widget and \textit{transpile} it to include only logical Clifford+$T$+$R_Z(\theta)$ gates, using exact gate equivalences. Here, the $\theta$ values represent non-Clifford angles. Later, during the consumption stage, when we want to perform single-qubit logical measurements, we need to further decompose each $R_Z(\theta)$ node into a sequence of Clifford+$T$ gates through gate synthesis \cite{GridSynth,Kliuchnikov2022}. 

We perform transpilation to the logical Clifford+$T$+$R_Z(\theta)$ using RRE. Performing gate synthesis at the end of the consumption stage enables us to generate, cache, and reuse the set of subgraphs $\{g_0,\dots,g_{n_\text{widgets}-1}\}$ and their attributes independently of the logical and hardware-level configurations, $\{\text{config}\}$.

Once transpilation is complete, the next step is to generate the unified MBQC-based set of graph states $\{g\}$ and schedules $\{S^{g}_\text{prep}\}$, $\{S^{g}_\text{consump}\}$, and $\{S^{g}_\text{meas}\}$, and other essential intermediary lists detailed below, which yields $\Tilde{S}_\text{est}$ sufficient for resource estimation of $U$.

\subsubsection{Cabaliser: compiling to graphs and consumption schedules}
\label{sub:cabalizer}

RRE uses the stabilizer tableau-based FT compiler Cabaliser~\cite{Cabaliser} as a crucial intermediate layer to transform the set of individual logical widgets, $\{U\}$, into corresponding MBQC-based subgraphs, consumption schedules, “locally simulated” quantum states, and other essential information. Cabaliser's outputs are locally simulated quantum states derived from exact calculations. Cabaliser effectively maps an input state $\ket{\text{state}}_{i-1}$ to $\ket{\text{state}}^\text{Cabaliser}_i$: it applies $U_i$, Pauli corrections, the consumption subschedule, and additional logical operations (such as Hadamards) to the ($i-1$)th graph state to create a faithful vector representation of the corresponding universal subgraph at step $i$. It is guaranteed that $\ket{\text{output}} \equiv \ket{\text{state}}^\text{Cabaliser}_{n_\text{widget}-1}$. This vector representation helps verify the unitaritiness of RRE and Cabaliser’s outputs for small $n_\text{input}$. We present a systematic method for verifying the graphs and schedules produced by RRE, accounting for widgetization and stitching in \cref{supp:unitary-verification}. For complete details on scalability and various compilation components of Cabaliser, we refer readers to \cite{Cabaliser, Paler2017, vijayan2024}.

For a given individual subcircuit $U_i$ that includes only Clifford+$T$+$R_Z(\theta)$, Cabaliser can generate the following set of quantities:
\begin{equation}
\begin{split}
    U_i(\{\text{Clifford},T,&T^\dagger,R_z(\theta)\}) \xrightarrow[]{\text{Cabaliser}} \\ 
    &\{g_i, S_\text{frames}^{g_i}, S_\text{input}^{g_i}, S_\text{output}^{g_i}, n^{g_i}_\text{logical}, \\ 
    &S_\text{consump}^{g_i},
    S_\text{meas}^{g_i}, \ket{\text{state}}^\text{Cabaliser}_i\}~.
\label{eq:Cabaliser}
\end{split}
\end{equation}
In this equation, $S_\text{frames}^{g_i}$ is a set that tracks the required Pauli corrections (frames) to be applied to measurements at the end. $S_\text{output}^{g_i}$ and $S_\text{input}^{g_i}$ represent the sets (or maps) of output and input nodes in the graph state, respectively. Because we decompose the entire algorithm in the time direction with a uniform width that covers all logical qubits, the number of intermediate input and output nodes remains the same. $n^{g_i}_\text{logical}$ denotes the quantum memory needed to process $U_i$. $S_\text{consump}^{g_i}$ identifies the graph consumption schedule for performing individual logical qubit measurements. $S_\text{meas}^{g_i}$ identifies which rotational basis to use for each measurement during the consumption stage. Finally, $\ket{\text{state}}^\text{Cabaliser}_i$ represents the locally simulated quantum state extracted from the outputs generated up to the $i$th step.

We repeat the process of \cref{eq:Cabaliser} for all $n^*_\text{widgets}$ distinct widgets, providing the user with the information required for the approximate estimation set in \cref{eq:approx_set}, excluding the preparation schedules, $\{S^{g}_\text{prep}\}$, which we generate using another tool detailed below.

\subsubsection{Substrate Scheduler: generating preparation schedules}
\label{sub:scheduler}

RRE uses the Substrate Scheduler~\cite{scheduler} to map the graph states $\{g\}$, computed by Cabiliser in \cref{eq:Cabaliser}, to a set of graph preparation sub-schedules $\{S_\text{prep}^g\}$. These preparation sub-schedules are optimized for the logical memory register and auxiliary bus in the superconducting hardware layout shown in \cref{fig:log-layout}.

For the Substrate Scheduler calculations, we assume a dual-rail or bilinear quantum memory and auxiliary bus pattern consisting of linear arrangements of rotated surface code logical qubits for storing graphs and mediating logical measurements on the nodes, as shown in \cref{fig:log-layout}. The pattern has a fixed size $\Tilde{n}_\text{logical} := \max_{g_i}(n^{g_i}_\text{logical})$ per rail. This assumption ensures no memory constraints when executing the preparation and consumption stages consecutively. Although the intra-module micro-architecture used for resource estimates follows the layout shown in \cref{fig:log-layout}, the Substrate Scheduler results can still be matched and used under the assumption of a wrapped-bilinear pattern. Currently, the Substrate Scheduler identifies an optimal schedule based on the individual mapping of each $g$ to this layout. The Substrate Scheduler's performance could be enhanced by incorporating additional information about node types and supporting customized bus architectures, which we plan to address in future work.

\subsubsection{Space-time-optimal resource estimation}
\label{section:t-gate-error}

\begin{table*}[htpb]
    \centering
    \begin{tabular}{ccccc} 
     \emph{\texttt{$T$-factory}}(superconducting) & $p_{out}$ & Phys.~Space($L_\text{width} \times L_\text{length}$) & Physical Qubits, $Q~$ & Cycles, $C$ \\
     \midrule
     (15-to-1)$_{17,7,7}$ & $4.5\times 10^{-8}$ & 64$\times$72 & 4620 & 42.6\\
     (15-to-1)$^6_{15,5,5}$ $\times$ (20-to-4)$_{23,11,13}$ & $1.4 \times 10^{-10}$ & 387$\times$155  & 43,300 & 130 \\
     (15-to-1)$^4_{13,5,5}$ $\times$ (20-to-4)$_{27,13,15}$ & $2.6 \times 10^{-11}$ & 382$\times$142 & 46,800 & 157 \\
     (15-to-1)$^6_{11,5,5}$ $\times$ (15-to-1)$_{25,11,11}$ & $2.7 \times 10^{-12}$ & 279$\times$117 & 30,700 & 82.5 \\
     (15-to-1)$^6_{13,5,5}$ $\times$ (15-to-1)$_{29,11,13}$ & $3.3 \times 10^{-14}$ & 292$\times$138 & 39,100 & 97.5 \\
     (15-to-1)$^6_{17,7,7}$ $\times$ (15-to-1)$_{41,17,17}$ & $4.5 \times 10^{-20}$ & 426$\times$181  & 73,400 & 128 \\
     (15-to-1)$^8_{23,9,9}$ $\times$ (15-to-1)$_{49,19,21}$ & $9.0 \times 10^{-23}$ & 696$\times$234  & 133,842 & 157.5 \\
     \bottomrule
    \end{tabular}
    \caption{$T$-factory Lookup Table: Litinski's $T$-factories that match our superconducting architecture with $p=10^{-3}$. Extended from \cite[table 1]{Litinski2019magicstate}.
    }
    \label{table:widget}
\end{table*}

In this section, we formulate the space-time volume requirements for graph state processing parts based on $\{\text{config}\}$, $\Tilde{S}_\text{est}$, and the architectures presented in \cref{sec:hardware-architecture}, with a spatially optimal size of the quantum register and time-optimal compilation strategies (given the spatial configurations), creating a ``middle-of-the-way" design.

At a macroscopic level, the architecture features a ladder-style structure with $n_{\text{per-leg}}^{\text{modules}}$ pairs of QPU modules connecting the legs, as illustrated in \cref{fig:2_modules}. We based this design on the assumption that superconducting platforms have a low syndrome extraction cycle and would benefit from minimizing the physical footprint while interleaving graph preparation and consumption. Consequently, we always create $\tilde{n}_{\text{logical}}$ rotated surface code patches for the logical memory register shared among all modules on a leg of the macro-architecture. 

At the intra-module level, we organize each module as a square grid of logical qubits (patches), the foundational building blocks for all components. The edge size is determined by $l_{\text{edge}} = \left \lfloor \sqrt{\frac{n_{\text{phys}}}{2d^2}} \right \rfloor$, where $n_{\text{phys}}=1,000,000$ represents the total number of physical qubits available per module. For simplicity in resource estimates, we assumed each rotated surface code patch contains $2d^2$ physical qubits.  

We performed a static allocation of intra-modular components, providing sufficient resources for the largest widgets, as illustrated in \cref{fig:log-layout}. Each module has \emph{five} main components, to which we allocate physical qubits. The unallocated or \enquote{wasted} space can still be used for ancillary operations. The first component is a quantum register of logical memory (graph-state) qubits. The second is a chain of logical auxiliary bus qubits of the same length, facilitating logical measurements on the memory qubits, and connects to the third component, i.e., the supplementary bus. The first two components are arranged in a snake-like bilinear pattern to fully occupy one QPU edge. The fourth component is the linear $T$-transfer bus, which supplies the auxiliary bus portion with $T$-states for graph consumption. The last components are $T$-factories. See \cref{sec:intra-arch} for details.

The space-time volume of the complete graph creation on the bilinear pattern before single-qubit logical measurements is 
\begin{equation}
\begin{split}
    V_\mathcal{G} = 2 \Tilde{n}_\text{logical} \mathbb{L}_\text{prep} d,
\end{split}
\end{equation}
where $\mathbb{L}_\text{prep}(\Tilde{S}_\text{prep}) = \sum_{\text{sub} \in \Tilde{S}_\text{prep}} L_\text{sub}$. Here, $L_\text{sub}$ is the number of distinct measurement steps in the preparation sub-schedule of $\text{sub}$. In other words, $\mathbb{L}_\text{prep}$ is the total number of distinct measurement steps required for the complete graph preparation.

We use the same $T$-factory, matched from an extended list of appropriate widgets, across all steps and modules. We build the extended $T$-factory list using a method proposed by Litinski (see the original table~1 in \cite{Litinski2019magicstate}). We consistently assume a homogeneous superconducting physical error rate of $p=10^{-3}$ for all quantum operations, which may originate from dominant error sources—whether from measurements or single- or two-qubit physical gates. We sourced the matching widgets from numerical simulations in \cite{Litinski2019magicstate} and simulated a larger $T$-factory, summarizing our findings in \cref{table:widget} as an extended \enquote{$T$-factory Lookup Table.} We recursively search this table to identify a suitable $T$-factory. Notably, \cref{table:widget} is part of the user-provided configuration set, $\{\text{config}\}$, and can be customized in RRE.  

In \enquote{$T$-factory Lookup Table}, \cref{table:widget}, the output probability, $p_{\text{out}}$, determines the logical error rate for the output state by $T$-factories and sets an upper limit on the error rate for any $T$-gate implemented using the auxiliary qubits. The specifications for the number of physical qubits, the dimensions of the 2D patch, and the number of physical cycle sweeps depend on the factory's construction. Our analysis focused exclusively on selecting $T$-factories that produce $T$-states, thus excluding any entries related to $CCZ$ from the original \cite[table 1]{Litinski2019magicstate}.

Next, we must include the resource cost of decomposing non-Cliffords into Clifford+$T$ (gate synthesis) to consume graph-state nodes via single-qubit measurements. After completing a preparation sub-schedule, at each sub-step of the consumption schedule, we use one or more purified $T$-resources to perform measurements in the desired basis. The FTQC must move $T$-states from $T$-factories to the auxiliary bus through patch deformations\cite{Litinski2019GOSC,Horsman2012-ua} at negligible time cost $O(1)$. We assume that graph nodes either require arbitrary-angle $R_Z$-basis measurements or a single $T$-basis measurement. Measurements involving $R_Z$-axis rotations are decomposed into sequences of $T$-gates (and Cliffords), with a fixed worst-case length of $L_{\varepsilon} = \lceil c_{0}\log(1/\varepsilon) + c_{1} \rceil$. We determine this length using the gate synthesis methods available to RRE. Here, $\varepsilon$ refers to the diamond-norm error for decompositions into Clifford+$T$—see \cite{GridSynth,Kliuchnikov2022} for further details.

Generating each distilled $T$-state in the $T$-factories requires $C$ cycles, and we need one further surface code's $d$-cycle tock to interact with the graph node and measure it in either the $X$ or $Z$ basis. In this work, we ignore the time cost of any operation that requires $O(1)$ tocks (idling, deforming surface code patches, etc.), assuming we can perform them in parallel with other operations during sub-steps at no additional cost or with negligible delay. Therefore, for each non-Clifford measurement of memory nodes, assuming an $R_Z$-basis as an example, the volume increases as follows:
\begin{equation}
\begin{split}
    2\Tilde{n}_\text{logical}  \mathbb{L}_\text{prep} d \rightarrow &2\Tilde{n}_\text{logical}  \mathbb{L}_\text{prep} d + \\
    &(2\Tilde{n}_\text{logical}+n_{\text{per-leg}}^{\text{modules}}l_\text{transfer-bus})(C+d)L_\varepsilon~.
    \label{eq:vol_increase}
\end{split}
\end{equation}
Above, the second term reflects the fact that both graph node consumption and $T$-state distillation occur over a space of $2\tilde{n}_\text{logical} +n_{\text{per-leg}}^{\text{modules}} l_\text{transfer-bus}$. All nodes with distinct measurement basis types, $T$ and $R_Z$, must be measured sequentially.

If we assume the worst-case sequential execution, the upper bound on the full space-time volume becomes:
\begin{equation}
\begin{split}
    V^\text{sequential}_\text{total} &= 
        2\Tilde{n}_\text{logical} \mathbb{L}_\text{prep} d +\\
        &(2\Tilde{n}_\text{logical}+l_\text{transfer-bus}) (N^\text{seq}_\text{consump}d + N^\text{seq}_\text{distill} C)~.
\label{eq:V_seq}
\end{split}
\end{equation}
We define the total number of sequential $T$-states for graph consumption as
$N^\text{seq}_\text{consump} = \left \lceil \frac{n^T_\text{init}}{n'} \right \rceil + L_\varepsilon\left \lceil \frac{n^{R_z}_\text{init}}{n'} \right \rceil$. Recall that we assumed a simplified $T$-routing scheme that allows the auxiliary bus to serve up to $n'$ simultaneous magic states. FTQC can only consume distinct $T$ and $R_Z$ nodes in parallel. The number of sequential $T$-states to distill is
$N^\text{seq}_\text{distill} =  \left \lceil \frac{N^{T}_{tot}}{n'} \right \rceil$, where the total $T$-state measurements required is defined as $N^{T}_{tot} := n^{R_Z}_\text{init} L_\varepsilon + n^T_\text{init}$. The quantity $n^{T}_{tot}$ is the proxy we use for $T$-count and typically becomes larger than $N_\mathcal{G}$. We do not include $T$-factory space-time volumes, as their errors subsume into the output probability $p_{out}$ of the distilled state \cite[fig.~1]{Litinski2019magicstate}. If the select widget from \cref{table:widget} has a second (20-to-4) layer (the second and third widgets), then we must replace $n' \rightarrow 4n'$ in all space-time equations.

As detailed in \cref{sec:MWPM}, a single-cycle step of the surface code (with multiple physical operation layers) has a failure probability $p_C$ that characterizes the relationship between the physical and logical error rates, as given in \cref{eq:log_to_phys}. Consequently, we can express the logical failure probability of a $d$-cycle tock as:
\begin{equation}
    p_\text{logical} = 1 - (1 - p_C)^d.
\label{eq:logcell_to_C}
\end{equation}
The failure rate for a total volume of these $d$-cycle units, denoted as $V^\text{sequential}_\text{total}$, must satisfy the following condition:
\begin{equation*}
    p_{\text{algo-fail}} > 1 - (1 - p_\text{logical})^{V^\text{sequential}_\text{total}},
\end{equation*}
where $p_{\text{algo-fail}}$ represents the overall algorithm failure rate or error budget for executing the complete circuit $U$. This budget encompasses the entire space-time volume determined by the user within the parameter set $\{\text{config}\}$.

Assuming $p_C \ll 1$, we can approximate $(1 - p_C)^d \approx 1 - p_C d$, leading us to conclude that $p_\text{log-cell} \approx p_C d$.
Consequently,
\begin{equation*}
 p_{\text{algo-fail}} \gtrsim 1 - \left(1 - p_C d\right)^{V^\text{sequential}_\text{total}}.
\label{eq:p_algo_fail}
\end{equation*}
This equation highlights the relationship between the failure rates and the characteristic space-time volume for the complete QEC process.

Subtracting each side of \cref{eq:p_algo_fail} from 1 and taking the logarithm of both sides, we get
\begin{eqnarray}
    \log \left(1 - p_{\text{algo-fail}} \right ) \lesssim &V^{\text{sequential}}_{\text{total}} \log \left (1 - p_Cd \right) \nonumber
    \\ \lesssim & - V^{\text{sequential}}_{\text{total}} p_Cd. 
    \label{eq:p_algo}
\end{eqnarray}
If we substitute $V^{\text{sequential}}_{\text{total}}$ from \cref{eq:V_seq} into \cref{eq:p_algo}, the space-time logarithmic inequality to solve can be expressed as
\begin{equation}
\begin{split}
    \kappa d &\left(\frac{p}{p_{\text{thresh}}}\right)^{(d~+ 1)/2} \big(
    2\Tilde{n}_{\text{logical}} \mathbb{L}_{\text{prep}} d~+ (2\Tilde{n}_{\text{logical}} \\ 
    &+ n_{\text{per-leg}}^{\text{modules}}l_{\text{transfer-bus}}) (N^{\text{seq}}_{\text{consump}}d + N^{\text{seq}}_{\text{distill}} C)
    \big) < -J_1
\label{eq:logarithmic-eq}
\end{split}
\end{equation}
where $J_1 = \log(1-p_{\text{algo-fail}})$. The minimum code distance required is the smallest integer $d$ that satisfies \cref{eq:logarithmic-eq}. The only free variables are the cycle time of the $T$-factory, i.e.~$C$, and $\varepsilon$. We previously extracted $\mathbb{L}_{\text{prep}}$ and $\Tilde{n}_{\text{logical}}$ from the Cabaliser and Substrate Scheduler.  

To determine $C$ and $\varepsilon$, we perform nested loops to iterate through widgets and then compute $\varepsilon$ of the gate synthesis. The inner loop feeds back into the Clifford+$T$ decomposition, where the failure rate of a $d$-cycle tock, $p_{\text{logical}}$, sets the approximation error as $\varepsilon < p_{\text{logical}}$. By selecting a specific widget, we fix $C$, which also determines the output error rate of the distilled state, $p_{\text{out}}$. 

In a compiled quantum algorithm, we aim to ensure that the distilled $T$-states have precisely the same error rates as any other gates in the compiled circuit. A Pauli initialization or measurement occupies a $d$-cycle unit in the space-time volume. Thus, its failure rate is governed by \cref{eq:logcell_to_C}. This error rate must be dominant. Therefore, the error output, $p_\text{out}$, from the distillation circuit should be less than the failure rate of this cell. That is, we must ensure that the distillation output is at least as good as a unit cell in the space-time volume. Consequently, we first select the smallest value of $C$ from \cref{table:widget} and solve \cref{eq:logarithmic-eq} for $d$. We then verify that, for a physical error rate of $p=10^{-3}$ and our calculated solution for $d$, the following holds for the corresponding $p_\text{out}$ entry,
\begin{equation}
    p_{\text{logical}} = 1-(1-p_C)^d > p_\text{out}
\end{equation}
If this condition is unmet, we must choose a better-performing $T$-factory in the outer loop. We then proceed to the next entry on the list and repeat until we identify a $T$-factory that performs at least as well as $p_{\text{logical}}$.  

Once a specific $T$-factory is selected, the total number of allocated physical qubits across all modules at any given time and the total intra-modular runtime (including the first subgraph preparation time at step-$0$, all consumption operations, and delays) can be calculated as follows:
\begin{equation}
\begin{split}
     n&_\text{alloc-phys} = \\
     &4 \Tilde{n}_\text{logical} d^2 + n^\text{modules}_\text{per-leg} (n_\text{T-factories} Q + 2d^2l_\text{transfer-bus}), \\
     t&^\text{total}_{\text{consump}} = \\
     &8td(L^0_\text{prep} + N^\text{seq}_\text{consump}) + t^\text{delay}_\text{distill} + t^\text{delay}_\text{prep}~.
\label{eq:N_req_phys}
\end{split}
\end{equation}

In \cref{eq:N_req_phys}, $t$ denotes the characteristic gate time, which sets the architecture's quantum tock for all intra-modular operations. The factor of 8 arises because, in our architecture, each surface code sweep for stabilizer syndrome extraction circuits requires at most one initialization, two Hadamards, four entangling gates, and one measurement, all assumed to take time $t$ (see, for instance, \cite[fig.~9]{Fowler2012}). The temporal cost of initializing the first subgraph $g_0$ with $L^0_\text{prep}(S^0_\text{prep})$ is always nonzero and was explicitly included. FTQC performs the remaining graph preparations and $T$-state distillation operations in parallel with consumption operations. Overall delays may accumulate across time steps, defined as $t^\text{delay}_\text{distill} = \sum_i t^\text{distill-delay}_\text{i}$ and $t^\text{delay}_\text{prep} = \sum_i t^\text{prep-delay}_\text{i}$, which we detail below. 

To execute each consumption sub-schedule $i$, a specific number of distilled magic states must be present on the module's $T$-transfer bus. We must introduce a delay if we require more magic states than were accumulated on the $T$-transfer bus during the prior preparation step, $i-1$. The number of $T$ states per module after a preparation step is
\begin{equation}
\begin{split}
     n^{T-\text{per-module}}_i = \max \big(  \lfloor &\frac{n_\text{T-factories} t^\text{prep}_i}{8tC} \rfloor,  l_\text{transfer-bus} \big ) \\ &\text{for}~i\in \{0,\dots,n_\text{widgets}-1\}.
\end{split}
\end{equation}
Here, we have defined the hybrid (intra- and inter-modular) time required to prepare the subgraph for preparation step $i$ as $t^\text{prep}_i = 8d( N_i^\text{intra-comms}t + N_i^\text{cross-module} t_\text{inter} )$. Where 
$N_i^\text{intra-comms} = \sum_{j \in L_{i,j}^\text{prep}} \left \lfloor \frac{\mathbb{L}_i^\text{prep}}{\sum_{k \in L_{i,j}^\text{prep}} \lfloor d^\text{max}_{i,j,k}/\tilde{n}_\text{logical} \rfloor} \right \rfloor~\text{for}~n^\text{modules}_\text{per-leg}>1$ is the count of intra-module operations, while $N_i^\text{cross-module} = \sum_{j \in L_i^\text{prep}} \left \lceil \frac{\sum_{k \in L_{i,j}^\text{prep}} \lfloor d^\text{max}_{i,j,k}/\Tilde{n}_\text{logical} \rfloor}{n_\text{inter-pipes}} \right \rceil ~\text{for}~n^\text{modules}_\text{per-leg}>1$ represents the number of cross-module operations in the preparation sub-schedule $i$ (only in the vertical direction along a ladder leg). We assume that $t_\text{inter} \gg t$ is the characteristic physical timescale for all inter-module communications. Moreover, we define $d^\text{max}_{i,j,k}$ as the maximum distance between any pairs of graph nodes in the tuple $k$ of sub-list $j$ of preparation sub-schedule $i$ (see \cite{scheduler} for details), and $n_\text{inter-pipes}$ is the number of inter-modular pipes connecting each pair of modules in the macro-architecture. 

We can now define the distillation delay for step $i \in \{0,\dots,n_\text{widgets}-2\}$ as
\[
t^\text{distill-delay}_\text{i} = 
\begin{cases}
    8tC&\left \lceil \frac{ L^\text{max-seq}_{\text{consump},i} - n^{T-\text{per-module}}_i}{n_\text{T-factories}} \right\rceil, \\ &\text{if } L^\text{max-seq}_{\text{consump},i} > n^{T-\text{per-module}}_i \\
    0,              & \text{otherwise}
\end{cases}
\]
where $L^\text{max-seq}_{\text{consump},i} = n^{\text{max}-T}_i+L_\varepsilon n^{\text{max}-R_Z}_i$ is the maximum sequential $T$-length per module, considering all consumption nodes of step $i$. We have defined $n^{\text{max}-T}_i = \max_j(n^{T}_{i,j})~\text{for}~i\in S^\text{meas}_{i},~j\in \{0, \dots, n^\text{modules}_\text{per-leg}-1\}$ as the maximum number of $T$-basis measurement nodes in all modules specified by measurement sub-schedule $S^\text{meas}_{i}$. Similarly, $n^{\text{max}-R_Z}_i = \max_j(n^{R_z}_{i,j})~\text{for}~i\in S^\text{meas}_{i},~j\in \{0, \dots, n^\text{modules}_\text{per-leg}-1\}$.

Similarly, for step $i$, an idle delay must be added if preparation for the next subgraph $i+1$ is not yet complete, and this delay runs in parallel across the interleaving modules in the other leg of the ladder. An upper bound on the individual time for intra-modular node consumption in step $i$ is $t^\text{consump-intra}_i = 8td(\left \lceil \frac{n^{\text{max}-T}_i}{n_\text{T-factories}} \right \rceil + L_\varepsilon \left \lceil \frac{n^{\text{max}-Rz}_i}{n_\text{T-factories}} \right \rceil)$. Therefore, the preparation delay for step $i \in \{0,\dots,n_\text{widgets}-2\}$ becomes
\[
t^\text{prep-delay}_\text{i} = 
\begin{cases}
    t^\text{prep}_{i+1} &- t^\text{consump-intra}_i - t^\text{distill-delay}_\text{i}, \\
    &\text{if } t^\text{prep}_{i+1} >  t^\text{consump-intra}_i + t^\text{distill-delay}_\text{i} \\
    0.              & \text{otherwise}
\end{cases}
\]  

Lastly, we need to account for the total time required on the actual FTQC to stitch the output nodes of the graph at step $i$, located along one leg of the ladder, to the input nodes of the graph at step $i+1$, situated along the other leg of the ladder, for $i 
\neq n_\text{widget}-1$. We assume we can perform the hand-over process \textit{on-the-fly} by teleporting logical Bell pairs across the modules in a horizontal direction, which involves adding logical $CNOT$ or $CZ$ gates (or, for smaller problems, some equivalent $SWAP$ operations). We must incur a cost if FTQC needs to teleport graph nodes vertically within the same leg (which is infeasible in $O(1)$ operations). Therefore, an upper bound for this quantity can be expressed as
\begin{equation}
\begin{split}
t^\text{total}_\text{handover-inter} = 
8t_\text{inter}d\sum_{i \in \{0,\dots,n_\text{widget}-2\}} 
\left \lceil \frac{n^\text{cross-module-IO}_i}{n_\text{inter-pipes}} \right \rceil,
\end{split}
\end{equation}
where we define the total number of communications between cryostat modules needed to hand over output nodes of graph $i$ to $i+1$ as $n^\text{cross-module-IO}_i = \sum_{\Tilde{o}\in S^i_\text{output}, \Tilde{i} \in S^{i+1}_\text{input}} n^i_{\Tilde{o},\Tilde{i}}$; here, $n^i_{\Tilde{o},\Tilde{i}}$ is the number of crossings necessary to hand over node $\Tilde{o}$ of graph $i$ to node $\Tilde{i}$ of graph $i+1$ for $i \neq n_\text{widget}-1$.

Finally, the total FTQC hardware time can be expressed as
\begin{equation}
\begin{split}
    t_{\text{hardware}}^\text{total} = t^{\text{total}}_{\text{intra}} + t^{\text{total}}_{\text{handover-inter}} + t_{\text{decode-delay}}^{\text{total}}.
\end{split}
\end{equation}
In this equation, $t_{\text{decode-delay}}^{\text{total}}$ represents the overall delay incurred by performing decoding, entirely using classical methods, across all QEC cycles. We propose a method to estimate $t^{\text{decode-delay}}_{tot}$ in 
\cref{supp:RRE-outputs}, based on generic decoders.  

The following summarizes the workflow as algorithm~\ref{algo:algo1}.
\begin{algorithm}
 \caption{RRE's estimation methodology based on the fixed logical architectures of \cref{sections/graph-state-architecture}.}
 \label{algo:algo1}

 \SetKwData{Left}{left}\SetKwData{This}{this}\SetKwData{Up}{up}
 \SetKwFunction{Union}{Union}\SetKwFunction{FindCompress}{FindCompress}
 \SetKwInOut{Input}{input}\SetKwInOut{Output}{output}

  \Input{Logical quantum algorithm, \texttt{QuantumCircuit}, $U$}
  \Input{Look-up table: \texttt{$T$-factories} ($C$, $Q$, $p_\text{out}$)}
  \Output{Total allocated physical qubits, $n_\text{alloc-phys} = 4 \Tilde{n}_\text{logical} d^2 + n^\text{modules}_\text{per-leg} (n_\text{T-factories} Q + 2d^2l_\text{tansfer-bus})$}
  \Output{Total intra-modular time, $t^\text{total}_{\text{intra}} = 
     8td(L^0_\text{prep} + N^\text{seq}_\text{consump}) + t^\text{delay}_\text{distill} + t^\text{delay}_\text{prep}$} 

  \BlankLine

  \{$\Tilde{n}_\text{logical}$, $\Tilde{S}_\text{consump}$, $\Tilde{S}_\text{prep}$, $\Tilde{S}_\text{meas}$\} = Cabaliser-Tracker-Scheduler(\texttt{QuantumCircuit})\;

  \For{\texttt{widget} $\in$ \texttt{$T$-factories}}{
   \If{\texttt{widget} is (20-to-4)}{$n' \rightarrow 4n'$\;}
   \While{$M_\varepsilon == \text{False}$}{
    Solve $\kappa d \left(\frac{p}{p_\text{thresh}}\right)^{(d~+ 1)/2} \big(
    2\Tilde{n}_\text{logical} \mathbb{L}_\text{prep} d~+  (2\Tilde{n}_\text{logical}+n_{\text{per-leg}}^{\text{modules}}l_\text{transfer-bus}) (N^\text{seq}_\text{consump}d + 
    N^\text{seq}_\text{distill} C)
    \big) < -J$ for the smallest integer, $d$, where 
    $J = \log(1/(1-p_\text{algo-fail}))$, $p=10^{-3}$\;
    $p_\text{log-cell} = 1-(1-\kappa(p/p_\text{thresh})^{(d+1)/2})^d$\;
    $M_\varepsilon = p_\text{logical} > \epsilon$\;
    \If{$M_\varepsilon == \text{False}$}{Decrease $\varepsilon$\;}
   }
   $M_\text{logical} = p_\text{logical} > p_\text{out}$\;
   \If{$M_\text{log-cell} == \text{True}$}{Break loop\;}
  }

\end{algorithm}

%% file: sections/testcases.tex
To demonstrate resource estimation using the graph-state formalism, we generated circuits for selected pedagogical quantum algorithms that still target core computational challenges in real-world applications. Resource-efficient compilation necessarily involves hardware considerations in the synthesis methodology and the empirical evaluation of alternatives when analytical results are not readily available. As such, these circuits are then compiled into FT quantum computing patterns tailored to superconducting architectures presented in \cref{sections/graph-state-architecture}. This section summarizes the test cases selected for circuit synthesis to demonstrate our resource-estimation methodology.

\subsection{Quantum Fourier Transform}

The Quantum Fourier Transform (QFT) is a common component of quantum algorithms, including Quantum Phase Estimation. From the perspective of hardware-specific resource analysis, QFTs are likely to feature in application-relevant resource analysis, yet can be scaled down to a small handful of logical input gates, allowing us to trace and debug compilation workflows in our tool chain at the level of logical operations.  This makes QFT circuits ideally suited as the first test cases of our estimation framework. This work analyzes QFT circuits on $n_\text{input}$ logical qubits, where $n_\text{input} = 4, 10, 25, 50, 75, 100, 250, 500$ or $1000$.

\subsection{Hamiltonian simulations}
Simulation of lattice-based Hamiltonians encompasses a family of critical applications investigated by researchers in quantum chemistry and condensed matter physics. Moreover, simulating dynamics may be more useful for smaller hardware systems than methods that evaluate static properties, such as the ground state.

Hamiltonian simulation consists of evolving an initial state $\ket{\psi_0}$ to a state at time $t$, $\ket{\psi}$ according to,
\begin{equation}
\ket{\psi} = e^{-iHt} \ket{\psi_0}~,
\label{eq:hamiltonian-evolution}
\end{equation}
where $H$ is the Hamiltonian of the system of interest. We consider the simulation of two families of Hamiltonians: 2D Transverse-Ising models and 2D Fermi-Hubbard models. Each of these will be described in more detail below.

\subsubsection{Transverse-Ising Hamiltonian Simulations}
The Los Alamos National Laboratory (LANL) has created a public repository containing circuit generation and problem analysis for problems that may be amenable to quantum computing and are of interest to the scientific community \cite{LANLqc}.  Drawing inspiration from their investigation of magnetic lattices, we perform resource estimation for the simulation of a time-invariant Transverse-Ising Hamiltonian,
 \begin{equation}
    H = \sum_{\langle \textbf{jk} \rangle} \, Z_{\textbf{j}} \otimes Z_{\textbf{k}} + 0.1 \sum_{\textbf{j}} \, X_{\textbf{j}}.
    \label{eq:transverse-ising-ham}
\end{equation}

Here, $X_{\textbf{i}}$ and $Z_{\textbf{i}}$ represent the Pauli X and Z operators, respectively, acting on lattice site $\textbf{i}$, and angle brackets denote lattice sites connected by an edge.

Based on the studies documented by LANL, we carried out resource estimates of simulation of the Hamiltonian in \cref{eq:transverse-ising-ham} for a total time of $t=1$ (in the inverse Hamiltonian units) on a 2D triangular lattice (sometimes called a \textit{2D hexagonal Bravais lattice}) of size $N \times N$ where $N$ is an integer between $2$ and $19$.

\subsubsection{Fermi-Hubbard Hamiltonian Simulations}

We also investigated the Fermi-Hubbard simulation as a core computational capability for solving problems in room-temperature superconductivity \cite{Agrawal2024}. These problems typically involve studying the ground state energy, state configurations, and other observable system properties, including time-dependent quantities.

The Fermi-Hubbard Hamiltonian we consider is given by,
\begin{align}
    H = & -J\sum_{\langle \textbf{j},\textbf{k}\rangle,\sigma} \left[c^\dagger_{\textbf{j},\sigma} c_{\textbf{k},\sigma} + c^\dagger_{\textbf{k},\sigma} c_{\textbf{j},\sigma}\right] \nonumber \\ &
    +J'\sum_{\langle \langle \textbf{j},\textbf{k} \rangle \rangle,\sigma} \left[c^\dagger_{\textbf{j},\sigma} c_{\textbf{k},\sigma} + c^\dagger_{\textbf{k},\sigma} c_{\textbf{j},\sigma}\right] \nonumber \\ &
    + U \sum_{\textbf{j}} n_{\textbf{j},\uparrow} n_{\textbf{j},\downarrow} \nonumber \\ 
    &+ \frac{h_z}{2} \sum_{ \textbf{j}} (n_{\textbf{j},\uparrow} - n_{\textbf{j},\downarrow}) \nonumber\\
    &+ \mu  \sum_{\textbf{j}, \sigma} n_{\textbf{j},\sigma},
\end{align}
where single and double angled brackets denote all nearest-neighbor and next-nearest-neighbor lattice site pairs, respectively.  $c^\dagger_{\textbf{j},\sigma} ( c_{\textbf{j},\sigma}$) creation (annihilation) operators of a spin-$\sigma$ fermion on the $\textbf{j}$th lattice site where $\sigma \in \{\uparrow,\downarrow\}$. $J$ (primed or unprimed) and $U$ capture the magnitudes of the hopping and repulsion energies, respectively. $n_{\textbf{k},\sigma} = c^\dagger_{\textbf{k},\sigma} c_{\textbf{k},\sigma}$ is the number operator, $h_z$ is an external applied magnetic field, and $\mu$ is the chemical potential.

To motivate the specific Fermi-Hubbard simulations for which we will obtain resource estimates, we reviewed the literature to identify configurations of interest to the community. Broadly speaking, methods for simulating a Fermi-Hubbard system can be classified as stochastic or deterministic. As mentioned in \cite{schneider2021simulating}, deterministic methods do not \enquote{suffer from numerical issues seen in stochastic simulations when dealing with systems with non-zero chemical potential}. Moreover, this instability due to the \enquote{sign problem} becomes prohibitive for moderate $\mu \sim 0.3$. This suggests separating instances into those with $\mu = 0$ and those with $\mu > 0.3$. Here, we reference a few exemplary studies in the literature on stochastic and deterministic methods.

Some examples of stochastic studies include a study using Auxiliary-field quantum Monte Carlo (AFQMC) to estimate ground states on lattice sizes of $4 \times 4$ to $16 \times 16$, $U/J = 4, 8$, and $J' = 0$. \cite{qin2016benchmark}. In addition,  \cite{huang2019strange} used Determinant Quantum Monte Carlo (DQMC) to study doped ground state estimation using  $U/J = 6$, $J'/J = -.25$, and lattice sizes of $8 \times 8$, $12 \times 8$, $12 \times 12$, $16 \times 8$. 
In contrast, examples of deterministic studies include work on numerical linked-cluster expansions \cite{park2021thermodynamics}. In this study, researchers selected random repulsion strengths from an uncertainty \enquote{box} about $U_0$. Simulations were for a periodic $10 \times 10$ lattice with $U_0/J = 8$, $J'=0$. Other researchers used Projected Density Matrix Embedding (PDME) to perform a 2D Fermi-Hubbard study with $U/J = 2, 4, 6, 8$, $J'=0$ on a $40 \times 40$ lattice \cite{wu2019projected}. Finally, in \cite{schneider2021simulating}, the authors use Projected Entangled Pair States (PEPS) with a bond dimension of $D=10$, and claim that accuracy increases with $D$. Most of their results are from a $3 \times 4$ honeycomb lattice with $U/J \in [0,6]$ and $J'=0$. Still, the authors were able to calculate the ground state via imaginary time evolution on a $15 \times 15$ bipartite honeycomb lattice.

Of note is that in \cite{schneider2021simulating} it was determined that the time required to run these studies on a CPU varies like, 
$t_{\text CPU} \propto L_xL_y ( 2 \chi^3 D^4 d^2 + \chi^2 D^2 d^2 + \chi^2 D^4d^4),$
where $\chi$ is the truncated bond dimension and $d$ the problem dimension. Memory scaled as,
$2\chi^2 D^4d^2M_{\text num} + 2N\chi^2D^2dM_{\text num}$. The authors empirically found that $\chi = 3D$ was sufficiently accurate for their studies, with an error of about $1\%$.
These estimates provide a useful guidepost for the tipping point at which FTQCs may outperform classical computers on these types of problems.

Taking these studies as examples of the types of systems of interest to the community it suggests obtaining resource estimates for Fermi-Hubbard Hamiltonian simulations consisting of a 2D square lattice for total time $t=1$, with $J = 1$, $J' = 0$, $h_z = 0$, $\mu = 0$, $U=2$, and lattice sides of length $N=2$ to $20$, $30, 50, 75$ and $100$.

\subsubsection{Quantum algorithm and approaches for Hamiltonian simulations}
\newcommand{\epsqsp}[0]{\epsilon_{\text{qsp}}}

Owing to its optimal scaling characteristics for queries to the block encoding of the Hamiltonian \cite{low2017optimal}, we have chosen Quantum Signal Processing (QSP) as the Hamiltonian simulation algorithm for which we obtain a resource estimate. QSP is briefly described below, but more details can be found in \cite{low2017optimal}, \cite{low2019hamiltonian}, \cite{martyn2021grand}, and \cite{dalzell2023quantum}.

Given a Hamiltonian $H$ and simulation time $t$, QSP performs a sequence of single-qubit rotations, parameterized by a sequence of \textit{phase angles}, interlaced with a quantum walk-like operator. This allows one to apply a polynomial approximation of $e^{-iHt}$ to the input target state. The sequence of phase angles determines the polynomial approximation. It is computed offline, using a classical algorithm, with the desired error tolerance specified by the user-defined parameter $\epsqsp$. $\epsqsp$ is also inversely related to the probability of success of the QSP algorithm. Thus, for an accurate simulation, with a high probability of success, $\epsqsp$ may have to be set relatively small. For implementation details, we follow the examples accompanying the \verb|pyLIQTR| \cite{pyLIQTR} package \texttt{v1.3.3}, as well as the work done in \cite{martyn2023efficient,zeytinoglu2022}.

For demonstration purposes, we set the desired probability of success of our algorithm, $p_{\rm algo}$, to be equal to 0.95. In real-world applications, this is typically fixed to meet outcome requirements (e.g., accuracy and wall-time). Denoting the probability of success of a QSP circuit, $p_{\rm qsp}$, we therefore require,
\begin{equation}
p_{\rm algo-fail} \leq p_{\rm qsp}.
\label{eq:epsqsp-prob1}
\end{equation}
Since circuit size is inversely related to $\epsilon_{\text{qsp}}$, it is prudent to choose $\epsilon_{\text{qsp}}$ to be as large as possible while still guaranteeing that $p_{\rm qsp}$ exceeds the algorithm target of 0.95. According to \cite{low2017optimal} $(1 - 2 \epsilon_{\text{qsp}}) \le p_{\rm qsp}$. Thus, to ensure \cref{eq:epsqsp-prob1} is satisfied we set,
\begin{equation}
1 - 2 \epsilon_{\text{qsp}} = p_{\rm algo-fail},
\label{eq:epsqsp-prob2}
\end{equation}
and so in this work we choose $\epsilon_{\text{qsp}}$ to be equal to $0.025$.

\subsection{Simulation circuit synthesis}

For our Hamiltonian simulation circuits, we used \verb|pyLIQTR| \texttt{v1.3.3} to construct the Hamiltonians and synthesize the QSP circuits. The \verb|pyLIQTR| package provides tools that construct the Hamiltonians, calculate the necessary phase angles for QSP simulation, and generate QSP circuits. In addition, \verb|pyLIQTR| allows the user to request random phase angles rather than solving for the precise values required for an accurate simulation. In our work, we used random phase angles. This is appropriate because we only perform resource estimation, not numerically accurate simulations. The correct \textit{number} of phase angles is used in the circuits, but their numerical \textit{values} are random. This provides a conservative overestimate of the resource estimations without explicitly incurring the time-consuming step of solving for the phase angles.
For Fermi-Hubbard models, we utilized \verb|pyLIQTR|'s \texttt{FermiHubbard} model class, and employed the \texttt{FermiHubbardSquare} block encoding, which is based on \cite{babbush2018encoding}. 

The same overall approach was taken for the time evolution of the Transverse-Ising models. However, for these problems, we used \verb|pyLIQTR|'s \texttt{Heisenberg} model class, and \texttt{PauliLCU} block encodings. For QFT instances, we used the QFT circuits provided by the \texttt{qft} method in \verb|cirq|.

%% file: sections/results.tex
\begin{table*}[!tbh]
    \caption{Rigetti's resource estimation results on a superconducting fault-tolerant quantum computing architecture, \cref{sections/graph-state-architecture}, with space-optimized graph-state compilation. Here, ``\# Logical bus patches'' includes patches for the memory and auxiliary buses that mediate logical measurements in the bilinear pattern shown in \cref{fig:log-layout}. Likewise, ``\# $T$ patches'' includes logical patches for all $T$-factories and $T$-transfer Buses.}
    \centering
    \footnotesize
    \begin{tabular}{cc|c|c|c|c|c|c|c|c}
         \\
         \multicolumn{2}{c|}{Input circuit} & \multicolumn{2}{c|}{Logical resources} & \multicolumn{6}{c}{Superconducting FTQC resources}
         \\
         Instance & $\Tilde{n}_\text{logical}$ & \makecell{\# Logical \\bus patches} & \# T patches & \makecell{Total allocated phys.\\qubits, $n_\text{alloc-phys}$} & \makecell{Total modules,\\$2n^\text{modules}_\text{per-leg}$} & $t_{\text{tot}}^{\text{hardware}}$ & $P_{\text{Diss},\text{4K}}$ & $P_{\text{Diss},\text{20mK}}$ & $E_{\text{tot}}$ \\
        \midrule
\multicolumn{2}{l|}{\emph{Fermi Hubbard}}&&&&&&&&  \\[2pt]
2x2 & 36 & 280 & 4.05k & 1.32M & 2 & 4.09s &840W&168nW& 668Wh \\
4x4 & 74 & 332 & 3.32k & 1.21M & 2 & 35.1s &840W&168nW& 5.73kWh \\
8x8 & 182 & 728 & 2.07k & 1.38M & 2 & 9.49m &840W&168nW& 93kWh \\
16x16 & 578 & 2.31k & 1.25k & 3.30M & 4 & 6.9h &1.68kW&336nW& 8.11MWh \\
20x20 & 878 & 3.52k & 1.4k & 4.96M & 6 & 37.8h &2.52kW&504nW& 66.7MWh \\
50x50 & 5.05k & 20.2k & 5.54k & 36.1M & 42 & 7.14y &17.6kW&3.53$\mu$W& 772GWh \\
100x100& 10.1k & 40.7k & 19.3k & 114M & 132 & 63.3ky &55.4kW&11.1$\mu$W& 21.5TWh \\
\multicolumn{2}{l|}{\emph{QFT}}&&&&&&&&\\[2pt]
N-4& 4 & 88 & 6.38k & 1.10M & 2 & 2.8ms &840W&168nW& 457 mWH \\
N-10& 10 & 248 & 5.5k & 1.29M & 2 & 41.7ms &840W&168nW& 6.81Wh \\
N-25 & 25 & 2.02k & 3.75k & 1.31M & 2 & 427ms &840W&168nW& 69.7Wh \\
N-100 & 100 & 3.39k & 2.76k & 3.18M & 4 & 10.4s &1.68kW&336nW& 3.39kWh \\
N-500 & 500 & 18.4k & 4.5k & 17.5M & 20 & 2.03m &8.4kW&1.68$\mu$W& 199kWh \\
N-1000 & 1k & 37.2k & 9.57k & 43.4M & 52 & 10.6m &21.8kW&4.37$\mu$W& 2.69MWh \\
\multicolumn{2}{l|}{\emph{Transverse-Ising}}&&&&&&&&   \\[2pt]
2x2 & 67 & 652 & 2.43k & 1.41M & 2 & 64.4s &840W&168nW& 10.5kWh \\
5x5 & 212 & 2.98k & 816 & 4.39M & 6 & 6.64h &2.52kW&5.04nW& 6.42MWh \\
10x10 & 654 & 10.1k & 2.78k & 16.7M & 20 & 5.06d &8.4kW&1.68$\mu$W& 714MWh \\
19x19 & 2.21k & 41.1k & 13.5k & 87.1M & 110 & 82.6d &46.2kW&9.24$\mu$W& 64.1GWh \\
        \bottomrule

    \end{tabular}
    \label{tbl:RRE_results}
\end{table*}

Key physical resource requirements for the test cases described in \cref{sec:test_cases} across various problem sizes are summarized in \cref{tbl:RRE_results}. We demonstrate that our tool, RRE, can estimate resource requirements for problem sizes consistent with the literature and that extend beyond the scales of current studies.

Our methodology enables us to extend the state of the art in several ways. \textit{(1)} A full accounting for resources for both T counts, but also for executing the Clifford portions of the circuit as well. \textit{(2)} A breakdown of physical and temporal resources allocated to portions of the execution of the algorithm, namely: Clifford, $T$-state, and teleportation between modules. \textit{(3)} A quantitative estimation of the impact on total resources of the performance of subsystems.

\begin{figure}[!h]
    \centering
    \includegraphics[width=0.99\linewidth]{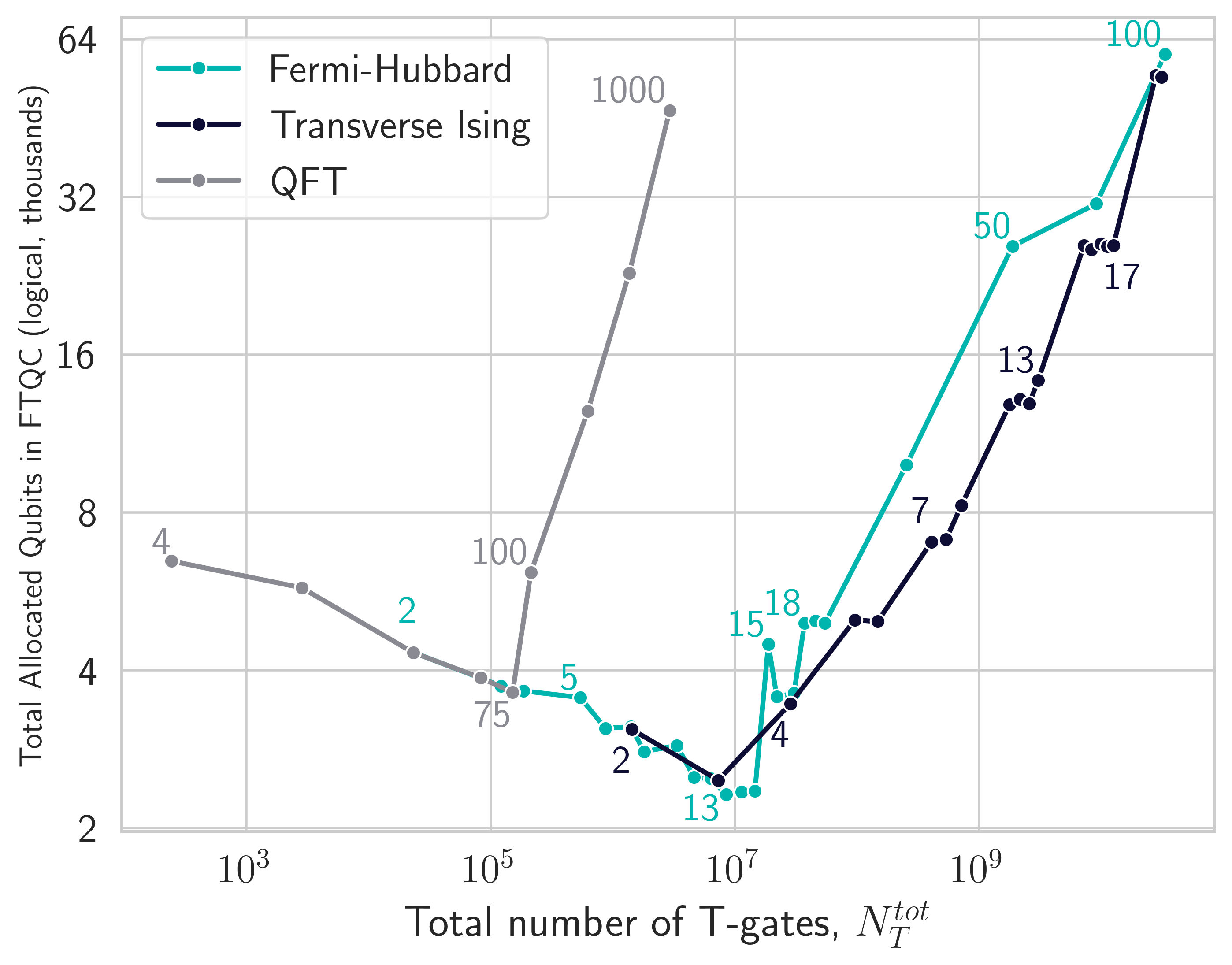}
    \caption{Total allocated logical qubits (in the bilinear pattern) and \texttt{T-count} for three sets of test cases. Grey indicates Quantum Fourier Transform results for sizes 4 to 1000. The black curve shows the Hamiltonian evolution of the Transverse-Ising model on triangular lattices ranging from 2x2 to 19x19. The teal line shows estimates from Hamiltonian simulations of a Fermi-Hubbard model on a square lattice, with sizes ranging from 2x2 to 100x100.}
    \label{fig:log_qubits_tcount}
\end{figure}

\begin{figure}[!h]
    \centering
    \includegraphics[width=0.99\linewidth]{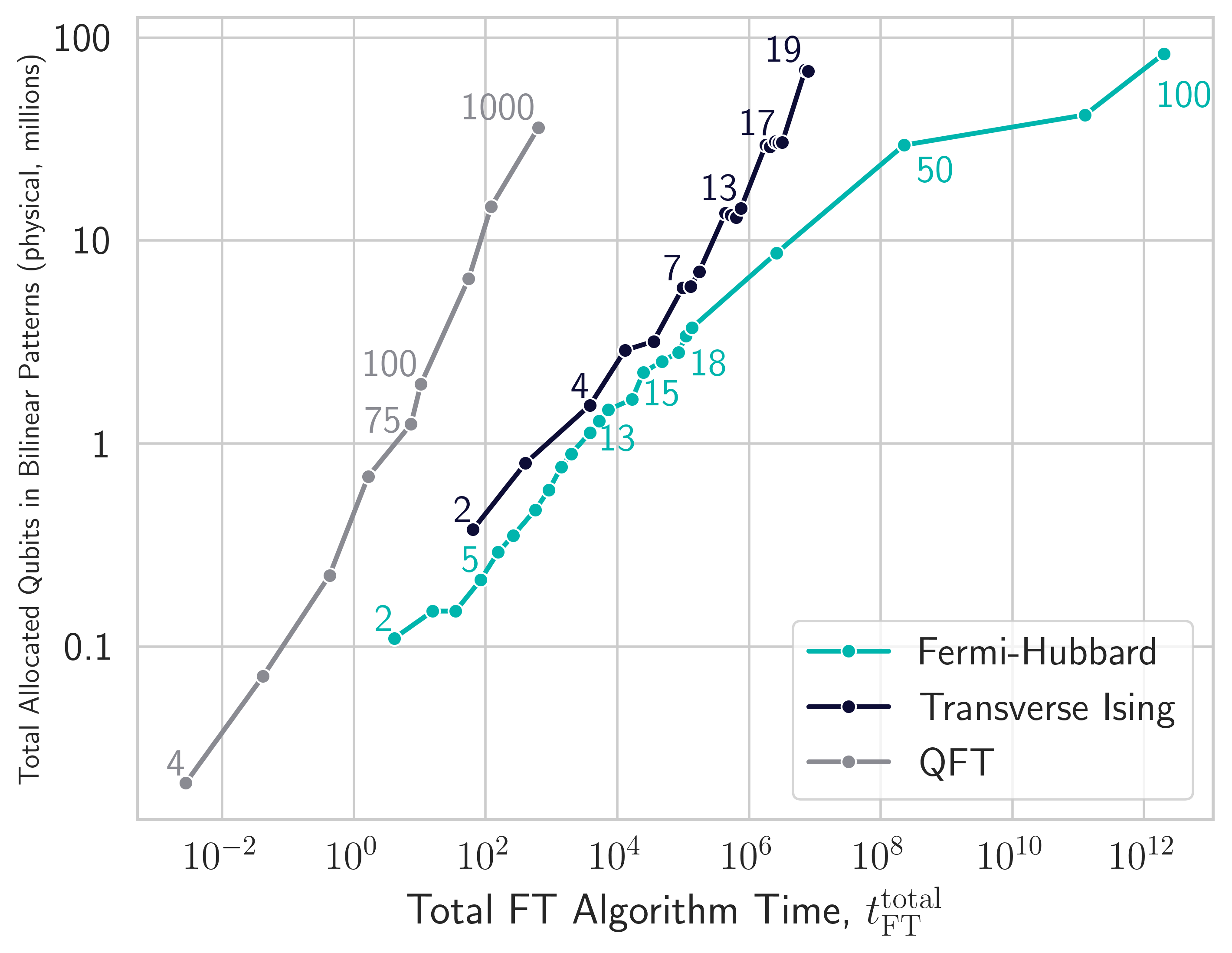}
    \caption{Total allocated physical qubits (in the bilinear pattern) and FT hardware runtime for three sets of test cases. Grey indicates Quantum Fourier Transform results for sizes 4 to 1000. The black curve shows the Hamiltonian evolution of the Transverse-Ising model on triangular lattices ranging from 2x2 to 19x19. The teal line is an estimate for the Hamiltonian simulation of a Fermi-Hubbard model on a square lattice with sizes ranging from 2x2 to 100x100. We set the Hamiltonian evolution time to 1 in units of $1/J$ for both the Transverse-Ising and Fermi-Hubbard models.}
    \label{fig:bus_runtime}
\end{figure}

When we visualize this ensemble of compiled logical resources and physical resource requirements (\cref{fig:log_qubits_tcount} and \cref{fig:bus_runtime}, respectively), we observe trajectories of runtime and auxiliary bus size that increase with problem size. This provides validation of the reasonableness of our methodology and its utility as a tool to support future studies. Interestingly, we observe a decreasing trend in the number of logical qubits for small-sized problems until the number of modules increases. As the number of modules increases, we observe a sharp rise in the number of logical qubits. This is because physical qubits are assigned to $T$-factories rather than to logical memory qubits or the auxiliary bus. As the problem size grows, the number of physical qubits used in the bilinear quantum memory and auxiliary bus pattern increases in the single-module regime. This reduces the number of $T$-factories that can be accommodated within the module. As the $T$-factories are large monolithic regions of logical qubits, removing them gives a net reduction in the number of logical qubits in use, even though the number of patches in the bilinear pattern is larger. The available storage space in a room can decrease even as smaller objects are added, provided that those smaller objects require removing larger objects to accommodate them.

\begin{figure*}
     \centering
     \begin{subfigure}[b]{0.3\textwidth}
         \centering
         \includegraphics[width=\textwidth]{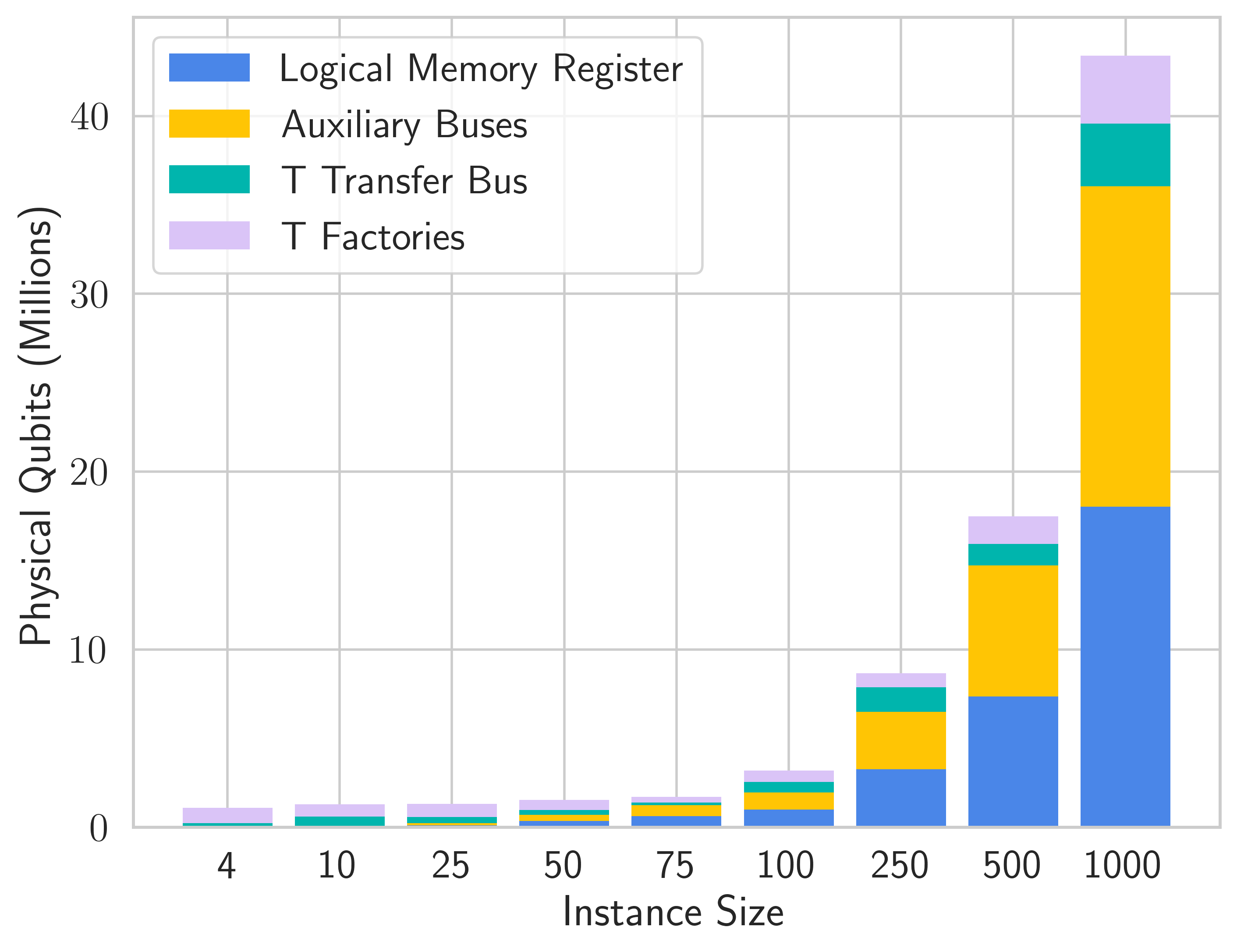}
         \caption{QFT}
         \label{fig:qubit_alloc_qft}
     \end{subfigure}
     \hfill
     \begin{subfigure}[b]{0.3\textwidth}
         \centering
         \includegraphics[width=\textwidth]{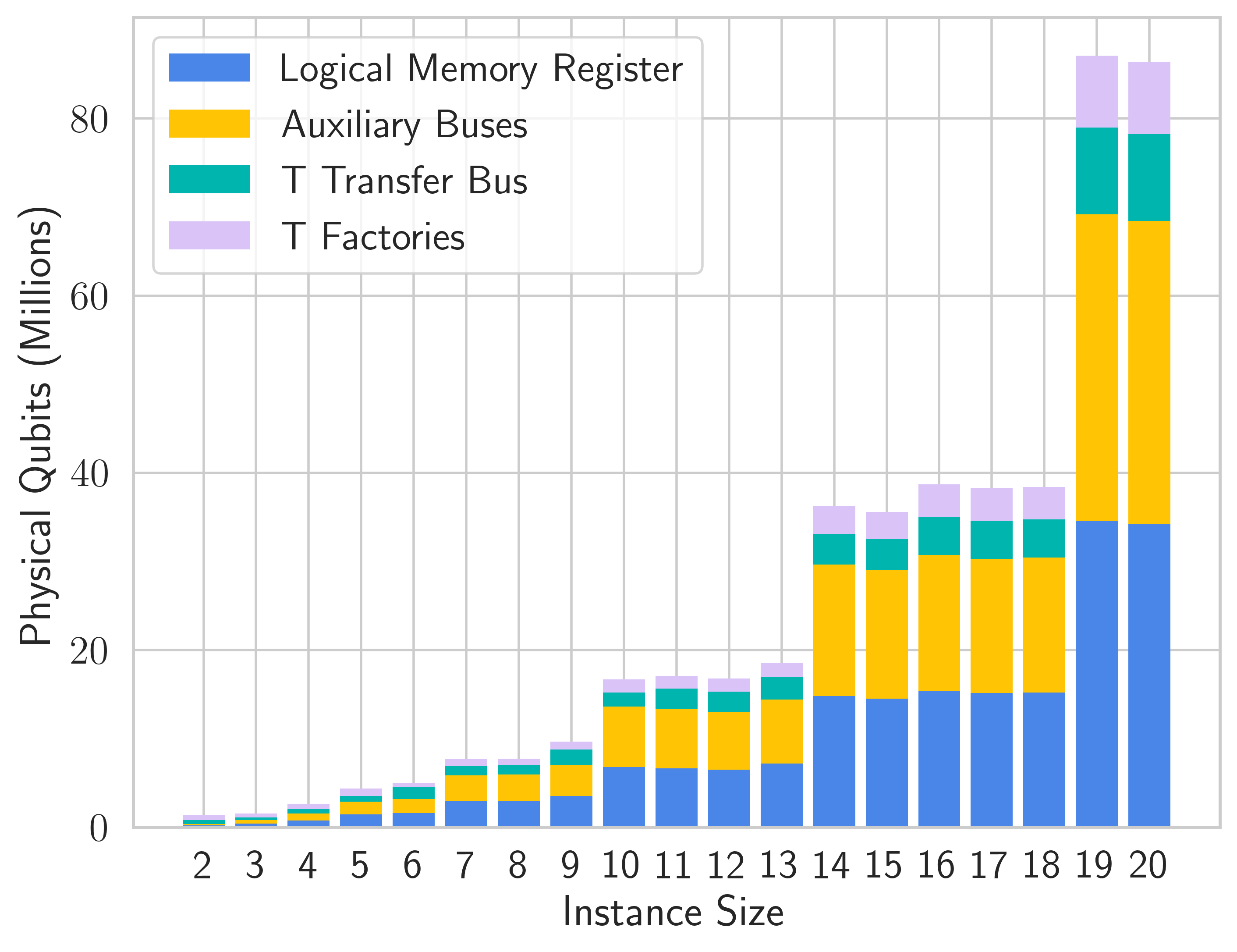}
         \caption{Transverse-Ising}
         \label{fig:qubit_alloc_ti}
     \end{subfigure}
     \hfill
     \begin{subfigure}[b]{0.3\textwidth}
         \centering
         \includegraphics[width=\textwidth]{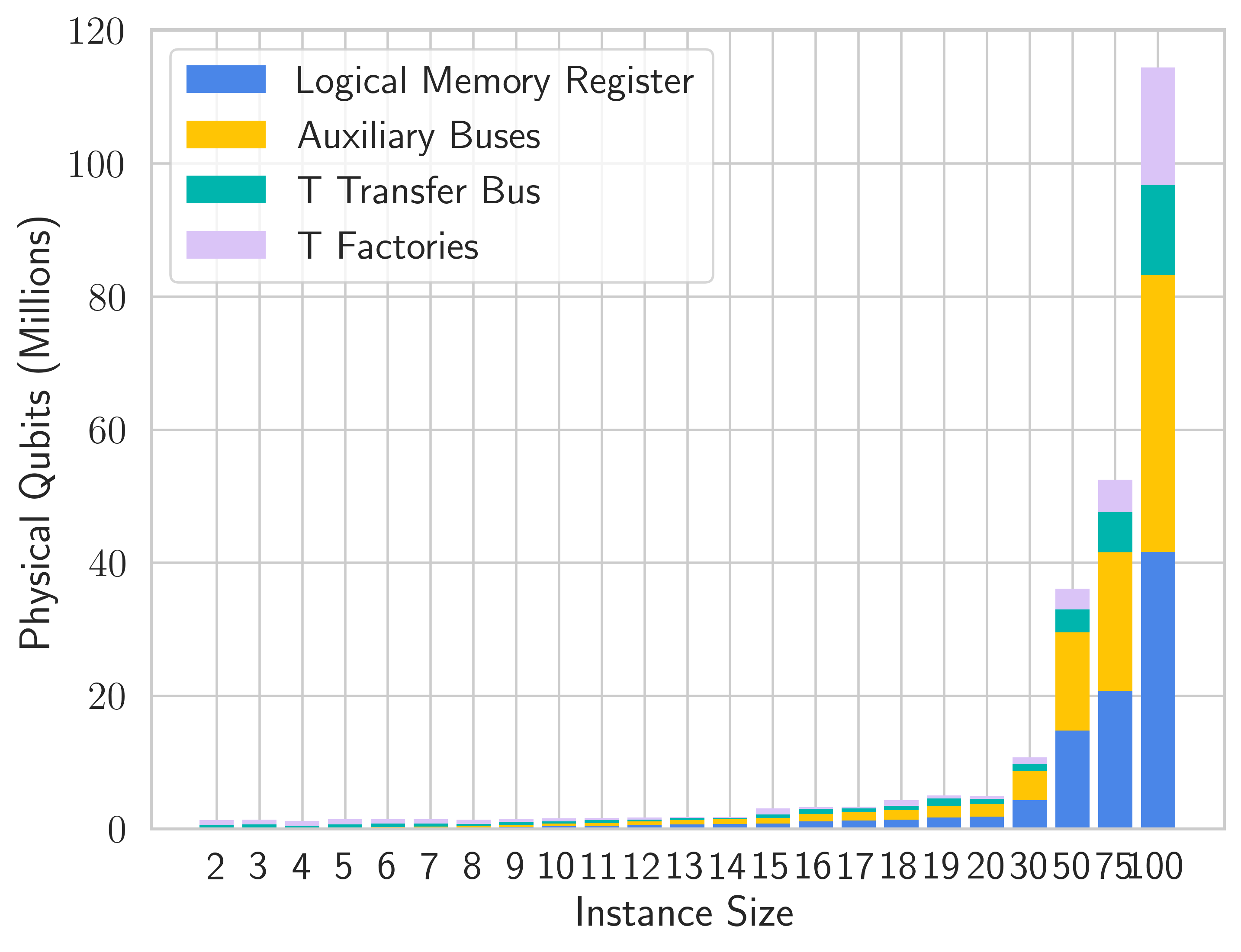}
         \caption{Fermi-Hubbard}
         \label{fig:qubit_alloc_fh}
     \end{subfigure}
        \caption{Physical qubit allocations for the types of logical qubits described in the logical micro-architecture shown in \cref{fig:log-layout}; those involved in the logical memory register, auxiliary buses (for single-qubit measurements, queuing, and routing), the $T$-transfer bus, and $T$-factories.}
        \label{fig:qubit_alloc}
\end{figure*}

The full accounting of both Clifford and $T$-state contributions to total resources, \textit{(1)}, is shown in \cref{fig:qubit_alloc} for all families of applications for which we estimated resources. It is interesting to note that conventional $T$-counting methodologies do \emph{not} often reveal resource competition between the Clifford and $T$-state generations of a quantum architecture. As problems scale, physical qubits must be allocated to meet the increased requirements of Clifford processing, thereby reducing the number of available physical qubits for MSD. This, in turn, reduces the number of $T$-factories that can be applied to a calculation, thereby increasing runtime in a non-linear fashion.

\begin{figure*}
     \centering
     \begin{subfigure}[b]{0.3\textwidth}
         \centering
         \includegraphics[width=\textwidth]{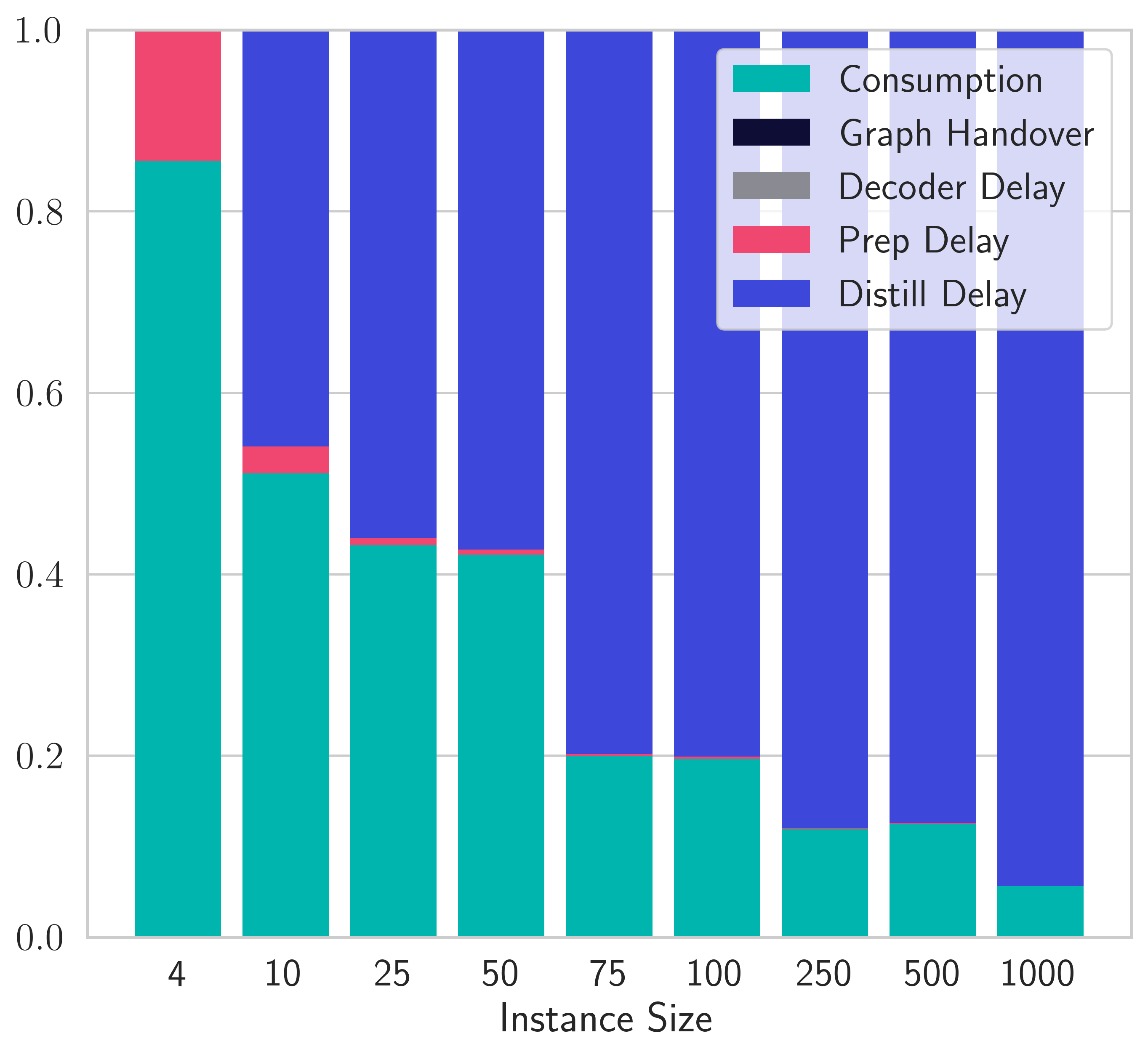}
         \caption{QFT}
         \label{fig:time_alloc_qft}
     \end{subfigure}
     \hfill
     \begin{subfigure}[b]{0.3\textwidth}
         \centering
         \includegraphics[width=\textwidth]{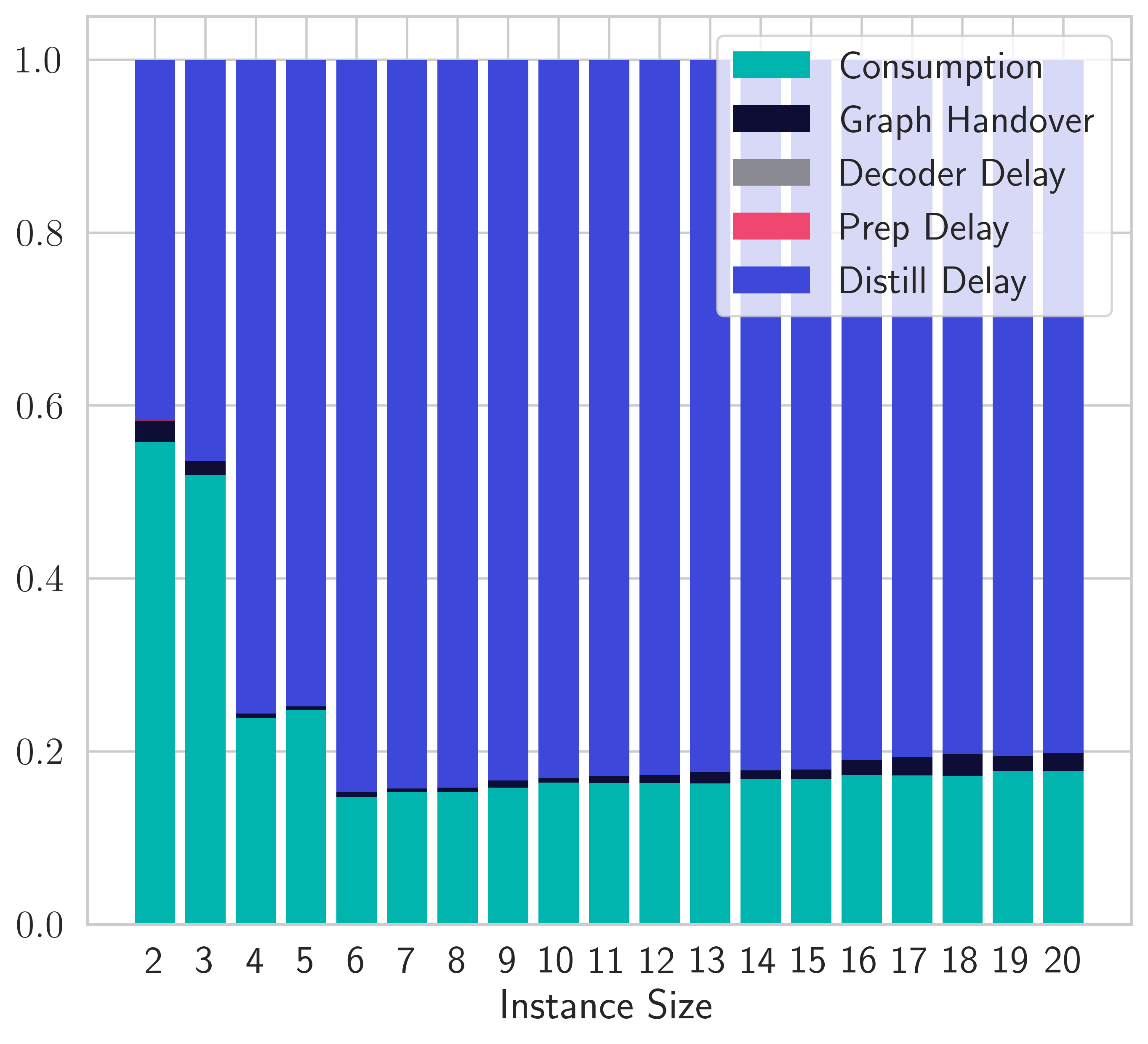}
         \caption{Transverse-Ising}
         \label{fig:time_alloc_ti}
     \end{subfigure}
     \hfill
     \begin{subfigure}[b]{0.3\textwidth}
         \centering
         \includegraphics[width=\textwidth]{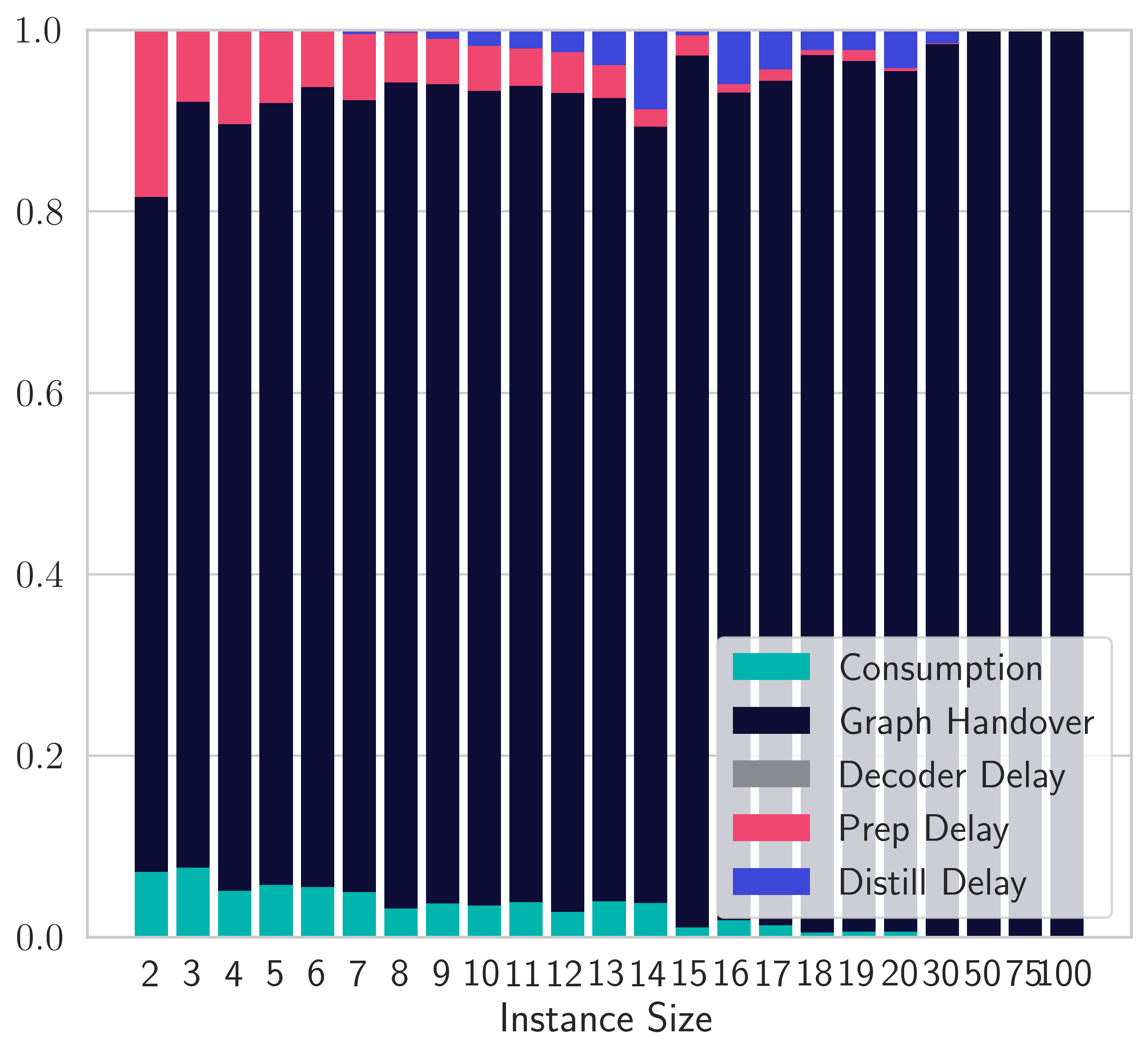}
         \caption{Fermi-Hubbard}
         \label{fig:time_alloc_fh}
     \end{subfigure}
        \caption{Proportion of FTQC's compute time allocated to graph state consumption (using distilled $T$-states), intermodule communications, distillation, and decoding delays for time evolution problems.}
        \label{fig:time_alloc}
\end{figure*}

\begin{figure}[!h]
    \centering
    \includegraphics[width=0.99\linewidth]{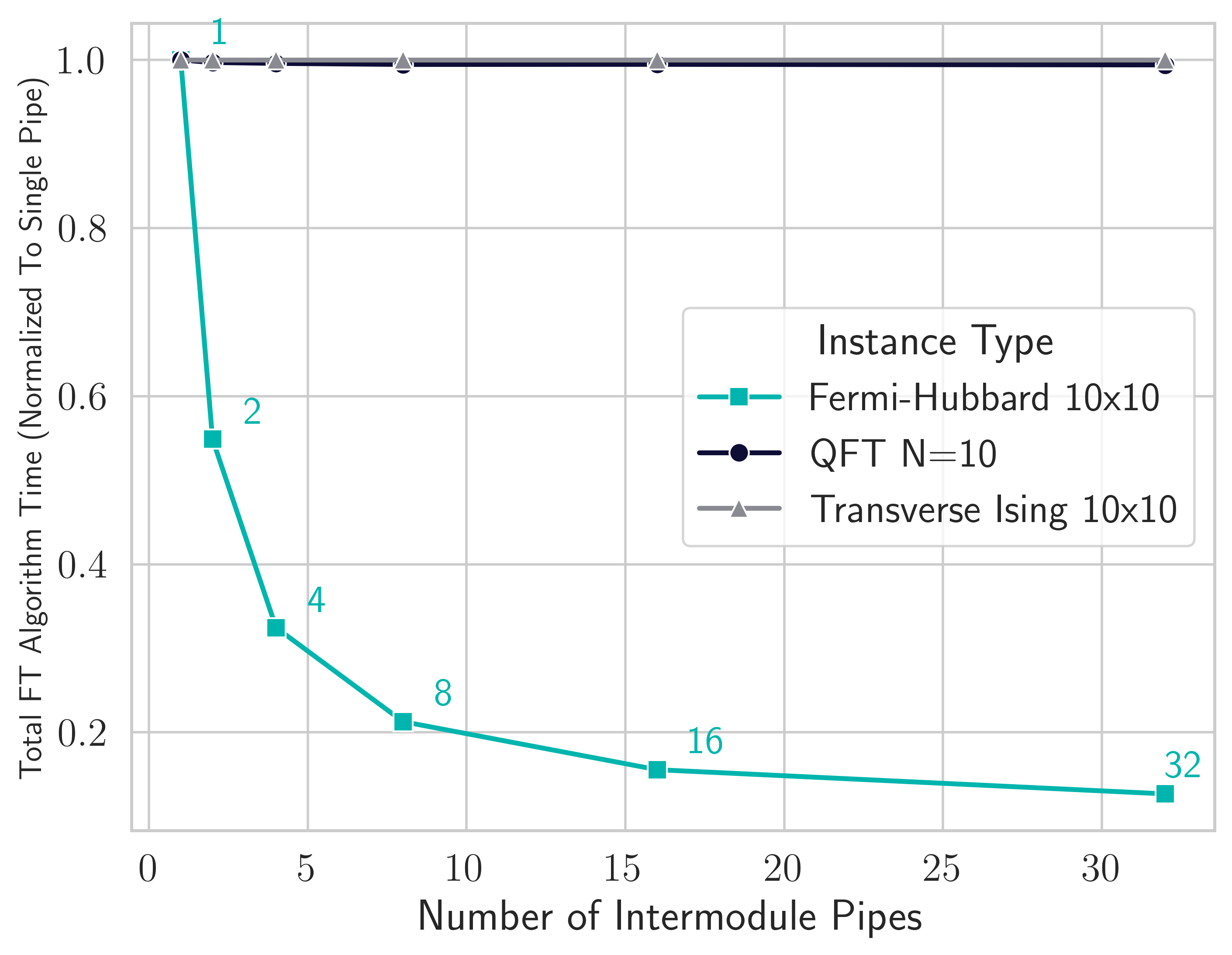}
    \caption{Comparison of normalized total FT runtime and number of intermodule pipes. The number of intermodule pipes varies from 1 to 32. Runtime is normalized to the corresponding value for one pipe. Black indicates Quantum Fourier Transform results of size 10. The grey plot is the Hamiltonian evolution of the Transverse-Ising model on a triangular lattice of size 10. The teal line represents estimates for the Hamiltonian simulation of a Fermi-Hubbard model on a 10x10 square lattice.}
    \label{fig:sweep}
\end{figure}

\begin{figure}
     \centering
         \centering
         \includegraphics[width=0.91\columnwidth]{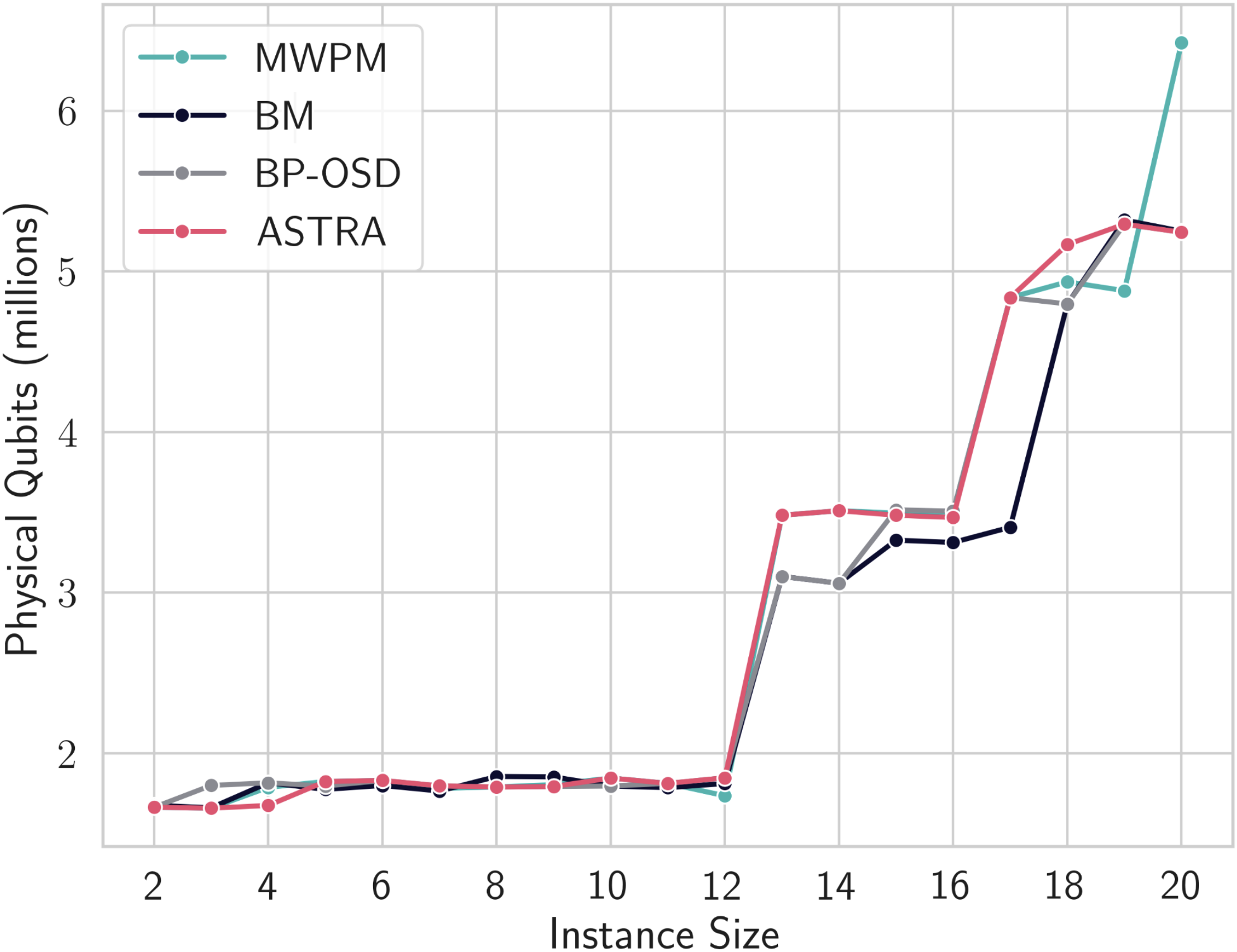}
         \caption{Total number of allocated physical qubits (parameter 35 of \cref{supp:RRE-outputs}) for Hamiltonian evolution of the Fermi-Hubbard of various sizes. Line colors denote different decoders; Teal indicates MWPM, black is the Belief Matching (BM) decoder, grey is Belief Propagation with ordered statistics decoding (BP-OSD), and magenta is Astra \cite{maan2024machine}.}
         \label{fig:decodersscale}
\end{figure}

This impact on the breakdown of total runtime, \textit{(2)}, is evident in \cref{fig:time_alloc}. Here, the contribution to graph consumption runtime (using distilled $T$-states) increases dramatically for the largest applications considered. This is because fewer physical qubits are allocated to create higher precision $T$-states, increasing the runtime penalty. The impact of runtime on teleportation in the distributed architecture is also evident in this figure. Despite a relatively slow inter-module interaction time (5x slower than the native code cycle), the contribution to overall runtime is application dependent. For QFT and transverse Ising applications, a relatively small portion of the overall runtime is spent performing this intermodule interaction, even at large problem sizes. This shows that computation can be partitioned across specific architectures and algorithms to enable distributed computing with relatively low impact. However, this appears to be algorithm-dependent, and for Fermi-Hubbard problems, the penalty for distributed quantum computation is high.

Moreover, we can test the sensitivity of runtime to parameters in this architectural model, \textit{(3)}. In \cref{fig:sweep}, we select an instance from each family, vary the number of inter-module connections, and calculate the impact on runtime. We define this in terms of a \textit{pipe} where the number of connections equals the required code distance for the algorithm. Additional pipes will enable parallel node-to-node teleportation between widgets. This is done to quantify the penalty associated with not having a fully connected lattice of physical qubits between modules.  Here, we observe that while runtime increases with fewer inter-module connections, the marginal benefit of adding more connections diminishes. Once again, this impact is highly application-dependent.

Similarly, we can directly quantify the impact of changing the decoders on the physical system size. Results for Fermi-Hubbard instances of increasing problem size are shown in \cref{fig:decodersscale} and demonstrate the utility and flexibility of our resource estimation tool for assessing the effects of changes across multiple subsystems within the overall architecture.

%% file: sections/future.tex
The results from Section~\ref{sec:results} can be improved by focusing on at least the following two aspects: a) using a decoder that is faster and may have better scaling than the standard choice of MPWM; b) optimizing the widget circuits by, for example, reducing circuit gate counts and depths. Below, we present the savings potential of these two approaches, along with additional details on the design of high-cooling-power cryostats.

\subsection{Replacing the MWPM decoder}

The MWPM decoder (e.g., \cite{paler2023pipelined}) is the de facto standard for resource estimation. Other decoders have been proposed, but MWPM remains the most efficient with respect to the decoding speed-physical error rate trade-off: it can handle high error rates while maintaining polynomial complexity. There are faster, lower-polynomial-complexity decoders (e.g., union-find variants or pure belief propagation) and decoders with linear complexity, which are challenging to scale to high distances (e.g., machine learning decoders). Nevertheless, to the best of our knowledge, no alternative decoder exists that has, at the same time, a lower degree of polynomial decoding complexity and a threshold higher than MWPM.

\begin{figure}[!h]
    \centering
    \includegraphics[width=0.99\linewidth]{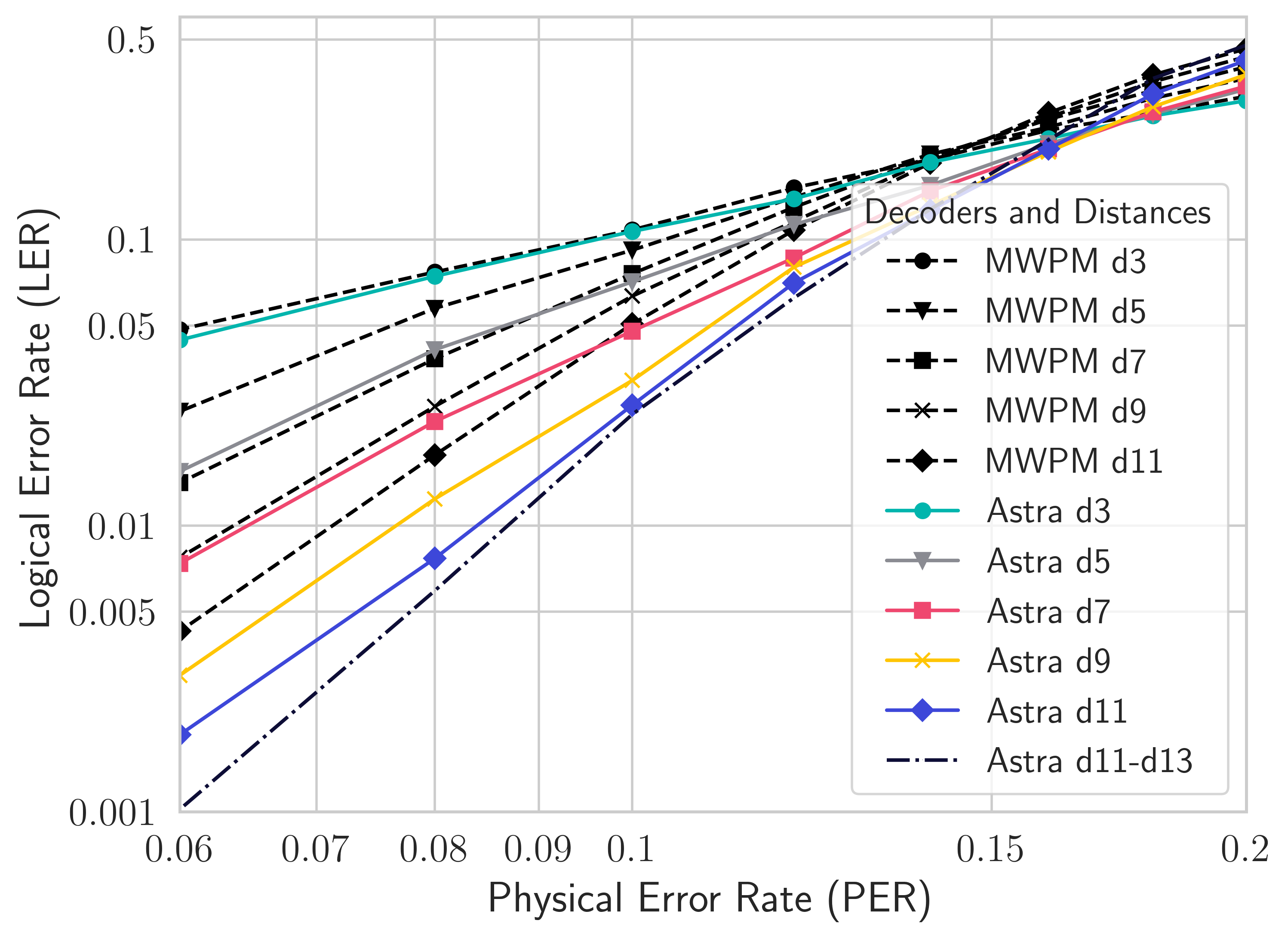}
    \caption{The Logical Error Rate (LER) for code capacity depolarising noise of Astra vs MWPM. Astra has a threshold of $\sim 17\%$, and MWPM has a threshold of $\sim 14\%$. Astra clearly outperforms MWPM in terms of LER. In particular, Astra d9's results are better than those of MWPM d11.}
    \label{fig:gnn}
\end{figure}

Herein, we report results in \cref{fig:gnn} for a machine-learning decoder that outperforms MWPM: it has a higher threshold and linear runtime complexity. Our machine learning decoder, Astra~\cite{maan2024machine}, uses graph neural networks (GNNs) in a novel way: we learn a message-passing algorithm on the Tanner graph of the code. We can train such decoders within hours on gaming GPUs for surface codes at distances up to 11. In contrast to other machine learning decoders, the training time of our decoder is significantly lower (e.g.~\cite{varsamopoulos2019comparing}) for comparable depths.

We train distinct GNN decoders for each distance using syndromes generated by passing the surface codes through a code-capacity noise channel. This channel assumes that the physical data qubits are affected by noise with a probability $p$ and that the syndrome measurements are perfect. For this work, we do not report the decoding of circuit-level correlated noise (e.g.,~\cite {paler2023pipelined}) and focus entirely on correcting depolarizing noise.

After training our decoders, we evaluate their decoding performance numerically and capture the decoder's performance through an equation of the type from Eq.~\ref{eq:log_to_phys}:
\begin{equation}
p[{\mathrm{GNN}}] = 0.56 \left (\frac{p}{0.17} \right )^{\frac{d + 1}{2}}~.
\label{eq:gnn_scaling}
\end{equation}

Typically, the performance-scaling equations for the surface code are obtained from simulations with circuit-level noise (e.g.,~\cref{eq:log_to_phys}) because this is considered the most accurate model. Nevertheless, the code capacity is more straightforward to benchmark for preliminary decoder versions. To enable comparison with Astra, we also benchmark MWPM with code capacity noise. The results are illustrated in Fig.~\ref{fig:gnn}. Based on the MWPM code capacity errors, we obtain the MWPM scaling equation as
\begin{equation}
  p[{\mathrm{MWPM}}] = 0.52 \left (\frac{p}{0.14} \right )^{\frac{d + 1}{2}}~.
\label{eq:mwpm_scaling}
\end{equation}

We observe that Astra has a higher threshold than MWPM ($17\% > 14\%$) and provides faster (better) logical error suppression ($0.56 > 0.52$). This implies that by replacing the MWPM with Astra, one can achieve the same logical error rate with lower-distance surface codes. After benchmarking its performance under circuit-level noise, future work will focus on quantifying the exact savings achieved by using the new decoder.

\subsection{Ultra-large scale circuit optimization}

The Fermi-Hubbard circuit instances analyzed in the previous sections are, for practical regimes, too large in terms of gate counts. Compiling, analyzing, and optimizing such circuits at graph-state or Clifford+T level is challenging: the state-of-the-art quantum software tools cannot handle ultra-large-scale circuits.

Widgets (detailed in~\cref{supp:decomp}) are subcircuits representing relevant parts of the larger circuits. Due to their size, widgets can be more easily analyzed and optimized. Nevertheless, even widgets are complicated to optimize for metrics such as \texttt{T-count} and $T$-depth. Optimization is a highly complex task, and most optimization heuristics are not designed (or do not perform as expected) for (sub)circuits of tens of logical qubits and hundreds of gates.

Tools like \texttt{pyLQTR} are built on top of other quantum software frameworks, such as Google Cirq, which were not designed for extreme scalability. We bridge this gap by developing and implementing a software tool that can easily handle ultra-large-scale circuits. Our tool, Pandora~\footnote{\url{https://github.com/ioanamoflic/pandora/}}, is based on PostgreSQL and leverages the capabilities of modern relational database software supporting multi-threaded operations on quantum circuits.

Pandora provides a simple API for developing new and more complex optimization methods. We interfaced our tool with Google Qualtran~\cite{harrigan2024expressing} and can easily extract large circuits for analysis and optimization. Currently, the circuits and widgets available in Qualtran have been manually optimized; therefore, automatic optimization is even more challenging. Therefore, the results from Table~\ref{tab:ulsqc} are far from optimal circuits.

Table~\ref{tab:ulsqc} reports preliminary results on optimizing arithmetic widgets as presented in~\cite{nam2018automated}. The latter are ubiquitous building blocks of practical application circuits, such as the Fermi-Hubbard model. Our circuit optimization procedure is not as complex as the one described in ref.~\cite{nam2018automated}. We implemented a heuristic that uses multiple threads for applying circuit template rewrites~\cite{nam2018automated}: 1) decomposition of gates (e.g., $TOFFOLI$); 2) canceling $H$, $T$, and $X$ gates; 3) canceling $CNOT$ gates; 4) commuting gates to the left and the right; 5) reversing directions of $CNOT$ gates.

To conclude, quantum software for ultra-large-scale circuit analysis and optimization can potentially lower the resources required for practical quantum computing applications. Pandora is a first step towards optimizing at scale.

\begin{table}[!t]
    \centering
    \footnotesize
    \begin{tabular}{r | r | r | r}
    Circuit &  Orig. \texttt{T-count} & Opt. \texttt{T-count} & Impr.\\
    \hline
    Adder8 & 266 & 236 & 11.27\% \\
    Adder16 & 602 & 536 & 10.96\% \\
    Adder32 & 1274 & 1144 & 10.20\% \\
    Adder64 & 2618 & 2368 & 9.54\% \\
    Adder128 & 5306 & 4648 & 11.64\% \\
    Adder256 & 10682 & 9258 & 13.33\% \\
    Adder512 & 21434 & 18562 & 13.39\% \\
    Adder1024 & 42938 & 38216 & 10.99\% \\
    Adder2048 & 85946 & 76056 & 11.50\%
    \end{tabular}
    \caption{Widget optimization results: addition circuits from Maslov~\cite{nam2018automated} are optimized for \texttt{T-count}. The original \texttt{T-count} is reported in the second column, and the optimized \texttt{T-count} in the third column. The optimization of Adder2048 required approx.~one hour.}
    \label{tab:ulsqc}
\end{table}

\subsection{Cryostat design}

The detailed design of high-cooling-power cryostats for operation at this scale is a significant challenge that will require dedicated effort over the coming years. For the resource estimates in this paper, we have produced total dissipation power (TDP) estimates for both the 4\,K and 20\,mK stages. These are presented in columns $P_{\text{Diss},\text{4K}}$ and $P_{\text{Diss},\text{20mK}}$ in \cref{tbl:RRE_results}, respectively. While the numbers appear daunting by today's standards, several facilities provide well in excess of kW of cooling power at 2\,K, such as the Linear Coherent Light Source II (LCLS II) at the Stanford Linear Accelerator Center (SLAC). This system can produce 4\,kW of cooling power at 2\,K and 18.5\,kW at 4.5\,K. While we consider a design for such a system beyond the scope of the current work, we believe it is feasible within the timescales required to develop it.

%% file: sections/RRE-outputs.tex
Building on the workflow of the graph-state-based fault-tolerant resource estimation method outlined in the main text, we summarize and reiterate the selected output parameters of RRE, providing additional details where needed. We detail $49$ logical architecture parameters or hardware-relevant outputs, which RRE writes to \texttt{stdout} and \texttt{CSV} files by default, and explain how to calculate them. The delivered \texttt{CSV} file can include results for additional output parameters. Users can add custom quantity classes for estimation in RRE using class templates from the \texttt{resources.py} module.

\begin{itemize}

\item \textbf{Parameter 1: Code distance, $d$}

$d$ denotes the surface code distance of logical patches. For the surface code, $\left\lfloor(d-1)/2\right\rfloor$ errors can be corrected before a logical operation fails. All non-distillation logical patches have size $2d^2$ and serve as foundational elements for architectural components in the FTQC ($T$-factories are parameterized by their own code-distance sets).

We need to calculate the minimum viable $d$ precisely and iteratively according to Algorithm \ref{algo:algo1}, which requires satisfying a logarithmic inequality in Step 6, also referred to in \cref{eq:logarithmic-eq}.

\item \textbf{Parameter 2: Number of logical qubits in the quantum memory register, $\Tilde{n}_\text{logical}$}

The term $\Tilde{n}_\text{logical}$ refers to the maximum number of logical nodes needed to process the complete graph state $\mathcal{G}$ according to the widgetization-based set $\Tilde{S}_\text{est}$ at any given time. In other words, this represents the maximum quantum memory required. Each complete quantum memory register chain, shared among the modules on the same leg of the macro-architecture, contains $\Tilde{n}_\text{logical}$ logical qubits. To prevent any quantum memory bottlenecks or delays, we allocate a fixed quantum memory and auxiliary bus space of size $2\lceil \Tilde{n}_\text{logical} / n^\text{per-leg}_\text{modules} \rceil$ to all modules and steps of the schedules. There may be more efficient methods for assigning the bilinear quantum register and auxiliary bus spaces. 

Note that the maximum quantum memory required for the complete $U$ is a distinct quantity, $n_{\text{logical}}$. For widgetized circuit input, we derive a proxy for this quantity from the subgraphs as $\Tilde{n}_{\text{logical}} = \max( n^{g_0}_{\text{logical}}, \cdots , n^{g_{n_{\text{widgets}}-1}}_{\text{logical}}) \approx n_{\text{logical}}$. Cabaliser can efficiently output each $n^{g}_{\text{logical}}$ for sufficiently small graph states.

\item \textbf{Parameter 3: Number of $T$-factories per module, $n_\text{T-factories}$}

The number of Litinski-style $T$-factories per module is calculated as $n_\text{T-factories} = 
\left \lfloor \frac{l_\text{edge}-1}{\left \lfloor\frac{l_\text{width}}{2d^2} \right \rfloor} \right \rfloor n^\text{col}_\text{T-factories}$. The factories send purified magic states to the $T$-transfer and auxiliary buses required for measurements on the logical qubits during the consumption stage. For more details, see \cref{section:t-gate-error}.

\item \textbf{Parameter 4: Number of \emph{logical memory} qubits per module}

The number of logical memory, or graph-state, or data qubits in the bilinear pattern in every module is equal to $\left \lceil \Tilde{n}_\text{logical} / n^\text{per-leg}_\text{modules} \right \rceil$. These correspond to the blue tiles in \cref{fig:log-layout}. In other words, every module on a leg of the macro-architecture contains the same number of logical memory qubits. Although the last module may require fewer active nodes for quantum operations, the same space is always allocated to this component.

\item \textbf{Parameter 5: Number of \emph{physical memory} qubits per module}

The number of memory, or graph-state, or data physical qubits in the bilinear pattern in every module is equal to $2d^2\lceil \Tilde{n}_\text{logical} / n^\text{per-leg}_\text{modules} \rceil$.

\item \textbf{Parameter 6: Number of \emph{logical} auxiliary qubits in the auxiliary bus per module}

The number of logical auxiliary qubits in the auxiliary bus section of each module is equal to $
\left \lceil \Tilde{n}_\text{logical} / n^\text{per-leg}_\text{modules} \right \rceil$. These correspond to the yellow tiles in \cref{fig:log-layout}, which FTQC uses to facilitate logical measurements on the logical graph-state qubits. Each module on a leg of the macro-architecture contains the same number of logical auxiliary qubits. Although the last module may need fewer active nodes for quantum operations, the same space is always allocated for this component.

\item \textbf{Parameter 7: Number of \emph{physical} qubits in the auxiliary bus per module}

The number of physical qubits in the auxiliary bus portion of every module is equal to $2d^2 \left \lceil \Tilde{n}_\text{logical} / n^\text{per-leg}_\text{modules} \right \rceil$.

\item \textbf{Parameter 8: Number of \emph{logical} qubits in the $T$-transfer bus per module}, $l_\text{transfer-bus}$

The number of logical qubits in the $T$-transfer bus portion of every module is calculated as $l_\text{transfer-bus} = (l_\text{edge}-l_\text{qbus} - n^\text{col}_\text{T-factories} \left \lceil\frac{L_\text{length}}{\sqrt2d} \right \rceil)l_\text{edge} + n^\text{col}_\text{T-factories} \left \lceil\frac{L_\text{Length}}{\sqrt2d} \right \rceil$. These correspond to the teal tiles in \cref{fig:log-layout}, which FTQC uses to actively store and transfer purified $T$-states from MSD factories to the auxiliary bus.

\item \textbf{Parameter 9: Number of \emph{physical} qubits in the $T$-transfer bus per module}

The number of physical qubits in each module's $T$-transfer bus portion is $2d^2l_\text{transfer-bus}$.

\item \textbf{Parameter 10: Number of logical qubits allocated to all $T$-factories in each module}

The number of logical nodes assigned to all $n_\text{T-factories}$ $T$-factories in each module can be calculated as $n_\text{T-factories}\left \lceil L_{\text{length}}/(\sqrt{2}d)\rceil\lceil L_{\text{width}}/(\sqrt{2}d) \right\rceil$. These correspond to purple tiles in \cref{fig:log-layout}.

\item \textbf{Parameter 11: Number of physical qubits allocated to all $T$-factories in each module}

The number of active physical qubits across all $n_\text{T-factories}$ factories per module can be computed as $n_\text{T-factories} Q$. After choosing a factory from the $T$-factory Look-up Table \ref{table:widget}, $Q$ becomes the number of physical qubits the factory requires for magic state distillation.

\item \textbf{Parameter 12: Total number of modules in both legs of macro-architecture}

The total number of interconnected QPU cryo-modules on both legs of the macro-architecture can be computed as $2 n^\text{per-leg}_\text{modules}$. The parameter $n^\text{per-leg}_\text{modules}$ is the number of modules per leg or the height of the ladder structure. See \cref{fig:2_modules} for details.

\item \textbf{Parameter 13: Total number of coherent interconnects in the macro-architecture}

The total number of width-$d$ coherent interconnects in the ladder macro-architecture can be computed as $n_\text{inter-pipes}\left(3 n^\text{per-leg}_\text{modules}-2\right )$. FTQC uses these connections to perform inter-module graph preparation operations on the same-leg or teleporting output to input nodes on the opposite-leg modules. See \cref{fig:2_modules} for details.

\item \textbf{Parameter 14: Number of allocated physical qubits per module}

The number of allocated physical qubits per module for all components can be computed as $2 \left \lceil n_\text{logical} / n^\text{per-leg}_\text{modules} \right \rceil d^2 + n_\text{T-factories}Q + 2l_\text{transfer-bus}d^2$.

\item \textbf{Parameter 15: Number of algorithmic logical qubits}, $n_\text{input}$

The number of algorithmic logical qubits or width, $n_\text{input}$, of the complete input circuit, $U$. For the time-sliced widgetization we perform on the input algorithm (cf.~\cref{fig:widgetization}), all widgets have the same number of input and output nodes, $n_\text{input}$.

\item \textbf{Parameter 16: Diamond-norm precision for gate synthesis, $\varepsilon$}

Extensive analytical and numerical studies (see, for example,~\cite{GridSynth, Kliuchnikov2022}) have established that a non-Clifford 1Q rotation gate can be systematically approximated as a chain of Clifford+$T$, a process known as gate synthesis. Here, the upper bound for $T$-length scales as $L_\varepsilon = \left \lceil c_0\log_2(1 / \varepsilon) + c_1 \right \rceil$. The diamond-norm precision, $\varepsilon$, indicates the rate at which FTQC utilizes the gate synthesis decomposition of arbitrary-angle rotations to perform $Rz$-basis measurements at the end, during each consumption step. We assume the use of the state-of-the-art decomposition approach of \emph{mixed-fallback} \cite{Kliuchnikov2022} by default, which offers $c_0=0.57, c_1=8.83$, representing an improvement over the previously employed \enquote{GridSynth} method \cite{GridSynth}, which had $c_0=3.0, c_1=0$.  

RRE internally processes non-Clifford angle decompositions, meaning the constructed graphs do \textit{not} depend on the $\varepsilon$-value and can be cached. The diamond-norm precision must be determined accurately and iteratively, following Algorithm 
\ref{algo:algo1}, which naturally involves solving a logarithmic equation in Step 6.

We argue that $\varepsilon$ should not be left for the user to set arbitrarily (although RRE allows users to impose a fixed value if needed). Extreme caution is required here because precisions \textit{are}, indeed, equivalent to the fault-tolerant resources required: if $\varepsilon$ is chosen too large, the rate condition in Step 6 of the algorithm \ref{algo:algo1} is not met, while if it's too small, the space-time resources will become wasteful and excessively large.

\item \textbf{Parameter 17: Total number of $T$-gates, or~\texttt{T-count}, or $N^{tot}_T$}

In the graph-state-based estimation method, our \emph{proxy} for the total \texttt{T-count}, or $N^{\text{tot}}_T$, is the total number of $T$-basis measurements required after the arbitrary-angle $R_z$'s are synthesized at the measurement points of the consumption stage. Therefore, we can write \texttt{T-count} as $N^{\text{tot}}_T = n^{Rz}_\text{init} L_\varepsilon + n^{T}_\text{init} = n^{Rz}_\text{init} \left \lceil c_0\log_2(1 / \varepsilon) + c_1 \right \rceil + n^{T}_\text{init}$.

\item \textbf{Parameter 18: Effective $T$-depth}

This parameter estimates the effective $T$-depth for the input logical circuit, $U$. This can be computed as $\left \lceil n^T_{tot} / \Tilde{n}_\text{logical} \right \rceil$.

\item \textbf{Parameter 19: Number of $Rz$-gates in the input circuit,} $n^\text{init}_{Rz}$

We denote the number of non-Clifford-rotation $Rz$-gates in the complete input algorithm, $U$, whether written explicitly or contained within other logical gates, as $n^\text{init}_{Rz}$. All non-Clifford $Rz$ gates must undergo gate-synthesis decomposition \cite{GridSynth,Kliuchnikov2022} to Clifford+$T$ at the end, during the consumption stage.

\item \textbf{Parameter 20: Number of $T$,$T^\dagger$-gates in the input circuit,} $n^\text{init}_{T}$

We denote the number of $T$ and $T^\dagger$ gates in the complete input algorithm, $U$, whether written explicitly or contained within other logical gates, as $n^\text{init}_{T}$. We perform this count
before performing gate-synthesis Clifford+$T$ decomposition of non-Clifford $R_z$-gates.

\item \textbf{Parameter 21: Number of Cliffords in the input circuit,} $n^\text{init}_\text{Clifford}$

The number of explicit Clifford gates in the complete input algorithm, $U$, which is denoted as $n^\text{init}_\text{Clifford}$.

\item \textbf{Parameter 22: Number of graph state nodes, $N$}

$N$ represents the total number of logical nodes in the complete graph state, $\mathcal{G}$, or the graph size. For widgetized circuits, we calculate a proxy for this quantity from subgraphs as $\Tilde{N} = \sum_{i=0,...,n_\text{widgets}-1} N_i + \sum_{i=0,...,n_\text{widgets}-2} n^\text{outputs}_i \approx N$. Here, $n^\text{outputs}_i = n_\text{input}$ denotes the number of output nodes at step $i$ that must be added to the graph stitched from subgraphs to enable Bell pair teleportations.

\item \textbf{Parameter 23: Total number of measurement steps in the consumption schedule}

The total number of sequential measurement steps in the consumption schedule over all widgets (repeated or not). In other words, the total number of subsets in the nested sets in the consumption schedule outputted by Cabaliser\cite{Cabaliser}. This quantity governs the runtime of the quantum operation for the graph consumption stage. 

\item \textbf{Parameter 24: Total number of measurement steps in the preparation schedule,} $\mathbb{L}^\text{prep}$

The total number of sequential measurement steps in the preparation schedule over all widgets (repeated or not). In other words, the total number of sub-sets in the nested sets in the preparation schedule outputted by Substrate Scheduler\cite{scheduler}. This quantity governs the runtime of the quantum operation for the graph preparation stage.

\item \textbf{Parameter 25: Total number of widgets or widgetization time steps}, $n_\text{widgets}$

The total number of widgets (repeated or not), which is equivalent to the number of time steps in $U$ given the time-sliced decomposition presented in \cref{fig:widgetization} and \cref{supp:decomp}.

\item \textbf{Parameter 26: Number of distinct widgets}, $n^*_\text{widgets}$

$n^*_\text{widgets}$ is the number of distinct widgets in set $[U_0,\dots,U_{n_\text{widgets}-1}]$ for the time-sliced widgetization of $U$. This means one can always write $n^*_\text{widgets} \leq n_\text{widgets}$. This is the number of widgets that FTQC needs to compile.

\item \textbf{Parameter 27: decoder tock}, $t^{\rm decoder}_{\rm tock}$

For our proposed superconducting FT architecture, we assume it is feasible to engineer generic decoders utilizing modern GPU or FPGA hardware and methods such as MWPM or neural networks.
$t^{\rm decoder}_{\rm tock}$ is the time required for the decoder to complete decoding $d$ QEC sweeps expressed in seconds. This can be approximated as $t^{\rm decoder}_{\rm tock} = dt_\text{decoder}$. We assume it is feasible to engineer generic GPU or FPGA-based decoders with a decoding tock or delay in the same order of magnitude as $t^{\rm decoder}_{\rm tock}$ for superconducting architecture. Here, we consider a characteristic decoder cycle time of $t_\text{decoder} = 1~\mu$s.

\item \textbf{Parameter 28: Intra-module quantum tock for graph processing, $t^{\rm intra}_{\rm tock}$}

The time needed for FTQC to sequentially perform $d$ QEC sweeps for all intra-module quantum operations, which prepare or consume the subgraphs, is computed as $t_{
\rm intra} = 8dt$. Here, $t$ represents the characteristic (dominant) gate time that sets the architectural tock for all intra-modular operations. The factor of 8 arises from the fact that in our physical architecture, each surface code sweep for syndrome extraction circuits necessitates one initialization, two Hadamards, four entangling gates, and one measurement, all of which are assumed to take a duration of $t$ (see, for example, Fig.~9 of 
\cite{Fowler2012}).

\item \textbf{Parameter 29: Intra-module quantum tock for $T$-factories}

The time required for $T$-factories to distill a $T$ state performing $d$ QEC sweeps sequentially. This can be computed as $8Ct$ for $T$-factories built from rotated surface code patches.

\item \textbf{Parameter 30: Total number of available physical qubits}, $n_{\rm avail-phys}$

Total number of physical qubits available in all $2n^\text{per-leg}_\text{modules}$ modules of the FTQC.

\item \textbf{Parameter 31: Number of available logical qubits per module,} $n_\text{avail-logical}$

The number of logical qubits available per module, which is allocated to different architectural components and quantum operations. This can be computed as $n_\text{avail-logical} = l_\text{edge}^2$.

\item \textbf{Parameter 32: Total number of \emph{unallocated logical} qubits}, $n_\text{unalloc-logical}$

Considering all modules and operations of the FTQC, the total number of unallocated \emph{logical} qubits. This parameter can be computed as $2n_\text{unalloc-logical} n^\text{per-leg}_\text{modules}$, where $n_\text{unalloc-logical} = l_\text{edge}^2 - 2\left \lceil\frac{\Tilde{n}_\text{logical}}{n^\text{per-leg}_\text{modules}} \right\rceil - l_\text{transfer-bus} - n_\text{T-factories}\left \lceil\frac{L_\text{length}}{\sqrt2d}\right \rceil \left \lceil\frac{L_\text{width}}{\sqrt2d}\right \rceil$.

\item \textbf{Parameter 33: Total number of \emph{unallocated physical} qubits}

Considering all modules and operations of the FTQC, the total number of unallocated \emph{physical} qubits. This parameter can be computed as $4n_\text{unalloc-logical} n^\text{per-leg}_\text{modules}d^2$.

\item \textbf{Parameter 34: Total number of \emph{allocated logical} qubits}

Considering all modules and operations of the FTQC, the total number of allocated \emph{logical} qubits to different architectural components. This parameter can be computed as $2n_\text{alloc-logical} n^\text{per-leg}_\text{modules}$, where $n_\text{alloc-logical} = 2\left \lceil\frac{\Tilde{n}_\text{logical}}{n^\text{per-leg}_\text{modules}}\right\rceil + l_\text{transfer-bus} + n_\text{T-factories}\left \lceil\frac{L_\text{length}}{\sqrt2d}\right \rceil\left \lceil\frac{L_\text{width}}{\sqrt2d}\right \rceil$.

\item \textbf{Parameter 35: Total number of \emph{allocated physical} qubits}

Considering all modules and operations of the FTQC, the total number of allocated \emph{physical} qubits to different architectural components. This parameter can be computed as $4n_\text{alloc-logical} n^\text{per-leg}_\text{modules}d^2$.

\item \textbf{Parameter 36: Number of concurrent decoding cores at consumption stage}, ${\rm cores}_{{\rm consump}}^{{\rm decoding}}$

In our superconducting architecture, we assume that generic reference classical decoders, equipped with state-of-the-art GPU or FPGA cores, perform decoding cycles. These receive the syndrome and output possible corrective actions, which can, for example, be performed at the same time as measurements at the ends of consumption steps.

Quantum operations do \emph{not} need to wait for the generic decoders we have assumed to finish their cycles to perform corrective measurements. One can always consider a staggered type of architecture, where classical and quantum units work together at each step. Moreover, since decoder tocks are typically much longer than quantum tocks, the best strategy is to let as many \emph{concurrent} decoding cores complete their cycles as possible at each consumption step. In this way, the decoding process only adds an overall delay to the consumption stage runtime on top of purely quantum operations. There would be a maximum number of cores that can run concurrently at any consumption step.

We define $\text{cores}_{\text{consump}}^{\text{decoding}}$ as the maximum number of concurrent decoding cores running at any consumption step. Hence, consumption steps often dominate intramodule runtime, so we set $\text{cores}_{\text{consump}}^{\text{decoding}}$ as the available number of concurrent cores for all decoding steps, including graph preparation, teleportation, consumption, or $T$ injection operations. This is calculated as $\text{cores}_{\text{consump}}^{\text{decoding}} = \left \lceil \frac{t_{\text{tock}}^{\text{decoder}}}{t^\text{intra}_\text{tock}} \right \rceil$.

\item \textbf{Parameter 37: Total QPU area}

This parameter represents the total area occupied by all QPU modules on both legs of the ladder macro-architecture.

\item \textbf{Parameter 38: Total number of couplers}

This parameter represents the total number of adjustable couplers required for all modules on both legs of the ladder macro-architecture.

\item \textbf{Parameter 39: Decoding power at consumption stage,} $\text{POW}^\text{decoding}_{\text{consump}}$

The Decoding power used during the consumption stage is based on a 100W reference decoding core (aligned with modern GPU or FPGA-based units). Therefore, we can write down $\text{POW}^\text{decoding}_{\text{consump}} = 100 \text{cores}^\text{decode}_{\text{consump}}$~W.

\item \textbf{Parameter 40: Power dissipation at 4K stage,} $P_{\text{Diss},\text{4K}}$

Power dissipation for the $4$-Kelvin stages of all (cryo-)modules in Watts.

\item \textbf{Parameter 41: Power dissipation at 20mK stage,} $P_{\text{Diss},\text{20mK}}$

Power dissipation for the $20$-milliKelvin stages of all (cryo-)modules in Watts.

\item \textbf{Parameter 42: Total graph consumption time, $t^{\rm tot}_{\rm consump}$}

The total time to consume subgraphs across all consumption schedule steps is expressed in seconds. $t^{\rm tot}_{\rm consump}$ includes the delays required to distill additional $T$-states or prepare the subgraphs, while excluding handover runtime and decoding delays. This is calculated as $t^{\rm tot}_{\rm consump} = 8td(L^0_\text{prep} + N^\text{seq}_\text{consump}) + t^\text{delay}_\text{distill} + t^\text{delay}_\text{prep}$, as detailed in \cref{section:t-gate-error}.

\item \textbf{Parameter 43: Total inter-modular graph handover time,} $t^\text{tot}_\text{handover-inter}$

The total inter-modular time for Bell-state teleportations to hand over (stitch) subgraphs across all scheduling steps. This is the time to teleport the output nodes on the current subgraph on the quantum register on one leg to the input nodes of the subgraph on the other leg for all steps. This is calculated as $t^\text{tot}_\text{handover-inter} = \sum_{i=0,\dots,n_\text{widgets}-2} t^i_\text{handover-inter}$ and was detailed in \cref{section:t-gate-error}.

\item \textbf{Parameter 44: Overall $T$-distillation delay,} $t^{\text{distill-delay}}_{tot}$

$t^{\text{distill-delay}}_{tot}$ is the overall delay included in the total consumption time, $t^{\rm tot}_{\rm consump}$, to distill additional $T$-states in the $T$-factories in the same modules to complete consumption-stage measurements. See \cref{section:t-gate-error} for details.

\item \textbf{Parameter 45: Overall graph preparation delay,} $t^{\text{prep-delay}}_{tot}$

$t^{\text{prep-delay}}_{tot}$ is the overall delay included in the total consumption time, $t^{\rm tot}_{\rm consump}$, to prepare the next subgraph in the schedule. The FTQC pipeline cannot proceed from the current consumption step until the next subgraph is prepared in the next set of modules on the other leg of the ladder. See \cref{section:t-gate-error} for details.

\item \textbf{Parameter 46: Overall decoding delay,} $t^{\text{decode-delay}}_{tot}$

$t^{\text{decode-delay}}_{tot}$ represents the total decoding delay accumulated from all graph consumption steps, which we must add to the overall FT hardware time. We assume there are \emph{no} decoying delays during subgraph hand-over operations due to $t_{\text{tock}}^{\text{decoding}} \leq 8t_{\text{inter}}d$. We can determine the total decoding delay, given that $t_{\text{tock}}^{\text{decoding}} > t_{\text{tock}}^{\text{quantum}}$ and $t_{\text{tock}}^{\text{decoding}} > 8Ct$, using $t^{\text{decode-delay}}_{\text{tot}} = \left (8td(L^0_\text{prep} + N^\text{seq}_\text{consump} ) + t^\text{delay}_\text{prep}\right) \left (t_{\text{tocks}}^{\text{decoding}} - t_{\text{tock}}^{\text{quantum}} \right) + \left \lceil\frac{t^\text{delay}_\text{distill}}{8Ct}\right \rceil \left (t_{\text{tocks}}^{\text{decoding}} - 8Ct \right)$.

\item \textbf{Parameter 47: Total wall-time over a single FT algorithm step,} $t_\text{hardware}^\text{total}$

$t_\text{hardware}^\text{total}$ is our estimated \emph{upper-bound} for the wall-time accumulated from all non-simultaneous components running in modules on both legs and coherent interconnects of the architecture of \cref{sec:hardware-architecture} over a single FT algorithm step, $U$. Algorithmic step should not be confused with $n_\text{widgets}$ widgetization steps of $U$. (An algorithmic step occurs when the user requires $n_\text{algo-reps}$ repetitions of $U$ on the FTQC.) This wall time includes all inter- and intra-tocks required at the consumption stages, delays for graph preparation and $T$-distillation, subgraph handover time, and decoding delay. In other words, this is our overall time cost for one step, which can be formulated as $t_{\text{hardware}}^\text{total} = t^\text{total}_{\text{intra}} + t^\text{total}_\text{handover-inter} + t_{\text{decode-delay}}^\text{total}$. See \cref{section:t-gate-error} for details.

\item \textbf{Parameter 48: Total FT algorithm time,} $t_{\text{FT}}^\text{total}$

RRE allows users to specify the number of times, $n_\text{algo-reps}$, with which they would like to repeat an exact copy of the complete algorithm $U$ back-to-back. For these cases, the total FT time is calculated as $t_\text{FT}^\text{total} = n_\text{algo-reps} t_\text{hardware}^\text{total}$. 

\item \textbf{Parameter 49: Total energy consumption,} $E_\text{tot}$

We report estimates for $E_\text{tot}$, the \emph{upper-bound} on the total energy usage of the FTQC, considering all modules, quantum and decoding operations, and algorithmic steps, which, to our knowledge, is unique to RRE.

We identify \emph{three} key power-consuming units that dissipate energy during FT computations. These units include decoding cores with a fixed TDP of $P^{\text{decoding}}_{\text{per-core}}=100\text{W}$ on duty during all QEC operations. We assume a fixed number of concurrent decoding cores equal to ${\rm cores}_{{\rm consump}}^{{\rm decoding}}$ are assigned to all decoding opertions, which includes graph preparation, consumption, handover, and $T$-distillation across full $t^{\text{FT}}_\text{total}$.  This means we can approximate an upper bound to decoding energy consumption per core as $P^{\text{decoding}}_{\text{per-core}} t^{\text{FT}}_\text{total}$.
 
We also include 4\,K units with a TDP of $P_{\text{dissip},\text{4K}}$ and assume a single cooling efficiency factor of $\eta_{\text{4K}}=500$ for this stage. Additionally, we have lower-temperature MXC (20\,mK) units with a TDP of $P_{\text{dissip},\text{20mK}}$ and assume a single cooling efficiency factor of $\eta_{\text{20mK}}=10^9$ for this stage. Each unit comprises several power-consuming stages, and we determine these values via thermal analysis of the signal chain (see \cref{sec:thermal} and \cite{RRE, Martinez_2010, Raicu_2025} for details). Overall, we obtain $E_{tot} = (P^{\text{decoding}}_{\text{per-core}} \text{cores}_{\text{distill}}^{\text{decoding}} + \eta_{\text{4K}} P_{\text{dissip},\text{4K}}$ $+ \eta_{\text{20mK}} P_{\text{dissip},\text{20mK}} )t^{\text{FT}}_\text{total}$.

\end{itemize}

%% file: sections/widgetization.tex
In \cref{sub:stictching}, we established that our resource estimation tooling supports the widgetization of \verb|Cirq| circuits. The primary reason for such a widgetization tool is that fully unrolling a large circuit into its single- and two-qubit operations for a utility-scale application can place severe pressure on memory and compute resources. Moreover, for resource estimation using RRE, it is unnecessary and inefficient to revisit, re-examine, and recompile subcircuits that have already been prepared for RRE's use. To understand the widgetization process, a few key concepts about the nature of \verb|Cirq| circuits need to be defined.

\subsubsection{Structure of \texttt{Cirq} circuits}

\paragraph*{Definitions} 
For this discussion, we will follow the definitions of quantum circuit objects in the \verb|Cirq| documentation. As such, we define a quantum gate as \enquote{an effect that can be applied to a collection of qubits} (logical qubits for our purpose). A quantum operation is determined to be a pair consisting of a quantum gate and a list of logical qubits that the gate acts on. A moment is a collection of operations, each acting on a disjoint set of logical qubits. If one imagines a time-ordered list of operations, a moment corresponds to a single time \enquote{slice} where operations are being carried out in parallel. Finally, a quantum circuit is a time-ordered sequence of operations that can, if desired, be organized into a time-ordered sequence of moments.

\paragraph*{Nested circuits}
In addition to a library of common gates, \verb|Cirq| allows users to define custom gates by providing either a matrix specifying the unitary matrix defining the effect of the gate on logical qubits, or by providing instructions on how to create an ordered list of operations from a list of qubits to act on. The latter of the two options can be thought of as instructions for \enquote{decomposing} a gate into constituent gates. 
By providing a method to define gates in terms of other gates, \verb|Cirq| naturally provides the infrastructure to create circuits that are a hierarchy of nested subcircuits. We will focus on circuits comprised of this latter type of \enquote{decomposable} gates.

\begin{figure*}[!htb]
\begin{subfigure}{.45\linewidth}
    \includegraphics[width=\linewidth]{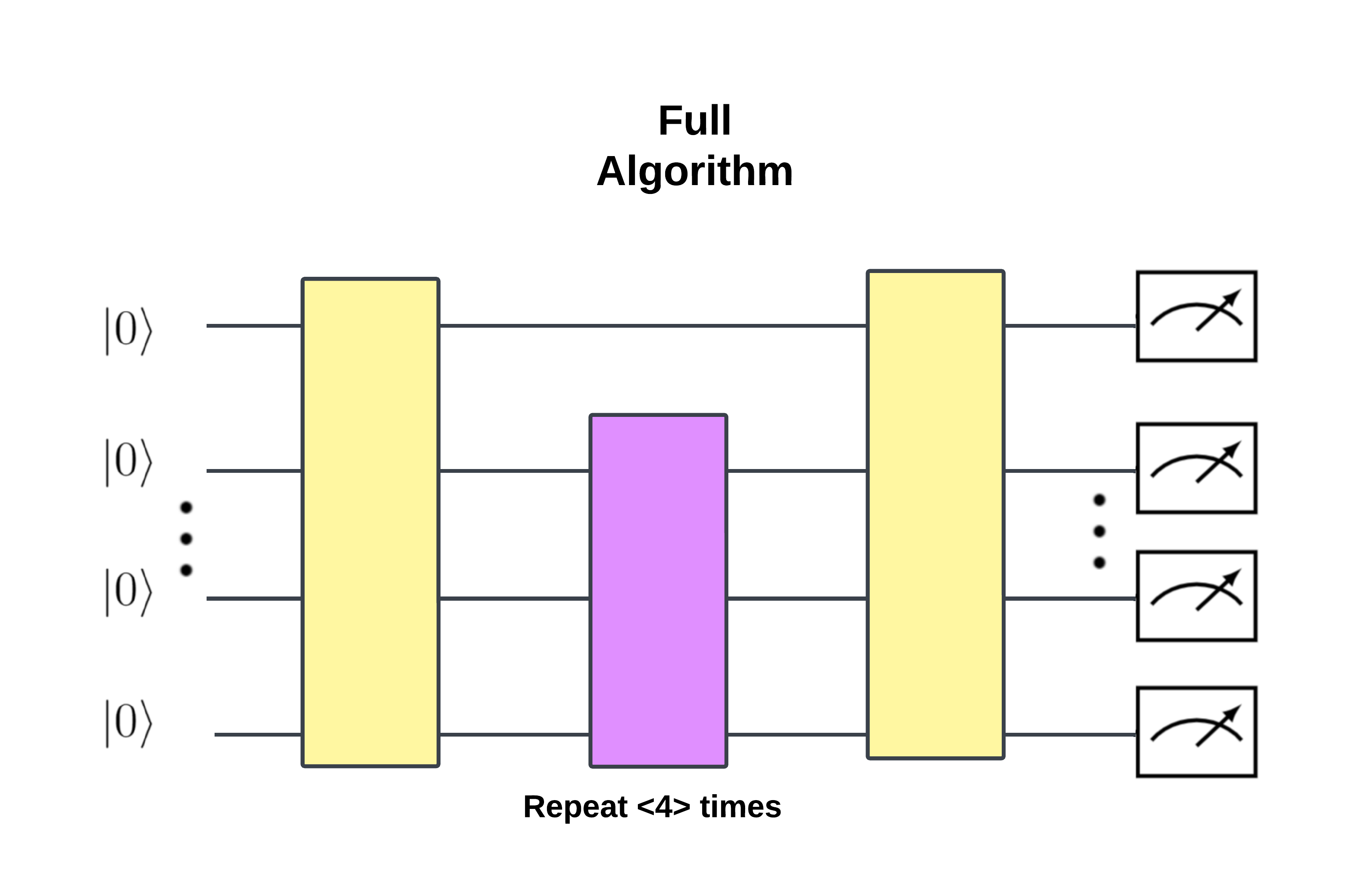}
\caption{Example algorithm consisting of three subcircuits. Subcircuit A (magenta) that is repeated $4$ times and Subcircuit B (yellow) that occurs before and after the $4$ repetitions of Subcircuit B.}
\label{fig:example_full_algorithm}
\end{subfigure}
\hfill
\begin{subfigure}{.45\linewidth}
    \includegraphics[width=\linewidth]{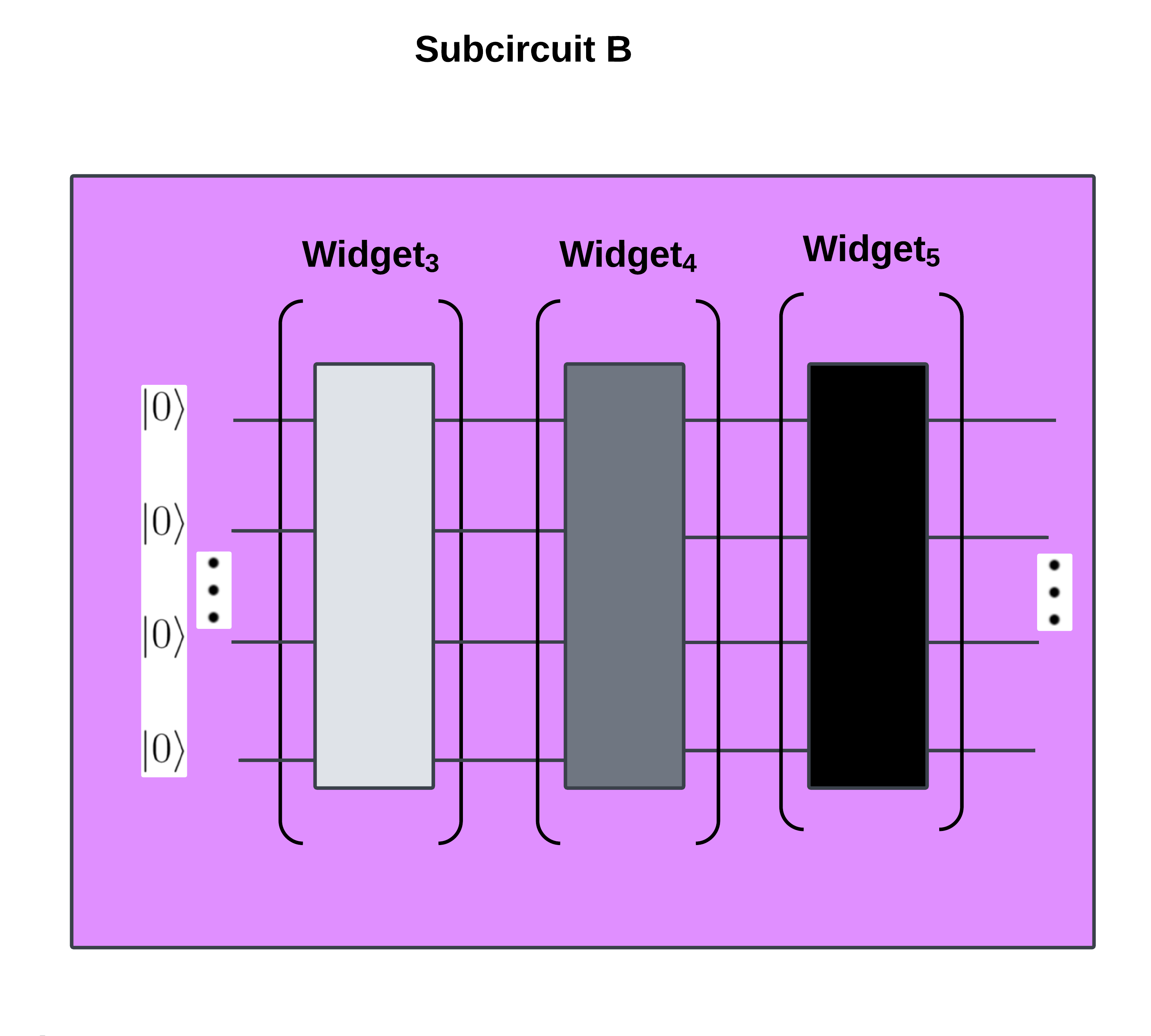}
\caption{Splitting of Subcircuit B into three smaller subcircuits via slicing. In this example, these slices meet the criterion for not requiring further splitting and, as such, are considered widgets.}
\label{fig:subcircuitB}
\end{subfigure}
\hfill
\\
\begin{subfigure}{.95\linewidth}
    \includegraphics[width=\linewidth]{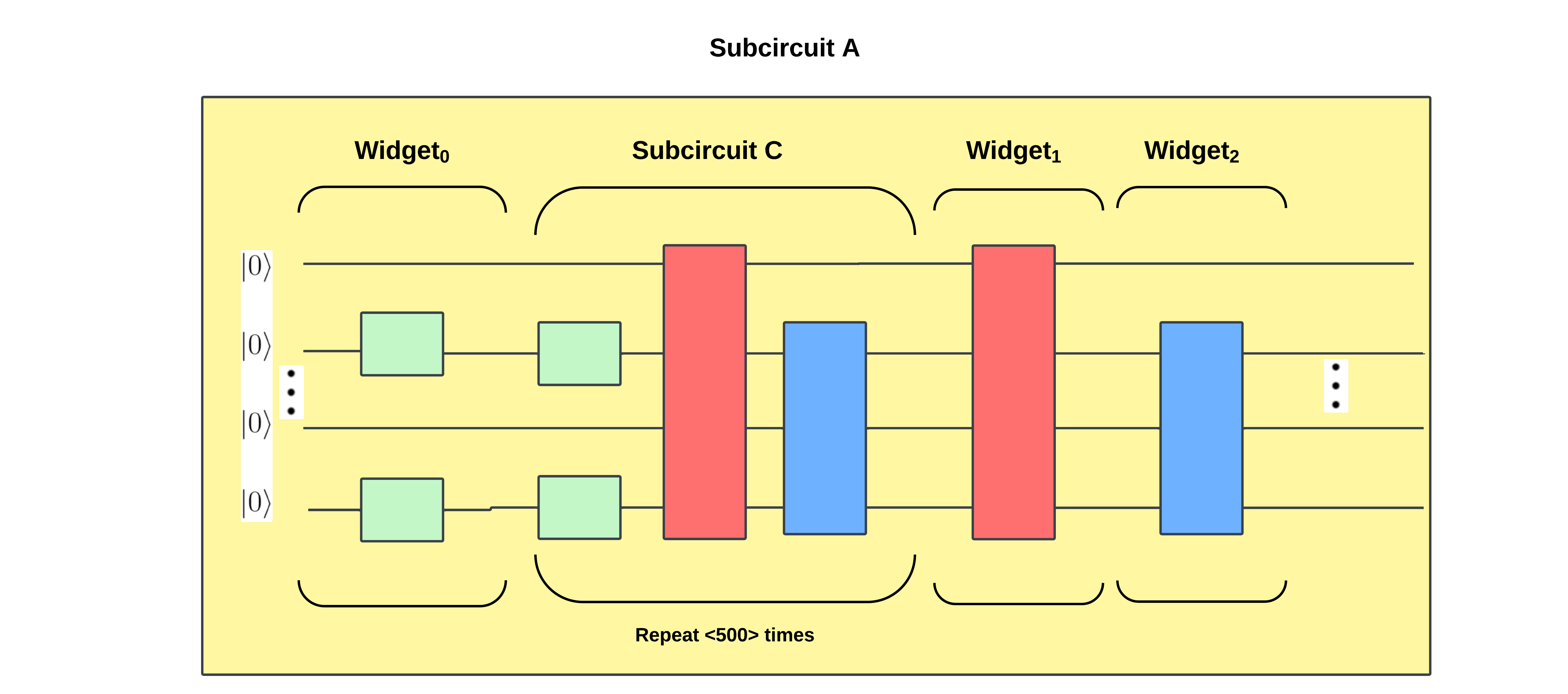}
    \caption{Division of Subcircuit A. Subcircuit A consists of Widget$_0$, $500$ repetitions of Subcircuit C, then Widget$_1$, and finally Widget$_2$.}
\label{fig:subcircuitA}
\end{subfigure}
\caption{Widgetization of an example circuit. The full circuit is shown in the panel
(\emph{a}), while its subcircuits and widgets are shown in
(\emph{b}) and (\emph{c}).}
\label{fig:example_circuits}
\end{figure*}

Utilizing this nested subcircuit structure, a user may construct a relatively simple-looking circuit that consists of billions, if not trillions, of single and two-qubit gates when fully unrolled to an alphabet of indivisible operations. This is particularly true where some subcircuits consist of many repetitions of smaller subcircuits. The strategy employed during widgetization is to identify moderately sized, recurring subcircuits (called \enquote{widgets}) by recursively decomposing and analyzing the nested subcircuit hierarchy. An example of a nested circuit structure is in \cref{fig:example_circuits}. 

\paragraph*{Division of subcircuits}
One may define a measure of subcircuit size, such as the number of gates or active logical qubits. Using this measure, one may decide that any particular subcircuit is \enquote{too large} to be a \enquote{widget}, and therefore the subcircuit needs to be divided into smaller constituent subcircuits. We find it helpful to define two distinct methods for dividing a subcircuit into its constituent parts. 

The first, and most natural, is to \enquote{decompose} the subcircuit into operations by using the decomposition instructions provided by the subcircuit's gates. This allows the subcircuit to be divided into constituent subcircuits, each comprised of a single operation. An example of this type of splitting is shown in \cref{fig:example_full_algorithm} and \cref{fig:subcircuitA}. 

The second type of division is to \enquote{slice} the subcircuit into collections of moments. The second method becomes essential when a subcircuit that involves many active logical qubits decomposes into many (thousands or tens of thousands) of operations, each acting on a few qubits. For example, a circuit consisting of thousands of Toffoli gates, each acting on 3 of a hundred logical qubits. In these cases, the sheer number of constituent operations slows analysis down and provides subcircuits that are far too small to be considered widgets. By slicing, the subcircuit can be divided into fewer constituent subcircuits of moderate size. An example of this type of division is shown in \cref{fig:subcircuitB}.

\paragraph*{Graph State Equivalence}

To accelerate the widgetization process, the user can provide rules that map subcircuits to unique identifiers, defining classes of \enquote{graph-state-equivalent} subcircuits. If two subcircuits are mapped to the same identifier, it is assumed that they will be compiled into graph states that consume equivalent resources. Only one representative of each graph state equivalence class must be split further and/or compiled. This yields substantial savings in classical processing time when performing both widgetization and resource estimation.

\subsubsection{Widgetization}

To widgetize $U$ without fully unrolling down to single and two-qubit operations, we perform the following steps:
\begin{enumerate}
\item Construction of a subcircuit dependency graph. This describes the circuit's nested structure as a directed graph.
\item Enumeration of widgets and \enquote{stitches} indicating when one widget is executed before another.
\end{enumerate}
Each of these steps will be described below.

\subsubsection{Construction of the Subcircuit Dependency Graph}

\begin{figure*}[!htb]
    \centering
    \includegraphics[width=0.9\textwidth]{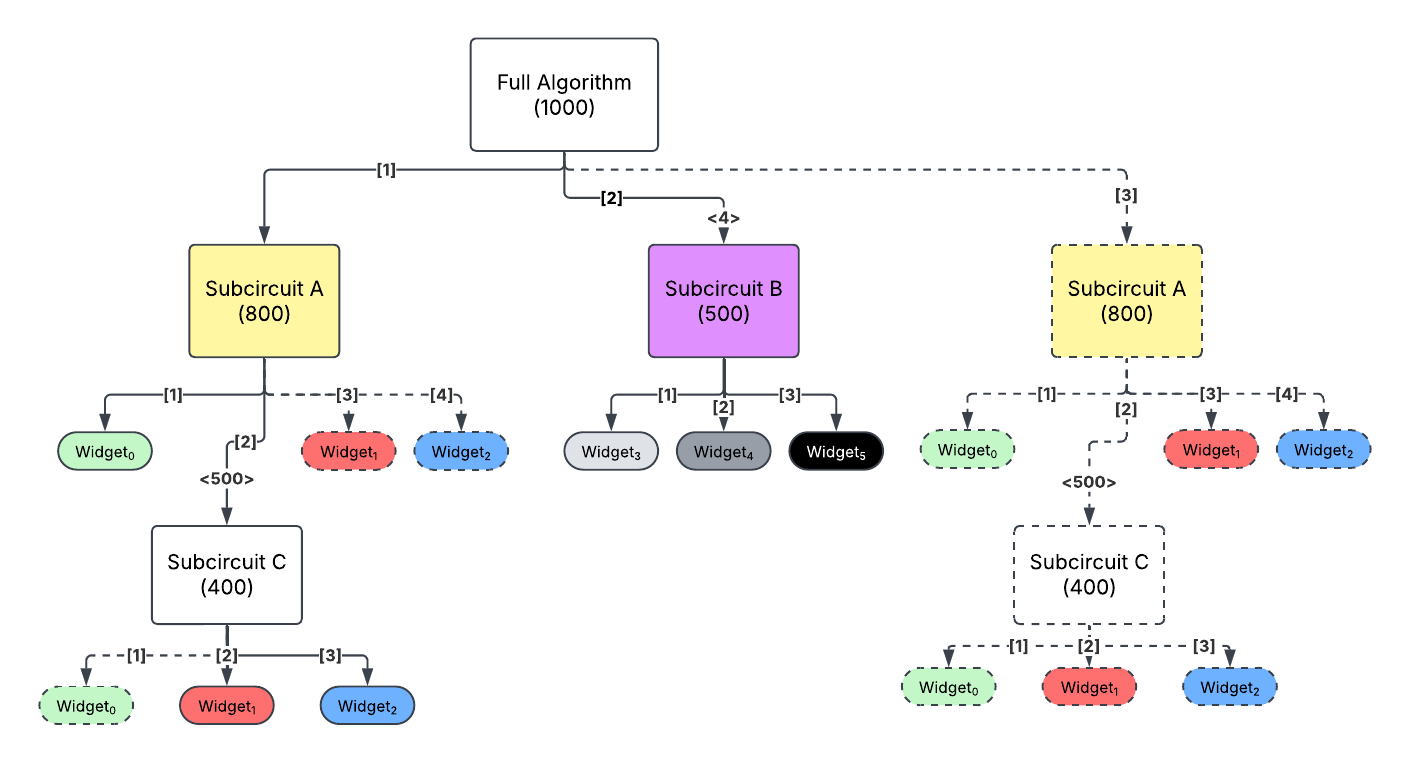}
    \caption{Subcircuit Dependency Graph of the example circuit described in \cref{fig:example_circuits}. Dashed lines indicate subgraphs that are graph state equivalent to a previously seen subcircuit and, therefore, are not directly stored in the dependency graph. Quantities in brackets are active qubits in the subcircuit. Numbers in square brackets indicate the ordering of the edges, while numbers in angle brackets indicate repetitions. An example of a splitting criterion that may have created this graph is to split a subcircuit if its number of active qubits is greater than or equal to $400$.}
    \label{fig:dependency_graph}
\end{figure*}

The first step in the widgetization process is constructing a directed graph (not to be confused with ASG states) that represents the original circuit's nested subcircuit structure. In this directed graph, the vertices represent subcircuits. A directed edge connects a parent vertex to a child vertex if the child vertex represents one of its parent's constituent subcircuits. Directed edges between parent and child vertices are \textit{ordered} to execute the subcircuits. This graph is created by recursively splitting subcircuits into smaller ones. This recursive splitting continues until a subcircuit meets specific user-defined criteria. The vertices that represent subcircuits and are not split are called leaf vertices. 

An example of a criterion that would prevent a subcircuit from being split is if its number of logical qubits or $T$-gates falls below a user-specified threshold. This recursive splitting is performed in a depth-first manner.Figure~\ref{fig:dependency_graph} shows an example of a graph created from the example circuit described in \cref{fig:example_circuits}.

\subsubsection{Enumeration of widgets and stitches}

Once a directed graph representing the nested structure of the original circuit is created, it is used to identify widgets and count how many times they appear during circuit execution. To account for teleportation overheads, we also count the number of times any ordered pair of widgets (called \enquote{stitches}) would be required during execution. 

Widgets are identified as unique leaf nodes in the Subcircuit Dependency Graph. Stitches are identified as ordered pairs of widgets if leaf nodes are written out during a depth-first search of the Subcircuit Dependency Graph while preserving the ordering of the edges connecting parent and child nodes.

From the standpoint of formal language theory, circuits constructed this way can be viewed as variables in a Context-Free Grammar, with the productions corresponding to the gate decomposition rules. At the same time, the Subcircuit Dependency Graph can be regarded as a parse tree. Widgets are viewed as terminals, and stitches are interpreted as ordered pairs of terminals. The number of widgets and stitches is then counted based on their occurrence in the yield of the parse tree.

%% file: sections/unitary-verification.tex
Robust verification of our compilation and estimation methodology across multiple layers was a priority in the design of RRE. To this end, we have developed an exact verification protocol and a series of tests 
to precisely check the correctness of individual subgraphs, $\{g\}$, and connected schedules, $\{S_\text{consump}\}$ and $\{S_\text{meas}\}$, as long as the widgets originated from small circuits with a few logical qubits. The protocol includes a step for exact simulation and matrix manipulation in the quantum algorithm, which imposes a widget-size restriction. Recall that every logical qubit involving nontrivial non-Clifford operations can correspond to 10's of physical qubits that need exact simulations (one may extend our protocol to 10's of logical qubits by performing near-exact tensor-network-based simulations).

This is how our verification protocol works step by step: The input logical unitary, $U = U_{n_\text{widgets}-1}\dots U_1U_0$, a small algorithm featuring a variety of logical operations (must involve non-Cliffords), is provided to the test module. $n_\text{widgets}$, $n_\text{input}$, and the number of non-Clifford operations per widget are kept small to keep the test simple and exactly simulatable. Given $U$, RRE will then run its usual estimation pipeline to generate the sets of 
\begin{equation}
\begin{split}
 &\{g_i, S_\text{frames}^{g_i}, S_\text{input}^{g_i}, S_\text{output}^{g_i}, n^{g_i}_\text{logical}, S_\text{consump}^{g_i},
 S_\text{meas}^{g_i}, \\
 &\ket{\text{state}}^\text{Cabaliser}_i\}~\text{for}~i\in\{0,\dots,n_\text{widgets}-1\}~.
\end{split}
\end{equation}

Recall that RRE, by default, initializes from an all-zero state as 
$\ket{\text{input}} \equiv \ket{s_0=0,s_1=0,\cdots,s_{n_\text{input}-1}=0}$.
Moreover, the RRE output estimation results for $U$ do not play a role in this exact verification protocol. Now, entirely independent of the operations of the compiler, one can construct the exact inverse of the original unitary $U$ for small-size cases through Kronecker matrix multiplications, namely~$U^\dagger$. (As an example, if the widget $U_1$ is the unitary for a logical \texttt{QFT4}, we then need to construct $U_1^\dagger$, i.e.~\texttt{InverseQFT4}, manually.)
If we now apply $U^\dagger$ to the final output state that RRE was generated through Cabaliser, $\ket{\text{output}} \equiv \ket{\text{state}}^\text{Cabaliser}_{n_\text{widget}-1}$, by the unitary principle, we should get back the initial state. That is $ U^\dagger  \ket{\text{state}}^\text{Cabaliser}_{n_\text{widgets}-1} = \ket{s_0=0,s_1=0,\cdots,s_{n_\text{input}-1}=0}$ will verify the correctness of graphs and schedules generated by RRE.  

In the RRE test modules and procedures, we simulated and verified the graphs and schedules generated by the software, following the protocol above for selected small input algorithms. These tests included widgets with non-Clifford instructions such as \texttt{QFT3} and \texttt{TOFFOLI3}, in addition to various Clifford operations.

%% file: sections/noise-and-scaling.tex
Section \ref{sec:MWPM} established that our resource estimation framework requires a physical-to-logical error scaling law to obtain resource counts. This error scaling law is given by, \cref{eq:log_to_phys},
$p_C = \kappa\left (\frac{p}{p_\text{thresh}} \right)^{(d + 1)/2},$ where $p_C$ is the logical error rate per cycle, $p$ is the homogeneous physical error rate, and $\kappa$ and $p_\text{thresh}$ are the coefficients to be fitted. Below, we describe the computational pipeline we used to determine the scaling coefficients, as well as the exemplar device noise model used for the results in this work.

\subsection{Device noise model}
For a preliminary demonstration of RRE, we chose a simple noise model that also shows the richness and flexibility the tool offers researchers when considering other devices, circuits, and algorithms. To this end, we split the noise model into two classes: one simple model for single-qubit gates and another, more realistic model for two-qubit gates. 

For single-qubit gates, we chose a depolarizing noise channel, parameterized by the homogeneous physical error rate $p$. Admittedly, this choice sacrifices realism for ease of implementation, but for an initial baseline demonstration, we considered it appropriate. The noise model for two-qubit gates was chosen to be weight-2 Pauli-error channels, with error rates derived from device-specific gate simulations. As mentioned in \cref{sec:MWPM}, this manuscript obtains resource counts for FT execution of circuits protected by a square planar $iSWAP$
surface code. As iSWAP gates were the only two-qubit gate in the surface code, the Pauli Error channel rates were derived from master equation simulations of parametric $iSWAP$ gates. The simulations were based on the two-qubit system with one tunable coupler described in \cite{sete2024error}. Process tomography was carried out to obtain the Pauli Error rates from the difference between the ideal $iSWAP$ Pauli Transfer Matrix and the simulated Pauli Transfer Matrix.

\subsection{Scaling coefficient pipeline}
\label{sec:scaling-coeff-pipeline}

\begin{figure}[!hbt]
    \centering
    \includegraphics[width=0.99\linewidth]{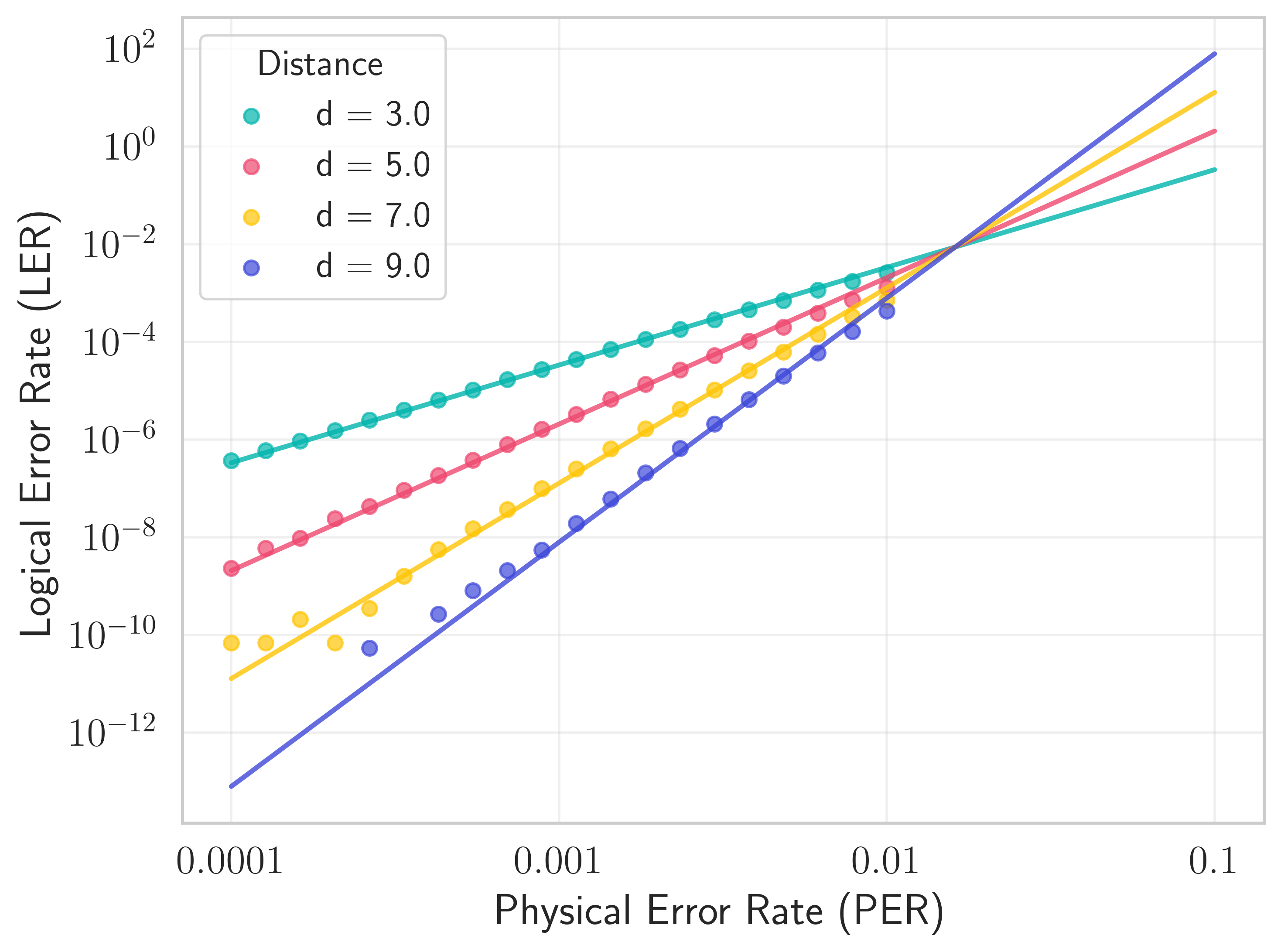}
    \caption{The Logical Error Rate (LER) to Physical Error Rate (PER) scaling obtained via the Scaling Coefficient pipeline in Appendix D \ref{sec:scaling-coeff-pipeline}. Dots are values calculated by the pipeline, and solid lines are lines of best fit, resulting in the power law given in \cref{eq:log_to_phys} with fit parameters of $\kappa[\text{MWPM}]=0.009$ and $p_\text{thresh}[\text{MWPM}]=0.016$.}
    \label{fig:ler-scaling}
\end{figure}

To determine the coefficients for the logical-to-physical error rate scaling law, \cref{eq:log_to_phys}, we performed a series of surface code simulations and coefficient fittings by using \texttt{Sinter} \cite{sinter} and \texttt{Stim} \cite{gidney2021stim}. 

Given the stochastic nature of the simulations, we conducted convergence studies to ensure the stability of the coefficients across a large number of shots. This was done by simulating and fitting coefficients for an increasing number of shots and a maximum number of errors, until the coefficients converged. The number of shots and maximum number of errors were $2^k \times 100,000$ and $2^k \times 10000$ respectively, where $k$ ranged from $1$ to $11$.

For each fixed number of shots and maximum number of errors, we swept over 20 values for $p$ ranging from $10^{-2}$ to $10^{-4}$ and distances of $d \in \{3,5,7, 9\}$. For each $p$ and distance combination, noisy surface code circuits were simulated for $10 \times d$ rounds (cycles). Noisy circuits were constructed by starting with a noiseless $iSWAP$ surface code circuit written in \texttt{Stim} provided by \cite{midout}. We then used the aforementioned noise model to add noise to this circuit in \texttt{Stim}. Syndromes were decoded using \texttt{pymatching} MWPM decoder \cite{pymatch}, followed by the calculations of the logical error rates (per cycle). Observed data were then fit to the two-coefficient scaling law. Fitted coefficients were observed to converge to two significant figures by $k=11$, and the coefficients for $k=11$ were used in all downstream resource estimates of \cref{sec:results}. The fit and logical error rates for $k=11$ are shown in \cref{fig:ler-scaling}.